\documentclass[superscriptaddress,
 amsmath,amssymb, twocolumn,
a4paper,10pt,
accepted=2024-08-29]{quantumarticle}
\pdfoutput=1
\usepackage[utf8]{inputenc}

\usepackage[numbers,sort&compress]{natbib}

\usepackage{amsmath, amsfonts, amsthm}
\usepackage{mathtools}
\usepackage{caption}
\usepackage{subcaption}
\usepackage{graphicx}
\usepackage[breaklinks, colorlinks=true, urlcolor=blue,
anchorcolor=blue, citecolor=blue, filecolor=blue, linkcolor=blue,
menucolor=blue, linktocpage=true, pdfproducer=medialab,
pdfa=true]{hyperref}
\usepackage{cases}
\usepackage{empheq}
\usepackage{latexsym}
\usepackage{amsbsy}
\usepackage{amssymb}
\usepackage{epstopdf}
\usepackage[usenames]{color}
\usepackage{float}
\usepackage[normalem]{ulem}
\usepackage{physics}
\usepackage{comment}
\usepackage{tikz}
\usepackage[vcentermath]{youngtab}
\usepackage{ytableau}
\ytableausetup{aligntableaux=center}
\usepackage{ragged2e}

\usepackage{tikz}
\usetikzlibrary{shapes.geometric, arrows}
\tikzstyle{process} = [rectangle, rounded corners, minimum width=2cm, minimum height=1cm,text centered, text width=2cm, draw=black, fill=red!20]
\tikzstyle{startstop} = [rectangle, rounded corners, minimum width=1cm, minimum height=1cm, text width=2.5cm, text centered, draw=black, fill=orange!30]
\tikzstyle{output} = [rectangle, rounded corners, minimum width=1cm, minimum height=1cm, text width=3cm, text centered, draw=black, fill=green!30]
\tikzstyle{decision} = [diamond, minimum width=3cm, minimum height=1cm, text centered, draw=black, fill=green!30]
\tikzstyle{arrow} = [thick,->,>=stealth]
\tikzstyle{question} = [rectangle, rounded corners, minimum width=2cm, minimum height=1cm,text centered, text width=2cm, draw=black, fill=blue!20]

\usepackage{pst-node}
\usepackage{tikz-cd}

\newsavebox{\measurebox}

\begin{document}

\title{Repeated measurements on non-replicable systems and their consequences for Unruh-DeWitt detectors}
\author{Nicola Pranzini}
\address{Department of Physics, P.O.Box 64, FIN-00014 University of Helsinki, Finland}
\address{QTF Centre of Excellence, Department of Physics, University of Helsinki, P.O. Box 43, FI-00014 Helsinki, Finland}
\address{InstituteQ - the Finnish Quantum Institute, University of Helsinki, Finland}
\email{nicola.pranzini@helsinki.fi}
\author{Guillermo García-Pérez}
\address{QTF Centre of Excellence, Department of Physics, University of Helsinki, P.O. Box 43, FI-00014 Helsinki, Finland}
\address{InstituteQ - the Finnish Quantum Institute, University of Helsinki, Finland}
\address{Algorithmiq Ltd, Kanavakatu 3C 00160 Helsinki, Finland}
\address{Complex Systems Research Group, Department of Mathematics and Statistics, University of Turku, FI-20014 Turun Yliopisto, Finland}
\author{Esko Keski-Vakkuri}
\address{Department of Physics, P.O.Box 64, FIN-00014 University of Helsinki, Finland}
\address{Helsinki Institute of Physics, P.O.Box 64, FIN-00014 University of Helsinki, Finland}
\address{InstituteQ - the Finnish Quantum Institute, University of Helsinki, Finland}
\author{Sabrina Maniscalco}
\address{QTF Centre of Excellence, Department of Physics, University of Helsinki, P.O. Box 43, FI-00014 Helsinki, Finland}
\address{InstituteQ - the Finnish Quantum Institute, University of Helsinki, Finland}
\address{Algorithmiq Ltd, Kanavakatu 3C 00160 Helsinki, Finland}
\address{QTF Centre of Excellence, Department of Applied Physics, School of Science, Aalto University, FI-00076 Aalto, Finland}
\address{InstituteQ - the Finnish Quantum Institute, Aalto University, Finland}

\begin{abstract}
The Born rule describes the probability of obtaining an outcome when measuring an observable of a quantum system. As it can only be tested by measuring many copies of the system under consideration, it does not hold for non-replicable systems. For these systems, we give a procedure to predict the future statistics of measurement outcomes through Repeated Measurements (RM). This is done by extending the validity of quantum mechanics to those systems admitting no replicas; we prove that if the statistics of the results acquired by performing RM on such systems is sufficiently similar to that obtained by the Born rule, the latter can be used effectively. We apply our framework to a repeatedly measured Unruh-DeWitt detector interacting with a massless scalar quantum field, which is an example of a system (detector) interacting with an uncontrollable environment (field) for which using RM is necessary. Analysing what an observer learns from the RM outcomes, we find a regime where history-dependent RM probabilities are close to the Born ones. Consequently, the latter can be used for all practical purposes. Finally, we numerically study inertial and accelerated detectors, showing that an observer can see the Unruh effect via RM.
\end{abstract}

\maketitle
\section{Introduction}
In quantum mechanics, the Born rule has an intrinsic frequentist meaning~\cite{Born26, NielsenC10}. Given a quantum system in the pure state $\ket{\psi}$ and POVM described by a set of effects $\{\hat{E}_m\}$, the Born rule tells that the probability of measuring the system and finding the outcome $m$ is
\begin{equation}
    p_m=\bra{\psi}\hat{E}_m\ket{\psi}~.
    \label{born}
\end{equation}
This means that assuming an observer has access to infinite identical copies of that system in the same state, the fraction of times they would obtain the outcome $m$ is $p_m$. Similarly, if we take a large-but-finite number of copies $N$ of the same experiment, the relative frequency of the outcome $m$ approaches $p_m$ as $N$ increases, and one can describe the system's state by the collection of the results obtained from these finitely many experiments~\cite{Araki99}.

In most cases, one can test the Born rule by taking a sufficiently large number of copies of the system, preparing them identically, and performing many identical measurements, making the frequentist interpretation effective. Once the rule is tested, one can apply it without the need to repeat measurements. However, if for any reason one cannot replicate a system, and hence test the Born rule, there is no \textit{a priori} motivation to take it as valid. At the same time, the precision to which QM has been tested over the last century makes it possible to conjecture that its mathematical framework describes a broader class of systems for which the above large-number approach has no right to exist, i.e. systems that admit no replicas. In particular, we assume all systems are fundamentally described by the postulates of quantum mechanics and suppose some systems cannot be replicated;  we call these systems \textit{non-replicable}. Notice that the idea of QM being the fundamental theory of reality can be found in most quantum-related disciplines: while we do not necessarily share this view, we take it as a working assumption throughout the paper.

Relevant to our work, examples of such systems are those typically studied by Quantum Cosmology (QC) and Quantum Field Theory (QFT). In QC, the system under consideration is the whole universe~\cite{DeWitt67,Hawking84}: as it is clear, testing the Born rule by making multiple copies is impossible in this case. Similarly, in QFT one often works under the assumption that experiments can be replicated by starting from some initial state (e.g. the vacuum) and that the field's state can be reset to some target state. However, such a degree of control on QFT degrees of freedom is, in practice, out of reach, at least due to causality reasons. Hence, there is no reason to deem using QM's postulates in these disciplines well-defined.

This paper discusses the possibility of extending QM to non-replicable systems. By extension we do not mean that the Born rule is changed in its formal expression, but rather that we change its domain of validity, i.e. we extended the set of systems for which it can be used. The logic is the following. First, we \textit{assume} all systems behave following the standard postulates of quantum mechanics. Then, we notice that this assumption is not enough to enable the usage of the Born rule in all settings. In other words, assuming a system is quantum is not enough for using the Born rule. Next, we notice that the Born rule trivially holds for all quantum systems admitting replicas (those for which QM was formulated) and search for a class of systems not admitting replicas for which the Born rule holds. We find that some such systems exist, hence providing an extension of the domain of validity of the Born rule: the outcomes of measurements performed on these systems resemble those performed on replicable ones. In particular, we study what statistics can emerge from measuring non-replicable systems many times and define conditions under which observers can still meaningfully talk about the Born rule. This is done by looking at what an observer performing Repeated Measurements (RM) on a non-replicable system could infer from a string of measurement outcomes.

As an example, we work out the details of RM performed on an Unruh-DeWitt (UDW) detector (e.g. a two-level system) interacting with a massless scalar field. This paradigmatic system provides an operational way to prove that the notion of particle is observer-dependent, which is a milestone result of QFT in non-inertial frames~\cite{BirrellD82,ParkerT09}. In particular, the detector model can be used to study the Unruh effect, which states that the quantum state observed to be empty of particles by inertial observers (the Minkowski vacuum) appears to be a thermal state for a class of uniformly accelerated observers (Rindler observers). A detector in uniform acceleration will excite and de-excite as if it would be absorbing and emitting particles in interaction with a thermal heat bath. An additional question is what happens when the state of the detector is measured. The measurement collapses the state of the detector, while the state of the composite detector-quantum field system factorizes to a product state. For an accelerated detector, establishing a thermal distribution for the state of the detector would require many measurements. However, the probabilistic interpretation assumes an \textit{i.i.d} setting, which becomes unrealistic by requiring either many identical copies of the detector and the quantum field, or the ability to reset the field state after the measurement. As an alternative, we resort to our repeated measurements protocol, and investigate whether a thermal distribution for the detector state can be reliably established. 

Summarising, the scope of this paper is two-fold. First, we propose the RM framework as a way of making predictions for a single run of an experiment performed on a non-replicable system, for which the standard frequentist interpretation of QM is not applicable. Second, we illustrate this approach for an UDW detector interacting many times with a quantum field; by doing so, we investigate the populations obtained by repeated measurements performed between the many interactions and see if and how much they differ from those obtained in the standard frequentist interpretation.

The paper is organized as follows. In Sec.~\ref{sec:2}, we introduce our RM scheme as a procedure for making predictions about measurements performed upon systems interacting with non-replicable environments. In Sec.~\ref{sec:3}, we review the theory of Unruh-DeWitt detectors and a recent proposal using them for defining quantum measurements in QFT. Then, we argue that an UDW detector interacting with a quantum field can be treated via RM and obtain the probability of observing any string of results, together with upper and lower bounds. Next, in Sec.~\ref{sec:4}, we numerically study the cases of UDW detectors moving along inertial and accelerated trajectories, and compare our results with those already present in the literature, obtained in the finite time interaction case within the frequentist interpretation. Finally, in Sec.~\ref{sec:5}, we present conclusions and discuss future prospects. 

Throughout this paper, we work in four-dimensional Minkowski spacetime with metric signature $(-,+,+,+)$. Four-vectors are denoted by uppercase letters, e.g. $X$, and their three-vectors spatial parts by lowercase bold letters, e.g. $\mathbf{x}$. We assume the units convention for which $\hbar=c=k_B=1$.

\section{Repeated measurements}
\label{sec:2}
\begin{figure*}
     \centering
     \begin{subfigure}[b]{\columnwidth}
         \centering
         \includegraphics[width=0.75\textwidth]{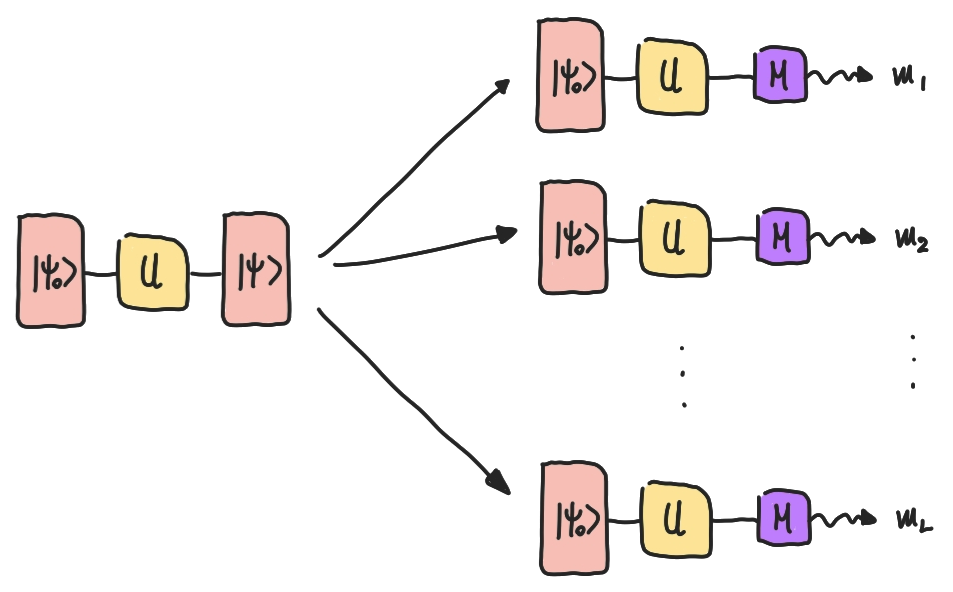}
         \caption{Standard approach for testing the Born rule.}
         \label{fig:standard_Born}
     \end{subfigure}
     \hfill
     \begin{subfigure}[b]{\columnwidth}
         \centering
         \includegraphics[width=.75\textwidth]{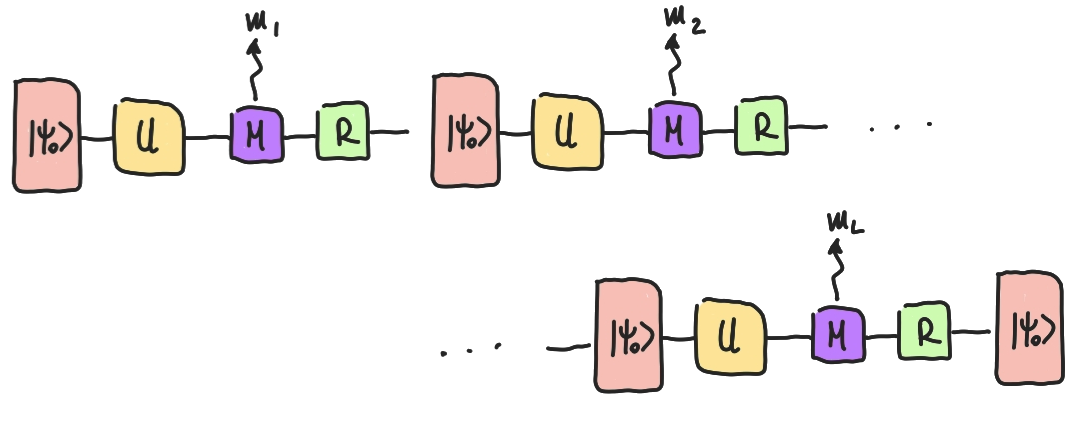}
         \caption{Alternative procedure for testing the Born rule via resets.}
         \label{fig:Born_from_one_system}
     \end{subfigure}
     \hfill
     \vspace{2em}
     \begin{subfigure}[b]{\columnwidth}
         \centering
         \includegraphics[width=.75\textwidth]{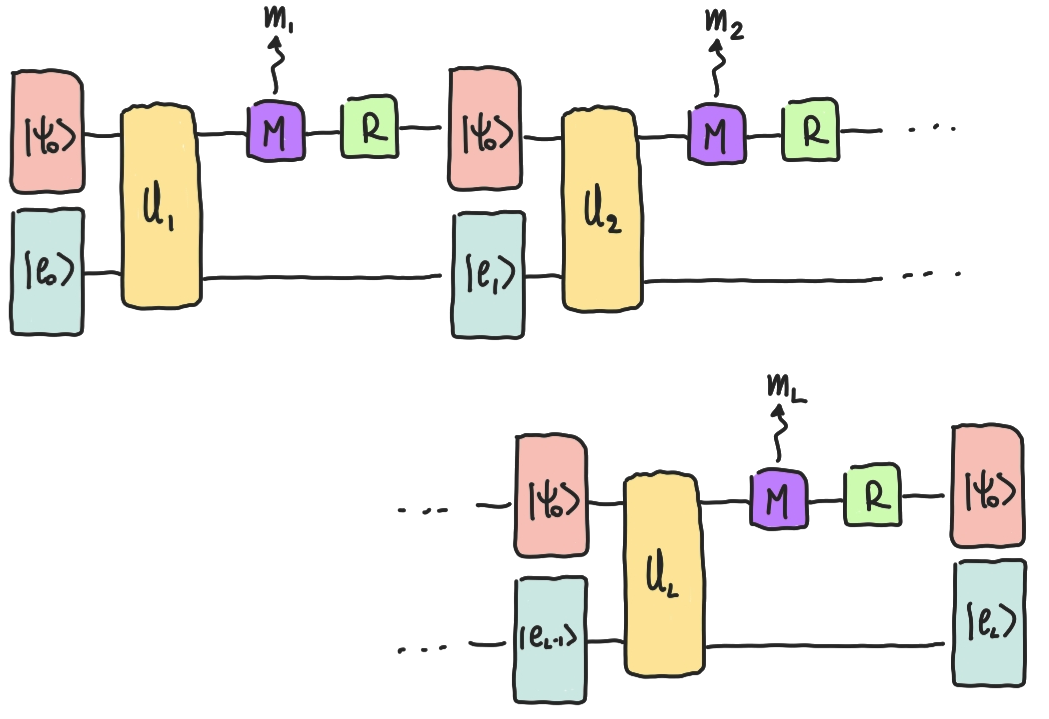}
         \caption{RM scenario.}
         \label{fig:RM_scheme}
     \end{subfigure}
        \caption{Procedures for getting many measurement outcomes from a system prepared in the state $\ket{\psi}=\hat{U}\ket{\psi_0}$, or in the RM setting. In panel (a), the standard procedure of replicating the system $L$ times is presented. In panel (b), an alternative way of getting the Born rule is considered. After the system is prepared and measured, its state is reset to the initial one, so that no replicas are needed. In panel (c), the RM scenario is presented; after each measurement $\hat{M}$ of panel (c) the global state of the system is separable due to the measurement update postulate.}
        \label{fig:measurements_procedures}
\end{figure*}

In this section, we present our scheme for making predictions about non-replicable systems. Let us first recall and formalize the standard frequentist approach to the Born rule. Suppose Alice investigates a system $S$ in some state $\ket{\psi}$, and that she can replicate it a large number of times $N$. In this way, she performs $N$ measurements of a selected observable getting $N$ outcomes, and builds statistics from these results~\footnote{Note that, as far as $\ket{\psi}$ is known, this procedure is not forbidden by the no-cloning theorem.}. This procedure is schematized in Fig.~\ref{fig:standard_Born}. Let us consider also the case where she can not produce replicas, yet has full control over the whole system, meaning she can apply any quantum operation to its state. Then, she can still test the Born rule by preparing the system in the initial target state, measuring it, and resetting it to the above state; the outcomes she obtains by this procedure are identical to those obtained by using replicas, as it is clear by Fig.~\ref{fig:Born_from_one_system}. Therefore, having full control over the system is enough to test the Born rule even without being able to produce replicas. In contrast, we call \textit{non-replicable} a system which, for any reason, we cannot prepare many times in the same state or we have no full control over it.

The relevance of the picture described by Fig.~\ref{fig:Born_from_one_system} is that it is easily generalized to describe the case of non-replicable systems. To do so, we restrict Alice to measure only a part of of the whole system, called detector and labelled by $\mathcal{D}$, while we call the rest of the system an environment, and label it with $\mathcal{E}$. Furthermore, we suppose she does not have full control over the system, meaning that she cannot apply quantum operations at will on the total state. In particular, we suppose she can measure and apply reset operators on $\mathcal{D}$, but all other operations she can use inevitably couple $\mathcal{D}$ and $\mathcal{E}$. Hence, all the measurement results she obtains from measuring $\mathcal{D}$ are perturbed by the state of $\mathcal{E}$, and no Born rule is expected to emerge from performing Repeated Measurements. This scenario is further explained by Fig.~\ref{fig:RM_scheme}, from which it is clear that each of Alice's measurements is testing the same system in a different state, and it is only by having full control of $\mathcal{E}$ that she would be able to observe the Born rule on $\mathcal{D}$.

It is important to note that calling the measured system \textit{detector} is not standard in quantum measurement theory, as this name is often reserved for the apparatus making the tested system decohere. Here we followed a different naming convention to be consistent with the literature regarding Unruh-DeWitt detectors, which we often reference in the second part of this work. To avoid confusion, we here stress that, in this work, measurements are idealized as instantaneous projective operators applied over the detector degree of freedom, and collapse is considered to happen as in the standard von Neumann prescription, without any mediator nor decoherence time needed.

\subsection{Statistical inference from RM}
\label{sec:2.statistical_inference}
Following the above discussion, let us consider a system $\mathcal{S}$ composed of two parts, one called the detector $\mathcal{D}$, and the other the environment $\mathcal{E}$. In particular, without loss of generality we study the case of $\mathcal{D}$ being a non-degenerate two-level system on which we can perform projective measurements, and take the total Hamiltonian of $\mathcal{S}$ to be
\begin{equation}
    \hat{H}_\mathcal{S}(t)=\hat{H}_\mathcal{D}+\hat{H}_\mathcal{E}+\lambda\hat{H}_{\mathrm{int}}(t)~,
    \label{general_RM_hamiltonian}
\end{equation}
where
\begin{equation}
    \hat{H}_{\mathrm{int}}(t)=\sum_{k=1}^N\chi_k(t)\hat{H}^{(k)}_{\mathrm{int}}(t)
    \label{H_with_chi_k}
\end{equation}
with $\{\chi_k(t)\}$ being a set of window functions with disjoint compact supports (i.e. $\textrm{Supp}(\chi_i)\cap \textrm{Supp}(\chi_j)=\emptyset$, $\forall i,j$), and $\lambda$ a small parameter modulating the strength of the interaction. Note that by using the interaction Hamiltonian \eqref{H_with_chi_k} we implicitly assume our ability to switch on and off the interaction, a task that might be hard to accomplish in practice. As a result of this setup, $\mathcal{D}$ and $\mathcal{E}$ interact $N$ times via as many possibly different Hamiltonians. Finally, let us take $\mathcal{S}$ to be initially in the separable state
\begin{equation}
    \ket{\psi_0}=\ket{0}\otimes\ket{e_0}~,
    \label{initialstate}
\end{equation}
where $\ket{0}$ is the lowest energy eigenstate of $\hat{H}_{\mathcal{D}}$, and $\ket{e_0}$ some environmental state. For the sake of simplicity, we take the detector's Hamiltonian to be
\begin{equation}
    \hat{H}_{\mathcal{D}}=\omega\ket{1}\bra{1}
    \label{H_D}
\end{equation}
with $\omega>0$, so that $\ket{0}$ has zero energy. Working in the interaction picture~\cite{BreuerP02}, the unitary operator describing the evolution generated by the $k$-th interaction has the form
\begin{equation}
    \hat{U}_k=\mathcal{T}\left\lbrace\exp\left(-i\lambda\int_{\textrm{Supp}(\chi_k)} \chi_k(t)\hat{H}^{(k)}_{\mathrm{int}}(t) dt\right)\right\rbrace~.
    \label{U_k}
\end{equation}
After the end of each interaction, we measure $\mathcal{D}$ via the PVM operators 
\begin{equation}
    \begin{dcases}
        \hat{M}_0=\ket{0}\bra{0}\\
        \hat{M}_1=\ket{1}\bra{1}~,
    \end{dcases}
    \label{PVM_definition}
\end{equation}
store the corresponding results in a bit string
\begin{equation}
    B_L=(b_1,b_2,\dots,b_L)~,
    \label{bit_string}
\end{equation}
where $L\leq N$, and reset the detector's state to $\ket{0}$ via some unspecified erasure procedure~\cite{PlenioV01}. Note that these measurements are always applied between the interactions, meaning that the measurement giving the bit string element $b_k$ is obtained after $\hat{U}_k$ ended, and before $\hat{U}_{k+1}$ starts. Moreover, the fact that $[\hat{M}_i,\hat{H}_\mathcal{D}]=0$ means that the times at which the measurements happen are not relevant.

Starting from a generic separable state $\ket{0}\otimes\ket{f}$, the $k$-th operator of the set \eqref{U_k} evolves the system to $\hat{U}_k\ket{0,f}$, and hence the outcome-dependent post-measurement state is
\begin{equation}
    \ket{\psi_b}=\frac{\hat{M}_b\hat{U}_k\ket{0,f}}{\sqrt{p_b(k)[f]}} ~,
\end{equation}
where
\begin{equation}
    p_b(k)[f]=\norm{\bra{b}\hat{U}_k\ket{0,f}}^2
    \label{conditioned_p}
\end{equation}
is the probability of getting $b$ after the $k$-th unitary is applied, given that the initial environmental state was $\ket{f}$. Therefore, since after each measurement the state of $\mathcal{D}$ is reset to $\ket{0}$, one can define the environmental state-dependent operators
\begin{equation}
    \hat{V}_b(k)[f]=\frac{\bra{b}\hat{U}_k\ket{0}}{\sqrt{p_b(k)[f]}}~,
    \label{V_operators}
\end{equation}
describing the collection of operations 1) application of $\mathcal{U}_k$, 2) measurement of $\mathcal{D}$ with outcome $b$, and 3) reset of $\mathcal{D}$ to $\ket{0}$; all together, these operations combine to give
\begin{equation}
    \ket{0}\otimes\ket{f}\xrightarrow[]{}\ket{0}\otimes\left(\hat{V}_b(k)[f]\ket{f}\right)~.
\end{equation}
When the RM procedure is repeated $L$ times upon the initial state \eqref{initialstate}, giving the bit string \eqref{bit_string}, the final state is
\begin{equation}
    \ket{\psi_L}=\ket{0}\otimes\left(\prod_{i=1}^L\hat{\mathcal{V}}_i\ket{e_0}\right)
    \label{final-state}
\end{equation}
where the operators $\hat{\mathcal{V}}_i$ are recursively defined as
\begin{equation}
    \begin{dcases}
        \hat{\mathcal{V}}_1=\hat{V}_{b_1}(1)[e_0]\\
        \hat{\mathcal{V}}_{n+1}=\hat{V}_{b_{n+1}}(n+1)\left[\prod_{i=1}^n\hat{\mathcal{V}}_i\ket{e_0}\right]
    \end{dcases}
    \label{op_recursive}
\end{equation}
and where the operator products appearing in Eq.s~\eqref{final-state} and \eqref{op_recursive} are ordered starting from the right by decreasing $i$. Therefore, each $V$-operator applied on the environment depends on all $V$-operators applied before it, making Eq.~\eqref{final-state} extremely hard to evaluate.

In order to make Eq.~\eqref{final-state} tractable, let us consider the simpler case of weak interaction and slowly-evolving environment, i.e. take the unitaries \eqref{U_k} to be
\begin{equation}
    \hat{U}_k=\hat{U}\otimes\mathbb{I}_\mathcal{E}+\epsilon\sum_l \hat{A}_{l}\otimes\hat{B}_{l}(k)+\epsilon^2\sum_l \hat{C}_{l}\otimes\hat{D}_{l}(k)+O(\epsilon^3)
    \label{U_Weak}
\end{equation}
where the set of operators $\{\hat{A}_{l},\hat{C}_{l}\}$ and $\{\hat{B}_{l}(k),\hat{D}_{l}(k)\}$ act on $\mathcal{H}_{\mathcal{D}}$ and $\mathcal{H}_\mathcal{E}$ respectively, and $\epsilon$ is a small parameter. First, let us briefly analyze the case of vanishing $\epsilon$, corresponding to having no interaction between $\mathcal{D}$ and $\mathcal{E}$, and applying many times the same $\hat{U}$ to the detector. Then, measuring $\mathcal{D}$, gathering data, and resetting it to $\ket{0}$ many times, means that all the measurements are performed on the same state $\hat{U}\ket{0}$, and the results strictly satisfy the Born rule \eqref{born} with $\hat{E}_m=\hat{M}_m$, i.e.
\begin{equation}
    p_m=\norm{\bra{m}\hat{U}\ket{0}}^2~.
    \label{p_born_part}
\end{equation}
In fact, this is equivalent to preparing many copies of $\mathcal{D}$ in $\hat{U}\ket{0}$ and measuring them, hence reproducing the standard prescription for getting the Born rule, as in the case pictured in  Fig.s~\ref{fig:standard_Born}-\ref{fig:Born_from_one_system}. Next, by taking $\epsilon$ to be small, and substituting the unitaries \eqref{U_Weak} in Eq.~\eqref{conditioned_p} one gets 
\begin{equation}
    p_m(k)[f]=p_m+\epsilon Q^{(1)}_m(k)[f]+\epsilon^2 Q^{(2)}_m(k)[f]+O(\epsilon^3)
    \label{U_Weak_Q_1_corrections}
\end{equation}
where
\begin{equation}
\begin{cases}
\begin{split}
     Q_m^{(1)}(k)[f]=\sum_l\langle\hat{A}^\dagger_{l}\hat{M}_m\hat{U}\rangle_0\langle\hat{B}_l^\dagger(k)\rangle_f+h.c.
\end{split}\\
    \begin{split}
         Q_m^{(2)}(k)[f]&=\sum_{l,l'}\langle\hat{A}^\dagger_l\hat{M}_m\hat{A}_{l'}\rangle_0\langle\hat{B}^\dagger_l(k)\hat{B}_{l'}(k)\rangle_f+\\&+\left[\sum_l\langle\hat{C}^\dagger_{l}\hat{M}_m\hat{U}\rangle_0\langle\hat{D}_l^\dagger(k)\rangle_f+h.c.\right]
    \end{split}
    \end{cases}
    \label{Qs_definitions}
\end{equation}
where we used the shorthand notation $\langle\hat{O}\rangle_\psi=\bra{\psi}\hat{O}\ket{\psi}$. Hence, for the generic state $\ket{0,f}$ the probability of obtaining the outcome $m$ is given by the Born rule of the $\epsilon=0$ case plus corrections at most of order $\epsilon$; yet, these corrections are as convoluted as Eq.~\eqref{final-state}, meaning that $Q_m(k)[e_k]$ will depend on all the states the environment explored since the RM procedure started, hence making any usage of Eq.~\eqref{U_Weak_Q_1_corrections} very complicated.

\subsubsection{Bayes' updates rules}
Inspired from the the vanishing $\epsilon$ case, one can try overlooking the contribution given by the memory-dependent part of the probabilities \eqref{U_Weak_Q_1_corrections}, and predicting the statistics of future outcomes as if the probabilities were well described by Eq.~\eqref{p_born_part}, i.e. as if the Born rule was valid. Clearly enough, observers following this strategy will deduce wrong probabilities. However, if the error they obtain can not be seen by confronting the predictions with the actual outcomes, one can still rely on the Born rule for all practical purposes (FAPP). To formalize this idea let us introduce Bayesian updating in this setting~\cite{vonToussaint11}.

Suppose we are given a set of data \textcolor{blue}{$\mathfrak{D}$} and have to decide between a set of hypotheses $\{\mathcal{H}_i\}$ which ones are correct. Moreover, suppose we have some prior knowledge about the hypotheses, summarized as probabilities of them being true, i.e. $P(\mathcal{H}_i)$. By observing the data, one can update these probabilities from their initial value to $P(\mathcal{H}_i|\mathfrak{D})$, which is the probability of $\mathcal{H}_i$ being true given the new knowledge acquired. This is done via the Bayes rule, i.e.
\begin{equation}
    P(\mathcal{H}_i|\mathfrak{D})=\frac{P(\mathfrak{D}|\mathcal{H}_i)P(\mathcal{H}_i)}{P(\mathfrak{D})}~,
    \label{Bayes}
\end{equation}
where $P(\mathfrak{D})$ is a normalization factor given by
\begin{equation}
    P(\mathfrak{D})=\sum_i P(\mathfrak{D}|\mathcal{H}_i)P(\mathcal{H}_i)~,
\end{equation}
and $P(\mathfrak{D}|\mathcal{H}_i)$ is the probability of obtaining the data assuming $\mathcal{H}_i$. A general update strategy then follows by applying Eq.~\eqref{Bayes} recursively: assuming we observed the data set $\mathfrak{D}$, our knowledge about the hypotheses is described by $P(\mathcal{H}_i|\mathfrak{D})$ which become our new prior, so that once we obtain the new data $\mathfrak{D}'$, all we have to do to update our current belief is to apply Eq.~\eqref{Bayes} again as
\begin{equation}
    P(\mathcal{H}_i|\mathfrak{D}')=\frac{P(\mathfrak{D}'|\mathcal{H}_i)P(\mathcal{H}_i|\mathfrak{D})}{P(\mathfrak{D}')}~.
    \label{Bayes}
\end{equation}
If, after enough data $\mathfrak{D}''$ it is
\begin{equation}
   P(\mathcal{H}_i|\mathfrak{D}'')\gg P(\mathcal{H}_j|\mathfrak{D}'')~,
   \label{threshold}
\end{equation}
for one $i$ and all $j\neq i$, then we claim $\mathcal{H}_i$ is the right hypothesis. If new data are later gathered in favour of other hypotheses, one must stop calling $\mathcal{H}_i$ the true one. If, at the same time, all hypotheses $\{\mathcal{H}_{i_1},\dots,\mathcal{H}_{i_R}\}$ satisfy Eq.~\eqref{threshold} for all $j\neq i_1,\dots,i_R$, yet between all these there is no one for which Eq.~\eqref{threshold} holds with respect to all other $i_1,\dots,i_R$, then all of them retain the same degree of trueness, meaning that we can interpret the data by means of any.

We are now ready to discuss if and how an observer can use the Born rule in situations where it does not apply strictly. Suppose Alice measures $L$ times the detector and gets the string of outcomes $B_L$. To interpret these data, Alice formulates the following set of hypotheses:
\begin{itemize}
    \item $\mathcal{H}_1(q)$: The Born rule holds as given by Eq.~\eqref{p_born_part}, and $p_1=q$.
    \item $\mathcal{H}_2(q)$: The Born rule holds approximately, with the Born part given as above and corrections of at most of order $\epsilon$, as in Eq.~\eqref{U_Weak_Q_1_corrections}.
\end{itemize}
Finally, let us suppose Alice has no previous knowledge or bias about the experiments, and therefore she assigns uniform priors to all hypotheses, i.e. $h=P(\mathcal{H}_1(q))=P(\mathcal{H}_2(q))=1/2$. Thanks to this choice, the priors are correctly normalized
\begin{equation}
    \sum_{i=1}^2\int_0^1 P( \mathcal{H}_i(\varrho))d\varrho=1~.
\end{equation}
Next, she looks at the data contained in $B_L$ and updates the probabilities about the hypotheses via
\begin{equation}
    P(\mathcal{H}_i(q)|B_L)=\frac{\frac{1}{2}P(B_L|\mathcal{H}_i(q))}{\frac{1}{2}\int_0^1 \left(2P_\varrho(B_L)+\epsilon\Delta P_\varrho(B_L)\right) d\varrho }~,
    \label{Bayes_strings}
\end{equation}
where we defined
\begin{equation}
    \begin{dcases}
        P(B_L|\mathcal{H}_1(q))=P_q(B_L)\\
        P(B_L|\mathcal{H}_2(q))=P_q(B_L)+\epsilon\Delta P_q(B_L)
    \end{dcases}
\end{equation}
and where
\begin{equation}
    P_q(B_L)=q^{n}(1-q)^{L-n}
    \label{born_string}
\end{equation}
with $n$ being the number of ones in $B_L$, and where $\Delta P_q(B_L)$ depends on the specific Hamiltonian and initial environmental state considered. When expanded, Eq.~\eqref{Bayes_strings} gives
\begin{widetext}
\begin{equation}
    \begin{dcases}
        P(\mathcal{H}_1(q)|B_L)=\frac{1}{2}\frac{P_{q}(B_L)}{\int_0^1 P_\varrho(B_L)d\varrho}-\frac{\epsilon}{4}\left[\frac{P_{q}(B_L)\int_0^1 \Delta P_\varrho(B_L)d\varrho}{\left(\int_0^1 P_\varrho(B_L)d\varrho\right)^2}\right]+O(\epsilon^2)\\
        P(\mathcal{H}_2(q)|B_L)=\frac{1}{2}\frac{P_{q}(B_L)}{\int_0^1 P_\varrho(B_L)d\varrho}+\frac{\epsilon}{4}\left[\frac{2\Delta P_{q}(B_L)\int_0^1  P_\varrho(B_L)d\varrho-P_{q}(B_L)\int_0^1 \Delta P_\varrho(B_L)d\varrho}{\left(\int_0^1 P_\varrho(B_L)d\varrho\right)^2}\right]+O(\epsilon^2)
    \end{dcases}
\end{equation}
\end{widetext}
making evident that:
\begin{enumerate}
    \item if $\Delta P_{q}(B_L)=0$, the posteriors are only modified to shift towards the correct Born rule, and no update is made for selecting between $i=1,2$.
    \item if $\Delta P_{q}(B_L)>0~(<0)$, the probability of $\mathcal{H}_2(q)$ becomes higher (lower) than that of the corresponding $\mathcal{H}_1(q)$, in accordance with the fact that observing $B_L$ was more (less) likely given the former hypothesis.
\end{enumerate}
The above equation provides a way to formalize the idea that we presented at the beginning of this paragraph, namely, as the question: \textit{given that the right hypotheses is $\mathcal{H}_2(q)$, can Alice use $B_L$ to rule out $\mathcal{H}_1(q)$ in favour of the former?}

Clearly, the answer to this question depends on the specific corrections considered; however, to select $\mathcal{H}_2(q)$ against $\mathcal{H}_1(q)$ it must generally be
\begin{equation}
    \frac{P(\mathcal{H}_2(q)|B_L)}{P(\mathcal{H}_1(q)|B_L)}\gg 1
\end{equation}
which, assuming $P_{q}(B_L)\neq 0$, is
\begin{equation}
    \frac{\Delta P_{q}(B_L)}{P_{q}(B_L)}\gg\frac{1}{\epsilon}~.
    \label{AlmostBorn_Validity}
\end{equation}
Moreover, it is easy to show that, if the first order corrections all vanish, then we can go at second order in $\epsilon$ and get
\begin{equation}
    \frac{\Delta^{(2)} P_{q}(B_L)}{P_{q}(B_L)}\gg\frac{1}{\epsilon^2}~,
\end{equation}
where the quantity at numerator is implicitly defined by
\begin{equation}
    P(B_L|\mathcal{H}_2(q))= P_q(B_L)+\epsilon^2\Delta^{(2)} P_q(B_L)+O(\epsilon^3)~.
\end{equation}

\subsubsection{Bit string probabilities}
\label{sec:2.bit_string_probabilities}
Let us now study the probability $P(B_L)$ of obtaining a certain bit string $B_L=(b_1,\dots,b_L)$. From the definition of conditional probability, we get
\begin{equation}
    P(B_L)=P(b_L|B_{L-1})P(B_{L-1})~,
\end{equation}
where $B_j$ with $j\leq L$ denotes the strings obtained by cutting $B_L$ at length $j$. Hence, by applying the above formula recursively we obtain that $P(B_L)$ can be expressed as a product of $L$ conditional probabilities having the form
\begin{equation}
    \begin{dcases}
        P(0|B_j)\\
        P(1|B_j)\\
    \end{dcases}~.
    \label{conditional_probabilities}
\end{equation}
For the sake of simplicity, assuming there are $n$ ones in $B_L$ we can completely determine it by its length $L$ and the location of ones in it, i.e. $(N_1,\dots,N_n)$; hence the two expressions
\begin{equation}
    B_L=(b_1,\dots,b_L)
    \label{B_L_string}
\end{equation}
and
\begin{equation}
    B_L=(L;N_1,\dots,N_n)
\end{equation}
contain the same information~\footnote{This notation is more economic whenever the number of ones in $B_L=(b_1,\dots,b_L)$, $n$, is much smaller than that of zeroes $L-n$. Conversely, if the number of ones is larger than that of zeros one can use a mapping where the $\{N_k\}$ denote the locations of the zeros and get a similar efficiency.}. Thanks to this notation, it is 
\begin{equation}
    P(B_L)=\prod_{\substack{j=1,\dots,L\\j\neq N_1,\dots,N_n}}P(0|B_{j-1})\prod_{j= N_1,\dots,N_n}P(1|B_{j-1})~.
    \label{P(B_L)}
\end{equation}
If the Born rule holds, the above expression simplifies to Eq.~\eqref{born_string}. On the contrary, in the case of small but finite $\epsilon$ and assuming the probabilities \eqref{conditional_probabilities} take the form \eqref{U_Weak_Q_1_corrections}, at first order in $\epsilon$ it is
\begin{equation}
    \tilde{P}_{q}(B_L)=P_q(B_L)+\epsilon\sum_{j=1}^L \prod_{j'\neq j}^LQ^{(1)}_{b_j}(j)[f_j] p_{b_{j'}}~,
\end{equation}
and hence
\begin{equation}
    \Delta P_q(B_L)=\sum_{j=1}^L \prod_{j'\neq j}^LQ^{(1)}_{b_j}(j)[f_j] p_{b_{j'}}~.
\end{equation}
This expression gives us an explicit way to test if the Born rule can be taken as valid in the case of RM. In fact, if Eq.~\eqref{AlmostBorn_Validity} does not hold, then the hypotheses $\mathcal{H}_1(q)$ and $\mathcal{H}_2(q)$ are equally valid, and one can use the former in place of the latter. In particular, if
\begin{equation}
    \frac{\tilde{P}_{q}(B_L)}{P_{q}(B_L)}\simeq 1~,
    \label{easy_threshold}
\end{equation}
is true, where $P_q(B_L)$ and $\tilde{P}_{q}(B_L)$ are the bit string probabilities obtained using hypothesis $\mathcal{H}_1(q)$ and $\mathcal{H}_2(q)$ respectively, then FAPP the Born rule can be used in place of the outcome dependent one. Also note that thanks to the notation \eqref{B_L_string}, the bit string update process can be viewed as a non-markovian random walk on an $L$-dimensional hypercube starting from $(0,\dots,0)$ by updating the $k$-th bit after the $k$-th measurement to the obtained measurement outcome (for related discussion of random walks on a hypercube see e.g. Ref.~\cite{StanislavW08}).

In conclusion, one should think of the RM approach as relative to an observer knowing neither if the system they are given to test is interacting with some environment nor the specific interaction occurring. Such an observer can perform RM and construct a probability distribution from the obtained outcomes. We claim that, in some cases, they need not know the details of the environment and interaction and yet use our result to apply the Born rule. To do so, experimenters having access to a single copy of a quantum system must apply the following strategy. First, they measure the system and collect data, e.g. in the form of a bit-string. Next, they make a reasonable list of hidden environments and interactions they might have overlooked, represented by the hypotheses $\{\mathcal{H}_i\}$. Then, by reading the gathered data, they assign to each hypothesis a probability, which plays the role of a prior for the updates induced by later readings. Then, the experimenter keeps measuring the system and updating their beliefs, following Eq.~\eqref{Bayes}; if Eq.~\eqref{easy_threshold} is satisfied for the hypothesis that the Born rule holds (possibly together with one or more hypotheses considering interactions and environments), the experimenter is entitled to the Born rule FAPP.

\begin{table*}[]
    \centering
    \begin{tikzpicture}[node distance=2cm]
        \node (start) [startstop] {$\mathcal{D}$ isolated?};
        \node (pro1a) [startstop, below of=start, xshift=-1.6cm] {$\mathcal{D}$ replicable?};
        \node (pro1b) [startstop, right of=pro1a, xshift=1.4cm] {$\mathcal{D}+\mathcal{E}$ replicable?};

        \node (result1a) [process, left of=pro1a, xshift=-1.7cm] {BR on $\{m_i\}$ (see Fig.~\ref{fig:standard_Born})};
        \node (result1b) [process, right of=pro1b, xshift=1.9cm] {BR on $\{M_I\}$};

        \node (pro2a) [startstop, below of=pro1a] {Can $\mathcal{D}$ be reset?};
        \node (pro2b) [startstop, below of=pro1b] {Can $\mathcal{D}+\mathcal{E}$ be reset?};

        \node (result2a) [process, left of=pro2a, xshift=-1.7cm] {BR on $\{m_i\}$ (see Fig.~\ref{fig:Born_from_one_system})};
        \node (result2b) [process, right of=pro2b, xshift=1.9cm] {BR on $\{M_I\}$};

        \node (pro3a) [question, below of=pro2a] {?};
        \node (pro3b) [startstop, below of=pro2b] {Are $\mathcal{D}$ and $\mathcal{E}$ weakly interacting?};

        \node (pro4a) [question, below of=pro3b] {?};
        \node (result3b) [output, right of=pro3b, xshift=1.9cm] {RM $\simeq$ BR on $\{m_i\}$\\ (see Fig.~\ref{fig:RM_scheme})};

        \draw [arrow] (start) -- node[anchor=east] {yes} (pro1a);
        \draw [arrow] (start) -- node[anchor=west] {no} (pro1b);

        \draw [arrow] (pro1a) -- node[anchor=south east, xshift=0.5cm] {yes} (result1a);
        \draw [arrow] (pro1b) -- node[anchor=south west, xshift=-0.6cm] {yes} (result1b);

        \draw [arrow] (pro1a) -- node[anchor=east] {no} (pro2a);
        \draw [arrow] (pro1b) -- node[anchor=west] {no} (pro2b);

        \draw [arrow] (pro2a) -- node[anchor=south east, xshift=0.5cm] {yes} (result2a);
        \draw [arrow] (pro2b) -- node[anchor=south west, xshift=-0.6cm] {yes} (result2b);

        \draw [arrow] (pro2a) -- node[anchor=east] {no} (pro3a);
        \draw [arrow] (pro2b) -- node[anchor=west] {no} (pro3b);

        \draw [arrow] (pro3b) -- node[anchor=south west, xshift=-0.4cm] {yes} (result3b);
        \draw [arrow] (pro3b) -- node[anchor=east] {no} (pro4a);
        
    \end{tikzpicture}
    \caption{Summary of the content of Sec.~\ref{sec:2.necessity}. The flowchart classifies all possible quantum systems one could consider. Even assuming the systems behave in accordance with the postulates of QM, the Born rule (RB in the flowchart) can be applied only to those systems represented by red boxes. On the contrary, blue boxes describe those systems for which it is unclear whether a Born-like rule can be found (and, in general, this is not the case). Finally, the systems represented by the green box admit a general strategy for finding an effective Born rule via RM.}
    \label{tab:Table_cases}
\end{table*}

\subsection{Applicability of the Born rule, necessity of RM}
\label{sec:2.necessity}
In this paragraph, we discuss the necessity and applicability RM in general. The discussion follows the logic represented as a flow chart in Table 1. To this end, we consider a quantum system $\mathcal{D}$ over which we can perform measurements and ask which rule (if any) we should apply to foresee the statistics of future measurements' outcomes. Moreover, we assume all systems behave following the standard postulates of quantum mechanics, meaning that the state of each isolated system is described by a normalized vector in some Hilbert space, its time evolution is described by some unitary operator, and if the system admits replicas then the Born rule holds.

First, suppose $\mathcal{D}$ is isolated. If $\mathcal{D}$ is also replicable, an experimenter can apply the standard procedure giving the Born rule (see Fig.~\ref{fig:standard_Born}) and get a list of outcomes $\{m_i\}$ with their related \textit{i.i.d.} probabilities $p(m_i)$, as in Eq.~\eqref{born}. If $\mathcal{D}$ is not replicable but it can be reset, then one can apply the procedure in Fig.~\ref{fig:Born_from_one_system} and still get $\{m_i\}$ and the relative $p(m_i)$. However, if $\mathcal{D}$ is not replicable and it admits no reset operation, then nothing can be said in general about the statistics of future outcomes from the result of the past ones: no Born rule generally exists in this case, even if we assume the system behaves by the quantum mechanical postulates\footnote{Notice that a Born-like rule may still emerge in some specific cases, either by chance or by choosing some very specific systems; these sporadic cases can be overlooked for the sake of simplicity. The same holds for non-replicable and non-resettable open systems, described by the right column of Table 1.}. For example, a system undergoing random different evolutions between any two repeated measurements allows no effective version of the Born rule.

A similar classification holds for open systems. Suppose $\mathcal{D}$ interacts with a quantum system $\mathcal{E}$ acting as an environment, such that $\mathcal{D}+\mathcal{E}$ is isolated. If $\mathcal{D}+\mathcal{E}$ is replicable, an experimenter can apply the standard procedure realising the Born rule to the composed system, as in the isolated case. Then, the outcomes of the replica experiments are all possible strings of outcomes $\{M_I\}$ of a given length $N$; when collected, these give the related \textit{i.i.d.} probabilities $p(M_I)$. Notice that, while the string probabilities are \textit{i.i.d.}, the occurrences of certain combinations inside a single are generally not \textit{i.i.d.}; an example of this is when $\mathcal{D}$ is a spin belonging to a spin chain. As in the isolated system, the above statistics can also be obtained when $\mathcal{D}+\mathcal{E}$ is non-replicable but overall resettable, and nothing can be said in general for non-replicable and non-resettable systems.  However, our analysis shows that, differently from its isolated counterpart, the non-isolated non-replicable system $\mathcal{D}$ sometimes allows using an effective Born rule for the single outputs of RM, even when $\mathcal{D}+\mathcal{E}$ is not replicable and non-resettable: such a rule is obtained by RM in the weak-coupling regime, as discussed in Sec.~\ref{sec:2.statistical_inference}.

Finally, we notice that the probabilities QM predicts for strings of outcomes obtained by measuring many copies of replicable weakly-interacting $\mathcal{D}+\mathcal{E}$ are exactly those obtained for their non-replicable non-resettable counterparts by the RM scenario, with the only difference that there is no possibility of empirical testing for the latter; yet, the formal expression for the probability $P(M)$ of obtaining a string $M$ repeatedly measuring many replicas of a replicable open system is the same we obtained for the unique string gathered by RM on a non-replicable one.

In the rest of this article, we explore the RM scenario for a two-level system interacting with a massless scalar field. As we will motivate in Sec.~\ref{sec:3.RM+UDW}, this system is non-replicable. According to the above classification, the two-level system is not isolated, and its larger environment is not replicable or resettable. Hence, a meaningful analysis is only possible when the two systems are weakly interacting: in such a case, RM gives a string of outcomes whose distribution is effectively Born-like. On the one hand, this example will serve as a tool to clarify our discussion about non-replicable systems; on the other hand, it will provide interesting results about relativistic quantum information and QFT in non-inertial frames and curved spacetimes.

\section{RM on Unruh-DeWitt detectors}
\label{sec:3}

\subsection{Unruh-DeWitt detectors}
\label{sec:3.unruh}
The Unruh-DeWitt (UDW) model for particle detectors was introduced as a tool for studying the notion of particle in QFT in curved and non-inertial spacetimes~\cite{Unruh76,DeWitt80}. In fact, thanks to this model one can abandon the definition of particles as field quanta in favour of a more operational approach, in which an observer can claim having detected a particle when observing a transition in the energy levels of a system they carry. Consequently, upwards and downwards transitions are respectively interpreted as absorptions and emissions of particles, and measurements on the detector as the only way to test if the field state contains particles or not. In practice, one often considers the idealized case in which, instead of studying the dynamics of one such detector, the focus is on the asymptotic populations of the detector's levels, and hence of a large number of replicas of identical detectors, interacting with as many independent quantum fields. Assuming the latter setting to be impossible or very hard to realize, this framework fits our definition of non-replicable systems, and we can apply the results of the previous section to see if an observer doing RM would get the same results of one applying the standard Born rule.

To this end, let us study a bipartite system composed of a a point-like non-degenerate two-level detector with free Hamiltonian given by Eq.~\eqref{H_D} and a real and massless scalar quantum field $\hat{\phi}$ in flat $(3+1)$-dimensional spacetime, playing the role of the environment $\mathcal{E}$. This idealized device is called an UDW detector. For a review of the theory of UDW detectors in flat and curved spacetime, see e.g.~\cite{CrispinoEtAl08, BirrellD82, Wald95}. The assumption of the detector being pointlike\footnote{An alternative option is that of taking the detector to have a spatial profile; this can be done by replacing the field operator with its spatially smeared version~\cite{Schlicht04, LoukoS06, Satz07}, where the smearing function describes the shape of the detector.} makes the discussion easier at the cost of being nonphysical and leading to divergences due to the small-distance fluctuations of the field that must be regulated~\cite{Takagi86}. Then, assuming the detector's position to be a classical degree of freedom we can write its world line as
\begin{equation}
    X(\tau)=(t(\tau),\mathbf{x}(\tau))~,
\end{equation}
where $\tau$ is the proper time. While the possibility of quantizing the detector's position and hence discussing UDW on superposed trajectories has recently gained attention~\cite{BarbadoEtAl20, FooEtAl20, Foo21}, we here limit our discussion to classical motions only, and postpone a more general treatment to future works. Hence, we can describe the interaction between the detector and the field via the proper time-dependent Hamiltonian
\begin{equation}
    \hat{H}_{\mathrm{int}}(\tau)=\chi(\tau)\hat{m}(\tau)\otimes\hat{\phi}(X(\tau))~,
    \label{H_int}
\end{equation}
where
\begin{equation}
    \hat{m}(\tau)=e^{-i\omega \tau}\ket{1}\bra{0}+e^{i\omega \tau}\ket{0}\bra{1}~,
\end{equation}
is the detector's monopole momentum operator in the interaction picture and $\chi(\tau)\in[0,1]$ is a smooth function describing the duration and the intensity of the interaction, called switching function. Physical switching functions have compact support, but it is often useful to relax this requirement. In order for the interaction to be weak, hence allowing to use perturbation theory, we multiply $\hat{H}_{\mathrm{int}}(\tau)$ by a small constant $\lambda$. Therefore, the full Hamiltonian is
\begin{equation}
    \hat{H}=\hat{H}_{\mathcal{D}}+\hat{H}_\phi+\lambda\hat{H}_{\mathrm{int}}(\tau)
    \label{H_tot}
\end{equation}
where
\begin{equation}
    \hat{H}_{\mathcal{D}}=\omega\ket{1}\bra{1}
\end{equation}
is the detector's free Hamiltonian and $\hat{H}_\phi$ is the field's one~\cite{BenattiF04}. Once the Hamiltonian \eqref{H_tot} is introduced, we interpret a detector's transition from the ground state $\ket{0}$ to the excited one $\ket{1}$ as the absorption of a field's particle, and the reverse as an emission. This enables to speak about particles even when global notions of particle number and vacuum state become ill-defined, such as in non-inertial and general relativistic settings~\cite{BirrellD82, CrispinoEtAl08, SvaiterS92}.

Since the Hamiltonian of our system is time-dependent, it is useful to work in the interaction picture~\cite{BreuerP02}. In this picture, the time evolution from an initial time $t_0$ to a final time $t$ is given by the time-ordered exponential
\begin{equation}
    \hat{U}(t_0,t)=\mathcal{T}\left[\exp\left(-i\lambda\int_{t_0}^t\hat{H}_{\mathrm{int}}(\tau)d\tau\right)\right]~.
    \label{U}
\end{equation}
Once we assume the initial state of the system to be in the separable state
\begin{equation}
    \ket{\psi(t_0)}=\ket{0}\otimes\ket{0_M}~,
    \label{Init_state}
\end{equation}
where $\ket{0_M}$ is the vacuum as described by an inertial observer, called Minkowski vacuum, and let the detector and the field interact via \eqref{H_int}, at time $t$ the system will be described by the entangled state
\begin{widetext}
\begin{equation}
    \ket{\psi(t)}=\hat{U}(t_0,t)\ket{\psi(t_0)}=\ket{0}\otimes\ket{0_M}-i\lambda\int_{t_0}^td\tau\chi(\tau)e^{-i\omega \tau}\ket{1}\otimes\hat{\phi}(X(\tau))\ket{0_M}+O(\lambda^2)~.
    \label{state}
\end{equation}
\end{widetext}
where the last equality is obtained by arresting the perturbation expansion of the exponential at first order in $\lambda$. From the Born rule, we get the probability of observing the detector in the excited state as
\begin{equation}
    q=\lambda^2\mathcal{F}_{\chi}(t_0;t)~,
    \label{p_e}
\end{equation}
where $\mathcal{F}_{\chi}(t_0;t)$ is the so-called response function, given by
\begin{equation}
    \mathcal{F}_{\chi}(t_0;t)=\int_{t_0}^td\tau \int_{t_0}^t d\tau' \chi(\tau)\chi(\tau') e^{-i\omega(\tau-\tau')}\mathcal{W}(\tau',\tau)~.
    \label{F}
\end{equation}
In the above definition, we used the distribution $\mathcal{W}(\tau',\tau)$; this is the pull-back to the the detector's world line of the $2$-point Wightman function $\mathcal{W}_2(X(\tau'),X(\tau))$, the latter being a special case of the more general $n$-point Wightman function
\begin{equation}
    \mathcal{W}_n(X(\tau_1),\dots,X(\tau_n))=\bra{0}\hat{\phi}(X(\tau_1))\dots\hat{\phi}(X(\tau_n))\ket{0} ~.
\end{equation}
Let us now fix $t=t_m>t_0$. As the $2$-point Wightman function is a well-defined distribution on $\mathbb{R}\times\mathbb{R}$~\cite{JunkerS02}, if the switching function is smooth and has compact support s.t. $\textrm{Supp}(\chi)\in[t_0,t_m]$, then Eq.~\eqref{F} gives an univocal result; in this case, Eq.~\eqref{p_e} is the probability of finding the detector in the excited state once the interaction ceased~\footnote{One could also consider the case in which the support of $\chi$ is larger than the integration interval. Then $q$ is the probability of finding the detector in the excited state when the interaction is still on. This case requires special care, as when the Wightman distribution is not integrated against a good test function Eq.~\eqref{p_e} depends on the chosen regularization procedure. As we will always consider integration intervals containing the support of $\chi$, we avoid discussing the problems arising in this case.}. 

Lastly, let us discuss the regularization we use to evaluate Eq.~\eqref{F}. As a function, $\mathcal{W}(\tau,\tau')$ diverges when the value of its arguments coincide. Therefore, using some regularization procedure is necessary to get finite results from the integrals of $\mathcal{W}$. Notice that this is only a mathematical trick as, since in this paper we are always interested in studying response functions for which the integration intervals include $\textrm{Supp}(\chi)$, our results does not depend on the specific choice of regularization. In practice, we regularize $\mathcal{W}(\tau',\tau)$ by a frequency cut-off, i.e. by substituting it with a function $\mathcal{W}_\epsilon(\tau',\tau)$ that is not divergent in $\tau=\tau'$, and performing the limit $\epsilon\to 0$ at the end of the calculation to recover a finite and sensible expression. Specifically, we will always consider either inertial trajectories, for which we substitute the Wightman function with 
\begin{equation}
    \mathcal{W}_\epsilon(\tau',\tau')=-\frac{1}{(2\pi)^2}\frac{1}{(\tau'-\tau-i\epsilon)^2-\abs{\mathbf{x}'-\mathbf{x}}^2}~,
    \label{ine_regular}
\end{equation}
or accelerated ones, for which we use
\begin{equation}
    \mathcal{W}_\epsilon(\tau',\tau)=-\frac{\alpha^2}{(4\pi)^2}\frac{1}{\sinh^2(\alpha(\tau'-\tau)/2-i\alpha\epsilon)}~,
    \label{acc_regular}
\end{equation}
where $\alpha$ is the detector's acceleration. In general, our discussion holds for all trajectories such that 
\begin{equation}
    \mathcal{W}(\tau',\tau)=w(\tau'-\tau)~.
    \label{A0}
\end{equation}

\subsection{Measurement-induced state collapse in QFT}
\label{sec:3.collapse}
Before describing RM on UDW detectors, we must ask if it is possible to use the standard the measurement postulate in relativistic QM. In fact, describing a measurement procedure in QFT is a longstanding problem~\cite{HellwigK70, Sorkin93, Lin14, BorstenEtAl21, Garcia-ChungEtAl21}. The main reason for this is that, if not extended carefully, the update procedure prescribed by the non-relativistic postulate may induce faster-than-light signalling, hence not being compatible with causality. To address this issue, many proposals have been put forward~\cite{HellwigK70,Fewster19,FewsterV20,BostelmannEtAl21}. Here, we follow the one recently appeared in Ref.~\cite{Polo-GomezEtAl21}: the reason for this choice is that the latter approach is the closest to an operational approach to quantum theory. There, the authors propose a measurement procedure using UDW detectors upon which a direct application of the non-relativistic measurement postulate is possible and compatible with causality. In particular, they show that if one accepts the field's state to be contextual (i.e. a subjective quantity, as opposed to a fundamental property of the system) causality is not violated by applying the measurement postulate on the UDW detector and updating the field's state consistently with the obtained outcome. Since this proposal is an essential piece of our construction, we here briefly review it with particular focus on the features we will need later.

Let the detector and the field initially be in the separable state
\begin{equation}
    \rho_0=\ket{0}\bra{0}\otimes\ket{\phi_0}\bra{\phi_0}~,
\end{equation}
and suppose the detector and the field interact for some finite time. For us, this interaction can be that obtained via the unitary operator \eqref{U} and by choosing the switching function to have compact support. Then, suppose that after the interaction ceased Alice performs a projective measurement (PVM) on the detector. For the sake of clarity, we take the PVM to be described by the measurement operators $\{\hat{M}_1,\dots,\hat{M}_n\}$. Assuming the non-relativistic update rule, if Alice finds the outcome $m_i$ associated with $\hat{M}_i$, then the total system collapses to the state
\begin{equation}
    \rho(m_i)=\frac{(\hat{M}_i\otimes\mathbb{I})\hat{U}\rho_0\hat{U}^\dagger(\hat{M}_i\otimes\mathbb{I})}{Tr[(\hat{M}_i\otimes\mathbb{I})\hat{U}\rho_0\hat{U}^\dagger]}
\end{equation}
with probability 
\begin{equation}
    p(m_i)=Tr[(\hat{M}_i\otimes\mathbb{I})\hat{U}\rho_0\hat{U}^\dagger]~,
\end{equation}
in accordance with the non-relativistic measurement postulate. By tracing over the Hilbert space of the detector, one gets the state of the field as
\begin{equation}
    \rho_\phi(m_i)=Tr_{\mathcal{D}}[\rho(m_i)]=\frac{Tr_{\mathcal{D}}[(\hat{M}_i\otimes\mathbb{I})\hat{U}\rho_0\hat{U}^\dagger(\hat{M}_i\otimes\mathbb{I})]}{Tr[(\hat{M}_i\otimes\mathbb{I})\hat{U}\rho_0\hat{U}^\dagger(\hat{M}_i\otimes\mathbb{I})]}~,
    \label{field_selective}
\end{equation}
which clearly depends on the specific outcome obtained. However, as $m_i$ is (classical) information that cannot be transmitted faster than light, imposing Eq.~\eqref{field_selective} to be the state of the field outside the causal future of the measurement would inevitably break causality. Specifically, by calling $\mathcal{I}$ the set of points for which the interaction and measurement happens, and $\mathcal{I}^+=\mathcal{J}^+(\mathcal{I})$ its causal future~\cite{HawkingE73}, then respecting causality means that no observer in in the complement of $\mathcal{I}^+$ can see the effect of the measurement performed by $A$ when performing any local operation. Hence, the prescription given in Ref.~\cite{Polo-GomezEtAl21} is that of using different update rules depending on the observer considered. In particular, their prescription is that the state of the system must be updated only for those observers having knowledge about the outcome and living in $\mathcal{I}^+$ (via the so-called selective update), or for those who know that the measurement was performed but not the outcome obtained (via the non-selective update). Hence, by calling $\rho^B_\phi$ the field's density operator as seen by an observer $B$, then:
\begin{enumerate}
    \item if $B$ is in $\mathcal{I}^+$ and knows the outcome obtained by $A$ (say $m_i$), $\rho^B_\phi=\rho_\phi(m_i)$ ;
    \item if $B$ is in $\mathcal{I}^+$ but does not know $A$'s measurement outcome, $\rho^B_\phi=\sum_ip(m_i)\rho_\phi(m_i)$;
    \item if $B$ is in the complement of $\mathcal{I}^+$ and knows $A$ performed a measurement, $\rho^B_\phi=\sum_ip(m_i)\rho_\phi(m_i)$;
    \item if $B$ is in the complement of $\mathcal{I}^+$ and doesn't know $A$ performed a measurement, $\rho^B_\phi=\ket{\phi_0}\bra{\phi_0}$.
\end{enumerate}
In particular, it is proven that the expectation values of any local field observable $\hat{O}$ having support outside $\mathcal{I}^+$ are equal when calculated via the density operators of the above points 3 and 4, meaning that the latter are the same FAPP.

\subsection{Effects of the switching function}
\label{sec:3.switching}
Despite appearing in Eq.~\eqref{p_e}, when integrated over infinitely long interaction times the switching function $\chi$ does not play a crucial role in describing the transition probability~\cite{SriramkumarP96,HiguchiEtAl93}. However, in the finite-time interaction case the details of the detector's response depends on the switching function describing the interaction, and hence deviate from those obtained in the infinite-time interaction case. This fact should not come as a surprise: it is a mere consequence of the energy-time uncertainty principle. In fact, no detector that is switched on for some finite time $T_{\textrm{on}}$ can measure energy differences with an accuracy greater than $1/T_{\textrm{on}}$. An example of this is given by the finite-time interaction between a scalar field in its ground state and an inertial detector. By preparing the two systems in their ground states and let them interact via the interaction Hamiltonian Eq.~\eqref{H_int} with $\chi(\tau)=1$ for all $\tau$, the states of the two systems do not change, consistently with the fact that all quantities are left invariant under time translations. On the contrary, by making $\chi(\tau)$ change over time, it is possible to find the detector in the excited state after the interaction took place. This is because turning on and off the interaction breaks the time translation invariance that caused the state not to evolve in the previous case.

Since we are interested in performing RM on one detector, and doing the measurements when the interaction is switched off, we are forced to consider finite-time interactions. Therefore, the results we will find must not be confronted with the idealized infinite-time interaction case, but the finite one. Moreover, to fit in the RM scenario we should choose the switching function to be smooth and having compact support, the latter coinciding with the time intervals in which the interaction is on. However, for the sake of simplicity let us consider a switching function $\chi(\tau)$ to be any smooth function $\chi(\tau,\Delta)$ such that
\begin{equation}
    \chi(\tau,\Delta)\simeq \begin{dcases}
        1 & \text{for}~\abs{\tau} \ll \Delta\\
        0 & \text{for}~\abs{\tau} \gg \Delta
    \end{dcases}
    \label{switching_condition}
\end{equation}
This describes an interaction happening around $\tau=0$ and having duration smaller than a given time scale $\Delta$. In particular, in what follows we will always use the Gaussian window function
\begin{equation}
    \chi_g(\tau,\sigma)=\exp\left(-\frac{\tau^2}{2\sigma^2}\right)~,
    \label{S^(1)}
\end{equation}
with $\Delta=4\sigma$. On the one hand, $\chi_g$ is zero FAPP outside the interaction times, and hence suitable for studying the effects of switching on and off of the interaction~\cite{SriramkumarP96}. On the other hand, it is not compactly supported and, as we aim at performing RM on the UDW by letting it interact many separate times with the field, the latter is a crucial property for our scope. To solve this issue one can follow two routes. First, one can consider a sufficiently small collection of non-compactly supported switching functions and build a sequence of switching by accepting to use a function that is only FAPP zero outside the interaction intervals. Second, one can consider bump switching functions~\cite{Tu11, Nestruev06} such as 
\begin{equation}
    \chi_b(\tau,\Delta)=\begin{dcases}
        \exp\left(\frac{\tau^2}{\tau^2-\Delta^2}\right) & \abs{\tau} < \Delta\\
        0 & \text{elsewhere}
    \end{dcases}
    \label{S^(3)}
\end{equation}
at the cost of having much harder analytic expression to solve for getting the probabilities we are interested in.

In the following, we decided to follow the first route, and left the second for future work. In particular, starting from any switching function $\chi$ satisfying the condition \eqref{switching_condition}, one can build a new switching function describing $N$ almost separate interactions. Hence, a repeated switching function is obtained by summing over the collection of switching functions $\{\chi_k\}$, each satisfying the condition \eqref{switching_condition} and translated to be centered in the middle of a time interval $I_k$ such that $I_k\cap I_{k'}=\emptyset,~\forall k,k'~s.t.~ k\neq k'$, i.e 
\begin{equation}
    \chi_{RM}(\tau)=\sum_{k=0}^N\chi_k(\tau)~.
    \label{chi_RM}
\end{equation}
It is clear from this contraction that the obtained $\chi_{RM}(\tau)$ is FAPP zero everywhere except in the times $\tau\in\cup_k I_k$. Notice that whenever the switching function $\chi$ is not taken to have compact support, $N$ must be smaller than
\begin{equation}
    N_{max}\ll \left[\max_{I_k}\{\chi(\tau)\}\right]^{-1}~.
    \label{chi_RM_limit}
\end{equation}
On the contrary, if $\chi$ has compact support then $N$ can be infinite and the sum in Eq.~\eqref{chi_RM} a series. As from an operational perspective Eq.~\eqref{chi_RM} describes an observer repeatedly turning on and off the detector, using a compactly supported $\chi$ has the additional advantage of leaving an observer the freedom to decide when to stop iterating the repeated interaction, while the milder condition \eqref{switching_condition} imposes a limit on the number of interactions from its beginning throughout Eq.~\eqref{chi_RM_limit}. Still, the simplicity obtained by using non-compact supported functions favour them despite the interpretational issues they come with.

\subsection{RM on UDW detectors}
\label{sec:3.RM+UDW}
According to the frequentist interpretation of the Born rule, to test the transition probability obtained in Eq.~\eqref{p_e} one needs either 1) $N$ distinct detectors interacting with $N$ independent fields or 2) a way to reset both the states of the detector and the field after each one of $N$ repetitions of the measurement, where $N$ is sufficiently large to collect statistics. Without discussing the feasibility of the above possibilities, in this section we assume neither as achievable, and describe the RM scheme for a single UDW detector interacting many times with the same quantum field~\footnote{Notice that one could instead take many UDW detectors interacting at different times with the same quantum field and obtain the same result. In fact, as long as the state of the detector is reset as discussed in Sec.~\ref{sec:2}, there is no difference between using the same detector or a new one for each measurement; yet the necessity of RM comes from the inability of controlling the field state, making these two choices equivalent for what concerns our results.}.

Let us consider the detector and field described in Sec.~\ref{sec:3.unruh}, the former moving along a trajectory $X(\tau)$, where $\tau$ the detector's the proper time, and let the switching function appearing in the interaction Hamiltonian \eqref{H_int} be \eqref{chi_RM}. Then, by measuring the detector after each bump in $\chi_{RM}$ we obtain a RM procedure as it is described in Sec.~\ref{sec:2}. By construction all the measurements are performed along a single world line, meaning that a time order between them can be introduced so that each can be said to happened either in the causal future or past of any other. Hence, if we suppose an observer carrying the detector and having full knowledge about preceding results, she is always entitled to use the selective measurement rule as defined in Sec.~\ref{sec:3.collapse}. For this reason, we will here refer to the state of the field as if it was uniquely defined and use the non-relativistic update rule on it, while what we really mean is the state of the field as seen by the observer, and what we update is the contextual state described by her.

In order to realize the RM procedure for UDW detectors, and obtain something close to the Born rule, we must select the interaction intervals $I_k$ appearing in Sec.~\ref{sec:3.switching} to be all alike. To this end, we define a \textit{repetition interval} $R_k$ as a proper time interval of duration $T=T_{\textrm{on}}+T_{\textrm{off}}$ made of an \textit{interaction part} $I_k$ (having duration $T_{\textrm{on}}$), during which the interaction Hamiltonian is switched on, and a \textit{measurement one} (of length $T_{\textrm{off}}$), for which there is no interaction and the observer can perform measurements. Then, we glue together many of these repetition intervals and label them by increasing natural numbers $R_k$. Hence, assuming the first repetition interval to begin at proper time $\tau=0$ the intervals take the form $R_k=[Tk,T(k+1))$, which can be split as
\begin{equation}
    R_k=\underbrace{[Tk,Tk+T_{\textrm{on}}]}_{I_k}\cup\underbrace{(Tk+T_{\textrm{on}},Tk+T_{\textrm{on}}+T_{\textrm{off}})}_{k\textrm{-th ~measurement interval}}~,
\end{equation}
such that $\textrm{Supp}(\chi_k)\subset I_k$. Moreover, we take all $\chi_k$ to be equally shaped within their support. Finally, assuming taking the above prescriptions we obtain that the UDW detector interacts many times with the massless scalar field, and the two evolve under the unitary operator
\begin{equation}
    \hat{U}_k=\mathcal{T}\left[\exp\left(-i\lambda\int_{I_k}d\tau\hat{H}_{\mathrm{int}}(\tau)\right)\right]~,
\end{equation}
where $\hat{H}_{\mathrm{int}}$ defined in \eqref{H_int}, with the choice $\chi(\tau)=\chi_{RM}(\tau)$. Also, since within the $k$-th interaction interval $\chi_{RM}$ coincide with $\chi_k$, one can simply replace the latter in the integral and get the same result. Thanks to this construction, our discussion about UDW detectors fits the RM scheme presented in Sec.~\ref{sec:2}, with the quantum scalar field initialized in the Minkowski vacuum playing the role of the environment $\mathcal{E}$. Moreover, expanding $\hat{U}_k$ at zeroth order in $\lambda$ gives
\begin{equation}
    \hat{U}_k=\mathbb{I}_{\mathcal{D}}\otimes\mathbb{I}_\phi+O(\lambda)~,
    \label{U_for_UDW}
\end{equation}
meaning that the unitary operator $\hat{U}$ appearing in Eq.~\eqref{U_Weak} is the identity. Therefore, the Born rule we are searching for must describe an approximation to the trivial situation where the detector is always found in its ground state, as it is clear from Eq.~\eqref{p_e}. As the measurements on $\mathcal{D}$ can only give $0$ or $1$ as outcomes, the output of the RM procedure is a bit string $B_L$, where $L$ is the number of RM performed; hence we can use the notation introduced in Sec.~\ref{sec:2.bit_string_probabilities}. In particular, we recall that
\begin{equation}
    B_L=(b_1,\dots,b_L)=(L;N_1,\dots,N_n)~,
\end{equation}
where $n$ is the number of ones in $B_L$ and the $\{N_i\}$ label their locations. By expressing the probability $P(b_{L+1}=1|B_L)$ via the shorthand notation
\begin{equation}
    \mathcal{P}(L+1|N_1,\dots,N_n)~,
    \label{conditional_probailities}
\end{equation}
the probability $P(B_L)$ of obtaining the string $B_L$ is easily obtained from the definition of conditional probability as
\begin{widetext}
\begin{equation}
    P(B_L)=\begin{dcases}
        P\left(b_L=1|B_{L-1}\right)P(B_{L-1})=\mathcal{P}\left(L|N_1,\dots,N_{n-1}\right)P(B_{L-1})~&\textrm{if}~b_L=1\\
        P\left(b_L=0|B_{L-1}\right)P(B_{L-1})=\left[1-\mathcal{P}\left(L|N_1,\dots,N_n\right)\right]P(B_{L-1})~&\textrm{if}~b_L=0
    \end{dcases}
\end{equation}
\end{widetext}
In the rest of this section, we recursively apply the above formula to calculate the probability of obtaining any string of results from the quantities \eqref{conditional_probailities}.

\subsection{Transition probabilities for RM on UDW detectors}
\label{sec:3.probabilities}
As motivated in the previous section, the most relevant quantities for RM on UDW detectors are the probabilities of observing the detector in the state $\ket{b_L}$ in the measurement interval labelled by $L$, given that is was previously found in the excited state $n-1$ times during the measurement intervals labelled by $\{N_1,\dots, N_{n-1}\}$, namely
\begin{equation}
    \mathcal{P}(L|N_1,\dots,N_{n-1})~.
    \label{def_p}
\end{equation}
In what follows, we will often refer to the above quantity simply as to $\mathcal{P}_L$, implicitly assuming the dependence on the previous history. In this section we provide both a closed formula for $\mathcal{P}_L$ and a numerical strategy for finding upper and lower bounds to it. The utility of the latter comes from the fact that the $\mathcal{P}_L$ are generally hard to evaluate, and having a simplified quantity describing their magnitude can be useful. In particular, these bounds are obtained by assuming the probability $q$ of finding the detector in the excited state for the first time (i.e. Eq.~\eqref{p_e}), and proceeding from there under simplifying assumptions. Note that, from this section onward, we will take the switching function to be fixed, hence not making the functional dependence of $\mathcal{F}$ on it explict, and stop to first order in $\lambda$ in $U_k$ expansions and at $\lambda^2$ in all probabilities. Finally, throughout this section we will use the shorthand notations
\begin{equation}
    \int_J=\int_{TJ}^{TJ+T_{\textrm{on}}}d\tau_i\int_{TJ}^{TJ+T_{\textrm{on}}}d\tau'_i \chi(\tau_i)\chi(\tau'_i) e^{-i\omega(\tau-\tau')}
\end{equation}
and
\begin{equation}
    \int_{J} = 2\int_{TJ}^{TJ+T_{\textrm{on}}}du_i\int_{0}^{u_i-TJ}ds_i \chi(u_i)\chi(u_i-s_i) \cos(\omega s_i)~,
    \label{int}
\end{equation}
where the one used depends on the integration variables appearing in the specific expression considered.

\subsubsection{Analytic expressions}
\label{App2_1}
At first order in $\lambda$, the probability of observing the detector in the excited state $N_1$, given it was never found in the excited state before, is
\begin{equation}
    \mathcal{P}(L)=\lambda^2\mathcal{F}(TL,TL+T_{on})=\int_L\mathcal{W}(X(\tau'),X(\tau))~.
    \label{F_repeat}
\end{equation}
Thanks to Eq.~\eqref{A0}, $\mathcal{F}$ does not depend on the specific interval $L$ in which the first transition is observed, meaning that we can simply write $P(L)=q$, and the integrals in Eq.~\eqref{F_repeat} can be evaluated by picking any value of $L$. In particular, by taking $L=0$ one gets
\begin{equation}
    q=\lambda^2\int_0^{T_{\textrm{on}}}du \int_0^{u} ds \chi(u)\chi(u-s)\left(e^{-i\omega s}\mathcal{W}(s)+h.c.\right)~,
    \label{change_of_variables}
\end{equation}
where we also used the change of variables
\begin{equation}
\begin{dcases}
    \tau=u,~\tau-\tau'=s~~&\textrm{for}~ \tau>\tau'\\
    \tau'=u,~ \tau'-\tau=s~~ &\textrm{for}~ \tau>\tau'
\end{dcases}
    \label{c.o.v}
\end{equation}
as it is customary in the theory of UDW detectors~\cite{Schlicht04}. While the above expression can be further simplified by noting that the strong Huygens' principle guarantees $\mathcal{W}(s)$ to be a real-valued function~\footnote{The strong Huygens' principle states that, for massless scalar fields in $(3+1)$-dimensions, the commutator $[\hat{\phi}(x),\hat{\phi}(y)]$ vanishes whenever $x$ lays inside the future light cone of $y$ (and vice versa); hence it vanishes for any pair of points of a time-like trajectory~\cite{JonssonEtAl15, CzaporML08, McLenaghan74}.}, for the moment we simply rewrite it as
\begin{equation}
    q=2\lambda^2\int_0^{T{on}}du \int_0^{u} ds \chi(u)\chi(u-s)\mathfrak{Re}\left[e^{-i\omega s}\mathcal{W}(s)\right]~;
    \label{q}
\end{equation}
this is because, being $\mathcal{W}$ divergent for $s=0$, it must be regularized making it complex as prescribed by Eq.s~(\ref{ine_regular}-\ref{acc_regular}). Finally, assuming the detector is actually found in the excited state in some interval $N_1$, the state of field collapses to
\begin{equation}
    \ket{\phi_{N_1}}=\mathcal{N}\int_{T{N_1}}^{TN_1+T_{\textrm{on}}}d\tau\chi(\tau)e^{-i\omega\tau}\hat{\phi}(X(\tau))\ket{0}~,
    \label{1st state}
\end{equation}
where $\mathcal{N}=\lambda/\sqrt{q}$ is a normalization constant. As a consequence, later probabilities must be evaluated by taking the field to be in the $N_1$-dependent state \eqref{1st state} instead of the Minkowski vacuum~\cite{Garcia-ChungEtAl21}. Hence, the probability of observing the detector in the excited state at the $L$-th interval after having found it in the excited state for the first time in the $N_1$-th interval is
\begin{equation}
    \mathcal{P}_2=\frac{\lambda^4}{q}\int_{N_1,L} \mathcal{W}_4(X(\tau'_1),X(\tau'_2),X(\tau_2),X(\tau_1))~.
    \label{full_p_2}
\end{equation}
This expression can be simplified by modifying the regularization prescription so that each time we find a cosine multiplying a divergent $2$-point Wightman function inside an integral, we apply the substitution
\begin{equation}
    \cos(\omega s)\mathcal{W}(s)\longrightarrow\frac{1}{2}\mathfrak{Re}\left[e^{-i\omega s}\mathcal{W}_\epsilon(s)\right]
    \label{new_prescription}
\end{equation}
and take the $\epsilon\rightarrow 0$ limit at the end of the calculation. In this way, and we can write Eq.~\eqref{full_p_2} as
\begin{equation}
    \mathcal{P}_2=\frac{\lambda^4}{q}\int_{N_1,L} \mathcal{W}_4(X(u_2),X(u_2-s_2),X(u_1),X(u_1-s_1))~,
    \label{p_2}
\end{equation}
where we performed the same change of variables as in Eq.~\eqref{change_of_variables} for each pair of variables, used the strong Huygens principle to commute the fields evaluated in separate interaction intervals. In this way, all $2$-point Wightman functions appearing from Wick's decomposition that do not evaluate in the same interval are automatically real valued and finite, while all others must be regulated as prescribed in \eqref{new_prescription}. Again, after the second outcome giving one as a result happens in the interval $N_2$ the state of the field collapses to the $(N_1,N_2)$-dependent state
\begin{widetext}
\begin{equation}
    \ket{\phi_{N_2,N_1}}=\mathcal{N}_2\int_{T_{N_1}}^{T_{N_1}+T_{\textrm{on}}}\int_{T_{N_2}}^{T_{N_2}+T_{\textrm{on}}}d\tau_1d\tau_2\chi(\tau_1)\chi(\tau_2)e^{-i\omega(\tau_1+\tau_2)}\hat{\phi}(X(\tau_2))\hat{\phi}(X(\tau_1))\ket{0}~,
\end{equation}
\end{widetext}
where $\mathcal{N}_2=\lambda^2/\sqrt{q\mathcal{P}_2}$. Hence, recursively applying this procedure $n$ times gives the probability defined in Eq.~\eqref{def_p} as
\begin{equation}
            \mathcal{P}_n=\frac{\lambda^{2n}}{\prod_{j=1}^{n-1}\mathcal{P}_j}\int_{N_1,\dots,N_{n-1},L} \mathcal{W}_{2n}(X_n,X'_n,\dots,X_1,X'_1)
    \label{p_N}
\end{equation}
where $X_n=X(u_n)$ and $X'_n=X(u_n-s_n)$. Therefore, in order to calculate the above probabilities one needs to evaluate all $(2k)$-points Wightman functions for $k<n$, and integrate them against the function
\begin{equation}
    f(\tau_1,\tau'_1,\dots,\tau_k,\tau'_k)=\prod_{j=1}^{k}\chi(\tau_i)\chi(\tau_i)e^{-i\omega(\tau_i-\tau'_i)}~,
\end{equation}
where $\tau_i,\tau'_i\in I_{N_i},~\forall i<n$, over the intervals $(N_1,\dots,N_{n-1},L)$ corresponding to the obtained excited outcomes, plus the one over which we are currently considering the probability of getting $b_L=1$.

Before giving the general strategy for finding bounds for Eq.~\eqref{p_N}, let us give a alternative expression for $\mathcal{P}_n$. Starting from the case $n=2$, we apply Wick's theorem to Eq.~\eqref{p_2} and obtain
\begin{equation}
\begin{split}
    \mathcal{P}_2=\frac{\lambda^4}{q}\int_{N_1,L}&[\mathcal{W}(s_2)\mathcal{W}(s_1)+\\+&\mathcal{W}(u_2-u_1)\mathcal{W}(u_2-s_2-u_1+s_1)+\\+&\mathcal{W}(u_2-s_2-u_1)\mathcal{W}(u_2-u_1+s_1)]~.
\end{split}\label{p_2_Wick}
\end{equation}
For later convenience, let us define
\begin{equation}
    \begin{dcases}
            \mathcal{C}_{2,1}(J_1,J_2)=\int_{J_1,J_2} \mathcal{W}(u_2-s_2-u_1)\mathcal{W}(u_2-u_1+s_1)\\
            \mathcal{C}_{2,2}(J_1,J_2)=\int_{J_1,J_2} \mathcal{W}(u_2-u_1)\mathcal{W}(u_2-s_2-u_1+s_1)
    \end{dcases}
    \label{C_2_i}
\end{equation}
and
\begin{equation}
\mathcal{C}_2(N_1,L)=\mathcal{C}_{2,1}(N_1,L)+\mathcal{C}_{2,2}(N_1,L)~,
\label{C_2(N_1,L)}
\end{equation}
where the first lower index in Eq.~\eqref{C_2_i} is the number of Wightman functions inside the integral, and the second runs over all the possible ways of connecting the points between the two intervals. For later convenience, let us call  $\mathfrak{C}(2)$ the number of these combinations, i.e. $\mathfrak{C}(2)=2$. Moreover, we define 
\begin{equation}
\mathcal{Q}=\int_{J} \mathcal{W}(s_1)=\frac{q}{\lambda^2}~,
\label{Q}
\end{equation}
which is independent of $J$, as usual. Since the first term in Eq.~\eqref{p_2_Wick} only involves Wightman functions evaluated in points belonging to the same interaction interval, and hence it can be factorized into two separate integrals for which the prescription \eqref{new_prescription} applies, $\mathcal{P}_2$ can be written as
\begin{equation}
    \mathcal{P}_2=q\left(1+\frac{\mathcal{C}_2(N_1,L)}{\mathcal{Q}^2}\right)~,
    \label{p_2_with_defs}
\end{equation}
meaning that, in order to get Eq.~\eqref{p_2} for any possible bit string $M=(L;N_1)$, one only needs to evaluate the fraction
\begin{equation}
    F_{N_1,L}=\mathcal{C}_2(N_1,L)/\mathcal{Q}^2~.
    \label{F_2}
\end{equation}

We now generalize the above strategy for all $n$. To make the discussion easier, let us call $\mathcal{W}_s$ any $2$-point Wightman function evaluated within the same interaction interval, and $\mathcal{W}_d$ those evaluated in points belonging to different interaction intervals. This nomenclature allows us to classify the
\begin{equation}
    \mathcal{N}(n)=\frac{(2n)!}{2^nn!}
\end{equation}
terms appearing from the Wick's theorem decomposition of the $2n$-points Wightman function appearing in $\mathcal{P}_n$. Following the above logic, we recognise that one of these terms is always the integral of a product of $n$ $\mathcal{W}_s$ functions, and gives $\mathcal{Q}^n$. Then, for each choice of two intervals amongst the $n$ possible ones (hence giving $n(n-1)/2$ terms of this kind), we have a product of $(n-2)$ $\mathcal{W}_s$ functions and two possibilities for two $\mathcal{W}_d$ functions evaluated across the selected intervals (i.e. the terms appearing in Eq.~\eqref{C_2(N_1,L)}, with its arguments being the labels of the selected intervals). Next, for each choice of three intervals we have a product of $(n-3)$ $\mathcal{W}_s$ functions and eight possibilities of three $\mathcal{W}_d$ functions evaluated across the three selected intervals, and so on. To explicitly write the expression obtained by this procedure, in analogy with \eqref{C_2(N_1,L)} we start by defining the quantity
\begin{equation}
    \mathcal{C}_k(J_1,\dots,J_k)
    \label{C_k(N_1,...,N_k)}
\end{equation}
as the sum of all possible combinations of $k$ $\mathcal{W}_d$ functions integrated over the interaction intervals labelled by $(J_1,\dots, J_k)$. When expanded, $\mathcal{C}_k(J_1,\dots,J_k)$ is a sum of $\mathfrak{C}(k)$ terms, the latter being the number of possible combinations in which $2k$ paired objects can be re-paired so that none is associated with its initial partner, found via the recurrence relation~\cite{Margolius01, Brawner00, OEIS22}
\begin{equation}
\begin{dcases}
        \mathfrak{C}(k)=2(k-1)[\mathfrak{C}(k-1)+\mathfrak{C}(k-2)]\\
        \mathfrak{C}(0)=1\\
        \mathfrak{C}(1)=0
\end{dcases}
\end{equation}
Then, similarly to Eq.~\eqref{F_2}, we define
\begin{equation}
    F_{J_1,\dots,J_k}=\frac{\mathcal{C}_k(J_1,\dots,J_k)}{\mathcal{Q}^k}
    \label{Fn}
\end{equation}
and
\begin{equation}
\begin{split}
    \Sigma(J_n,\dots,J_1)&=\int_{J_1,\dots,J_n}\mathcal{W}_{2n}(X_k,X'_k,\dots,X_1,X'_1)\\
    &=\mathcal{Q}^n\left(1+\sum_{k=2}^n\sum_{\{i\}}F_{J_{i_1},\dots,J_{i_k}}\right)~,
\end{split}
    \label{sigma definition}
\end{equation}
where the label $\{i\}$ runs over the ${n \choose k}$ possible choices of $k$ intervals amongst the $n$ ones. Indeed, it is 
\begin{equation}
    \sum_{k=0}^n{n \choose k}\mathfrak{C}(k)=\mathcal{N}(n)~.
    \label{decomposition}
\end{equation}
Finally, thanks to Eq.~\eqref{sigma definition} it is easy to show that \eqref{p_N} can be expressed as
\begin{equation}
    \mathcal{P}_n=
    q\left(\frac{1+\sum_{k=2}^n\sum_{\{i\}}F_{J_{i_1},\dots,J_{i_k}}}{1+\sum_{k=2}^{n-1}\sum_{\{i\}}F_{J_{i_1},\dots,J_{i_k}}}\right)~;
    \label{p as sigmas}
\end{equation}
once again, the task of calculating $\mathcal{P}_n$ is reduced to that of evaluating all the fractions defined in Eq.~\eqref{Fn}. In addition to being more compact, the advantage of this equation when compared with Eq.~\eqref{p_N} is that it factorize all divergent integrals in $\mathcal{Q}$ which we suppose we can evaluate, hence giving a finite and manageable expression for $\mathcal{P}_n$. Moreover, Eq.~\eqref{p as sigmas} makes it explicit that $\mathcal{P}_n$ can be written as a correction to the standard Born rule given by \eqref{p_e}.

\subsubsection{Upper and lower bounds}
\label{App2_2}
Despite its compactness, Eq.~\eqref{p as sigmas} is still extremely hard to evaluate analytically. Therefore, in this section we give upper and lower bounds for $\mathcal{P}_n$ to have an estimate of the results one should expect to obtain. This is done under the following assumptions:
\begin{itemize}
    \item [\textbf{A0:} ] the trajectory $X(\tau)$ satisfies Eq.~\eqref{A0}.
    \item [\textbf{A1:} ] the $2$-point Wightman function is definite negative, monotonously increasing and such that $\lim_{s\to 0}\mathcal{W}(s)=-\infty$.
    \item [\textbf{A2:} ] $T_{\textrm{on}}\omega\leq \pi/2$.
\end{itemize}
Note that we have already assumed \textbf{A0} since the beginning of this section, and is here listed only for completeness. Moreover, we also make two supplementary assumptions:
\begin{itemize}
    \item [\textbf{B1:} ] $s\gg s'~\Rightarrow~\mathcal{W}(s)\gg\mathcal{W}(s')$.
    \item [\textbf{B2:} ] the switching function is selected so that $T_{\textrm{off}}\gg T_{\textrm{on}}$, meaning that the detector rests long times between each measurement.
\end{itemize}
While not being necessary, these later assumptions provide an easier yet useful picture for finding the upper and lower bounds we are interested in. 

\begin{figure}
    \centering
    \includegraphics[width=0.5\textwidth]{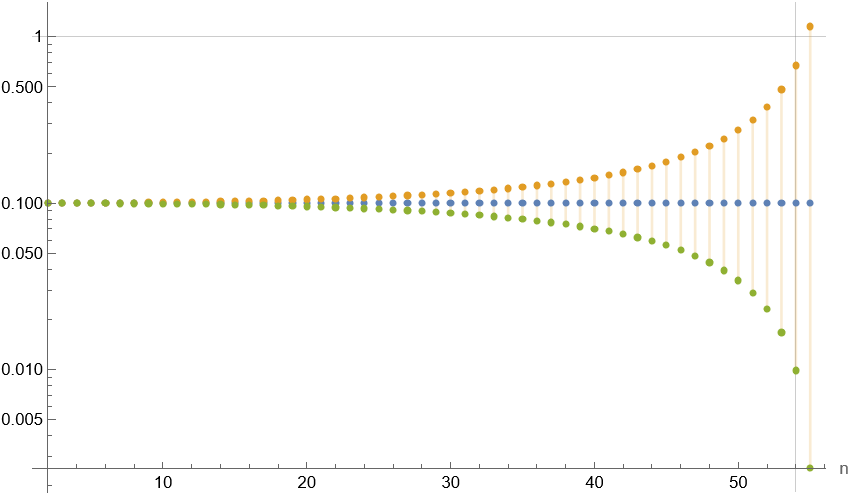}
    \caption{Log-plot of the loose upper (orange dots) and lower (green dots) bounds for $\mathcal{P}_n$, with $\gamma=0.01$ and $q=0.1$ (blue dots). The values of $\mathcal{P}_n$ for $n$ greater than $\mathcal{N}(q,\gamma)=55$ (vertical grey line) are clearly meaningless, as they give probabilities larger than 1 (horizontal grey line).}
    \label{fig:loose_bounds}
\end{figure}

While general strategies to get upper and lower bounds are given in App.~\ref{app:F} (where both tight and loose bounds are provided), we here only give the expression of a loose bound in term of the parameter \begin{equation}
    \gamma=\frac{\mathcal{W}(T_{\textrm{off}})}{\mathcal{W}(T_{\textrm{on}})}~,
    \label{gamma_loose}
\end{equation}
which, assuming \textbf{B1} and \textbf{B2}, satisfies
\begin{equation}
    \gamma\ll 1~.
\end{equation}
The details of the following discussion can be found in App.~\ref{app:F_loose}. By noticing that $\mathcal{C}_{k,i}(J_1,\dots,J_k)$ has sign $(-1)^{k}$, we can write is as
\begin{equation}
      \mathcal{C}_{k,i}(J_1,\dots,J_k)=(-1)^{k}\int_{J_1,\dots,J_k}\prod_{i=1}^k\abs{\mathcal{W}(\mathfrak{p}_i(\bold{u},\bold{s}))}~,
\end{equation}
where $\mathfrak{p}_i(\bold{u},\bold{s})$ are some linear combination of the components of $\bold{u}=(u_1,\dots,u_k)$ and $\bold{s}=(s_1,\dots,s_k)$. Thanks to \textbf{A2}, all functions appearing in the integral are now definite positive in the integration region, and hence
\begin{equation}
     0\leq\frac{(-1)^k\abs{\mathcal{C}_{k,i}(J_1,\dots,J_k)}}{\mathcal{Q}^k}\leq\gamma^k ~,
    \label{C_bound}
\end{equation}
which means
\begin{equation}
    0\leq \abs{F_{J_1,\dots,J_k}}\leq \mathfrak{C}(k)\gamma^k~.
    \label{F_bound}
\end{equation}
On the one hand, this expression permits us to find a bound for $\mathcal{P}_n$ as
\begin{equation}
     q(1-\delta_n)\leq\mathcal{P}_n(L;N_1,\dots,N_{n-1})\leq q(1+\epsilon_n) ~,
    \label{useful_bound}
\end{equation}
where 
\begin{equation}
    \epsilon_n,\delta_n\ll 1~,
    \label{epsilon_ndelta_n}
\end{equation}
for some $n$. On the other hand, Eq.\eqref{F_bound} makes it explicit that Eq.~\eqref{useful_bound} cannot hold for any $n$; indeed, since for large $k$ we have~\cite{OEIS22}
\begin{equation}
    \mathfrak{C}(k)\sim \frac{(2k-1)!!}{\sqrt{e}}~,
    \label{c(k) growth}
\end{equation}
the above bounds become meaningless for all $n$ larger than some limiting $\mathcal{N}(q,\gamma)$. This is better shown in Fig.~\ref{fig:loose_bounds}, where upper and lower loose bounds are obtained assuming the Wightman function \eqref{ine_regular}, $q=0.1$, and $\gamma=0.01$. The grey line in Fig.~\ref{fig:loose_bounds} represents the maximum $\mathcal{N}(q,\gamma)$ for which the bounds retain their validity. In particular, we will assume the bound \eqref{useful_bound} holds up to some
\begin{equation}
    \mathbf{n}<\mathcal{N}(q,\gamma)~.
    \label{bold_n}
\end{equation}
Finally, it is important to note that Eq.~\eqref{useful_bound} does not depend on the specific $(N_1,\dots,N_{n-1}\,L)$ considered. This property is to be ascribed to the looseness of the bound provided, and it is easily lost if considering tighter bounds (see App.~\ref{app:F_tight }).

Before concluding this section, let us summarize and discuss our results. Despite a Born rule is not generally expected to emerge from RM on one UDW, we found that the first few outcome probabilities (at most up to $\mathcal{N}(q,\gamma)$) do not deviate much from the Born one if $T_{\textrm{off}}$ is sufficiently longer than $T_{\textrm{on}}$. In particular, under these conditions we have
\begin{equation}
    \mathcal{P}_n(L;N_1,\dots,N_{n-1})\simeq q~,~~\forall n\leq \mathcal{N}(q,\gamma)~.
\end{equation}
Hence, despite $\mathcal{P}_n$ do not arise from a frequentist approach, since Eq.~\eqref{easy_threshold} is satisfied one can nonetheless extract information about later outcomes from earlier ones as they would do in the standard frequentist case, even without having access to a collection of copies of fields and UDW detectors.

\section{Explicit examples}
\label{sec:4}
\begin{figure*}
     \centering
     \begin{subfigure}[b]{0.49\textwidth}
         \centering
         \includegraphics[width=.7\textwidth]{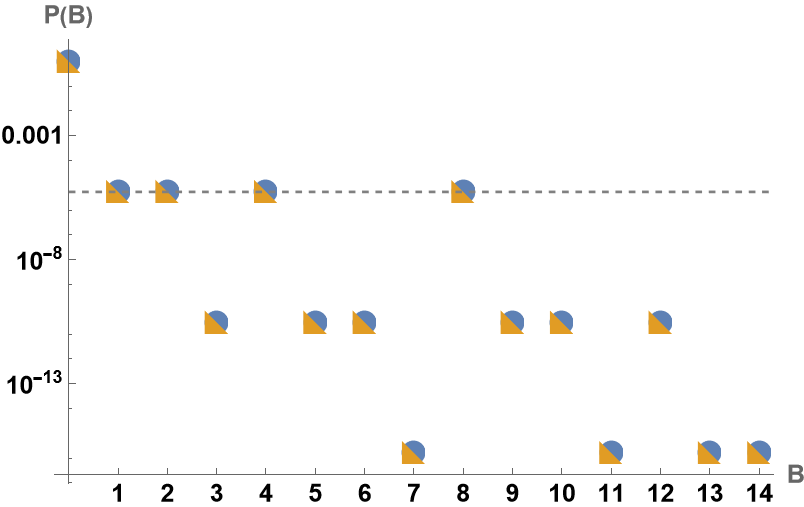}
         \caption{Results for the inertial trajectory, with $\omega=0.2$, $\sigma=1$, $T_{\textrm{on}}=T_{\textrm{off}}/10=8\sigma$, and $\lambda=10^{-2}$.}
         \label{fig:Inertial_Born_and_Almost}
     \end{subfigure}
     \hfill
     \begin{subfigure}[b]{0.49\textwidth}
         \centering
         \includegraphics[width=.7\textwidth]{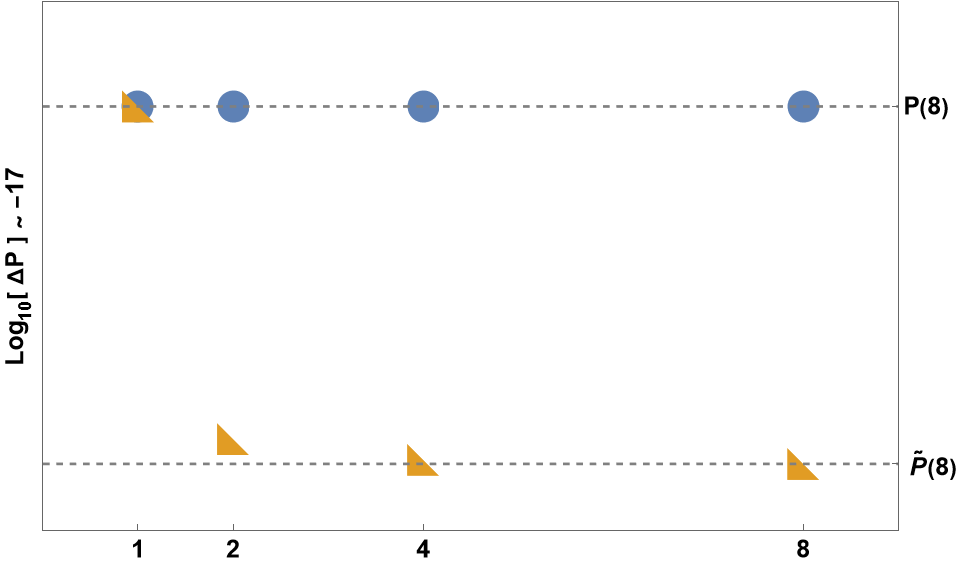}
         \caption{Magnification on the dashed line appearing on the left.\\~}
         \label{fig:Inertial_Born_and_Almost_Detail}
     \end{subfigure}
     \begin{subfigure}[b]{0.49\textwidth}
         \centering
         \includegraphics[width=.7\textwidth]{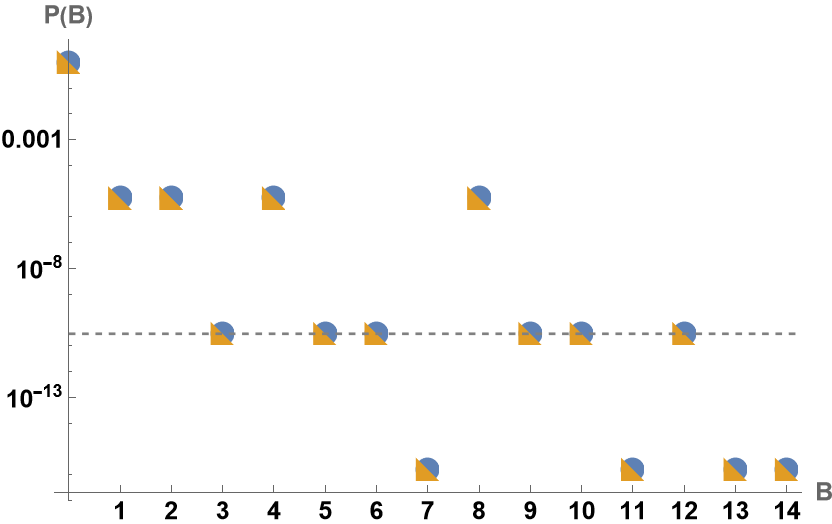}
         \caption{Results for the accelerated trajectory, with $\omega=0.2$, $\sigma=1$, $T_{\textrm{on}}=T_{\textrm{off}}/10=8\sigma$, $\lambda=10^{-2}$, and $g=0.1$.\\~}
         \label{fig:g=10_Born_and_Almost}
     \end{subfigure}
     \hfill
     \begin{subfigure}[b]{0.49\textwidth}
         \centering
         \includegraphics[width=.7\textwidth]{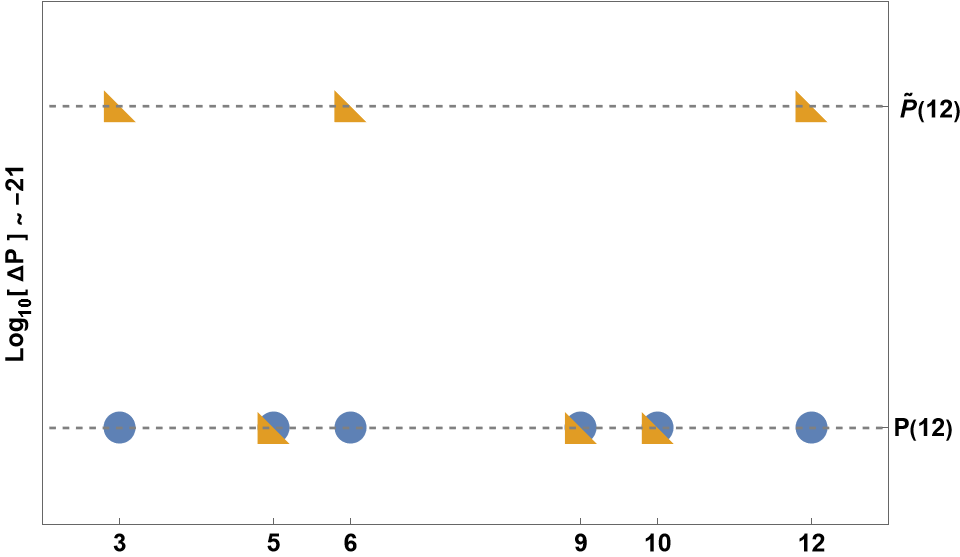}
         \caption{Magnification on the dashed line appearing on the left; orange triangles appearing having same height have differences in height of order $\sim 10^{-24}$.}
         \label{fig:g=10_Born_and_Almost_Detail}
     \end{subfigure}
         \caption{Logarithmic plots for the probabilities of obtaining the bit strings $B$, assuming that the Born rule holds (blue circles) and in the RM scenario (orange triangles), for the inertial (a) and accelerated (c) trajectories. On the horizontal axis, the bit strings are indicated in base $10$ representation. The (overlapping) gray lines correspond to the Born and RM probabilities of finding $B=(1,0,0,0)$, in panel (a), and $B=(1,1,0,0)$, in panel (c). In plots (b) and (d), these lines are magnified to show how the RM probabilities depend on the previous history (see main text for more details).}
        \label{fig:Born_and_Almost}
\end{figure*}
In this section we provide explicit results for the inertial and accelerated trajectories. In particular, in the coordinates system $(t,\bold{x})$ of an inertial observer, we choose
\begin{equation}
    X_{I}(\tau)=\begin{dcases}
        t(\tau)=\tau\\
        \bold{x}(\tau)=\bold{x}_0
    \end{dcases}
\end{equation}
for the inertial trajectory, and
\begin{equation}
    X_{A}(\tau)=\begin{dcases}
        t(\tau)=\alpha^{-1}\sinh(\alpha\tau)\\
        x(\tau)=\alpha^{-1}\cosh(\alpha\tau)\\
        y(\tau)=z(\tau)=const.
    \end{dcases}
\end{equation}
for the accelerated one. Next, we specify the switching function used and the results one would obtain if the Born rule was applicable. Specifically, we use the switching function obtained from using \eqref{S^(1)} with domain limited to a compact support as in Eq.~\eqref{chi_RM}, i.e.
\begin{equation}
    \chi_k(\tau)=\begin{dcases}
        e^{-(\tau-(kT+T_{\textrm{on}}/2))^2/(2\sigma^2)}~&\tau\in[kT,kT+T_{\textrm{on}}]\\
        0&~\textrm{otherwise}
    \end{dcases}
\end{equation}
where $T_{\textrm{on}}=8\sigma$. The results obtained assuming infinite interaction times (i.e. taking Eq.~\eqref{S^(1)} as $\chi$), are obtained in Ref.~\cite{SriramkumarP96} as
\begin{equation}
    q_{I}=\frac{\lambda^2}{4\pi}\left[e^{-\omega^2 \sigma^2}-\omega \sigma\Gamma\left(1/2,\omega^2 \sigma^2\right)\right]
    \label{gaussian_results_I}
\end{equation}
in the inertial case, where $\Gamma(a,x)$ is the upper incomplete gamma function defined by
\begin{equation}
    \Gamma(a,x)=\int_x^\infty t^{a-1}e^{-t} dt ~,
\end{equation}
and 
\begin{equation}
     q_{A}=\frac{\lambda^2}{2\pi}\sum_{n=-\infty}^\infty\left[e^{b_n^2/4\sigma^2+\omega b_n}\int_{-\Xi_n r_n}^\infty e^{-p^2}(p+\Xi_nr_n) dp\right]
    \label{gaussian_results_A}
\end{equation}
in the accelerated case, where
$b_n=-2\pi n/g$ and $r_n=\omega \sigma+b_n/2\sigma$, and where
\begin{equation}
    \Xi_n=\begin{dcases}
        -1~&\textrm{if}~n\leq 0\\
        +1~&\textrm{if}~n> 0
    \end{dcases}
\end{equation}
Assuming that limiting the support of $\chi_k$ to $[kT,kT+T_{\textrm{on}}]$ does little modification to the results obtained in Eq.s~\eqref{gaussian_results_I} and \eqref{gaussian_results_A}, we can take
\begin{equation}
    \mathcal{Q}_{I(A)}\simeq\frac{q_{I(A)}}{\lambda^2}
    \label{assumption_on_Q}
\end{equation}
in the inertial (accelerated) case. This provides us with everything we need for applying the tools developed in the previous section. 

Before calculating the probability of obtaining the bit string $B_L=(L;N_1,\dots,N_n)$ in the RM case, let us briefly discuss the results one obtains in the standard setting. In this case, $P(B_L)$ is the Born rule probability of observing the string of results $B_L$, where each $b_k\in B_L$ is the result of a measurement obtained from a different detector interacting with an different scalar field, but moving along the same trajectory labeled by $i=I,A$. In this case, probability of observing a one in $B_L$ is $q_{I(A)}$, independently of its location along the string; hence $P(B_L)$ reduces to
\begin{equation}
    P_{q_i}(B_L)=q_i^n(1-q_i)^{(L-n)}~.
    \label{StringBorn}
\end{equation}
The usual results are obtained assuming the detector reaches a state for which the detailed balance condition holds, and the most likely output from the above setup is the string for which the sampled ratio
\begin{equation}
    \mathcal{R}^{sampled}_i=\frac{n}{L-n}
\end{equation}
approaches the theoretical one
\begin{equation}
    \mathcal{R}_i= \frac{p_1}{p_0}~.
\end{equation}
In the case of infinite interaction times these are
\begin{align}
    \mathcal{R}^{\infty}_I&=0\\
    \mathcal{R}^{\infty}_A&=\exp(-\omega/T_U)\label{inf_rates}
\end{align}
for the inertial and accelerated trajectories respectively, where $T_U=\alpha/2\pi$ is the well-known Unruh temperature. These results hold any time one can perform measurements on replicas and under the usual assumptions and limitations imposed by QFT in non-inertial and curved backgrounds, i.e. when the length and time scales of the quantum processes under study are above their Planck values~\cite{BirrellD82}. However, since in the RM scenario we are forced to use finite time interactions, we must confront the results we get from RM with those obtained by using the Born rule and finitely long interaction, i.e. by using the probabilities \eqref{gaussian_results_I} and \eqref{gaussian_results_A} to get
\begin{equation}
    \begin{dcases}
        \mathcal{R}^{T_{\textrm{on}}}_I=\frac{q_I}{(1-q_I)}\\
        \mathcal{R}^{T_{\textrm{on}}}_A=\frac{q_A}{(1-q_A)}
    \end{dcases}
\end{equation}
By taking $T_{\textrm{on}}$ to be sufficiently long, these results reproduce those obtained from taking the infinite interaction limit~\cite{JuarezAubry19}. Note that in the finite-time interaction case the usual infinitesimal transition rates must be replaced with the entries of a stochastic matrix describing the integrated process. Therefore, while in the accelerated case the probabilities satisfy the condition given in \eqref{inf_rates}, the rates do not necessarily take the usual thermal form found in the literature.

Finally, let us find the probability of obtaining $B_L$ in the RM scenario. In this case, the Born rule gives way to the RM string probability rule obtained in Eq.~\eqref{p as sigmas}, and the probability of obtaining $B_L$ the more complicated expression
\begin{equation}
    \Tilde{P}_{q_i}(B_L)=\prod_{j=1}^n\mathcal{P}_{N_j}\prod_{\substack{j'=1\\j'\neq j}}^{L-n}\big(1-\mathcal{P}(N_{j'}|N_{j''},\dots,N_1)\big)~,
    \label{StringABorn}
\end{equation}
where, in the second product $j''$ is the highest value smaller than $j'$ for which $b_{j''}=1$. Assuming $L$ is smaller than the $\mathbf{n}$ defined in Eq.~\eqref{bold_n}, and defining
\begin{equation}
\begin{dcases}
    \mathbb{U}=\max_{n\leq\mathbf{n}}\{\epsilon_n\}\\
    \mathbb{L}=\max_{n\leq\mathbf{n}}\{\delta_n\}
\end{dcases}
\end{equation}
we obtain that, for short enough strings, one can use the Born rule in place of the effectively Born one, as discussed in Sec.~\ref{sec:2.bit_string_probabilities}. In fact, for typical values of the parameters $q_i$, $\gamma$, $\mathbf{n}$ and $L<\mathbf{n}$, the inequality
\begin{equation}
    (1+\mathbb{U})^n(1+\mathcal{R}_i\mathbb{L})^{L-n} \geq \frac{\tilde{P}_{q_i}(B_L)}{P_{q_i}(B_L)}\geq (1-\mathbb{L})^n(1-\mathcal{R}_i\mathbb{U})^{L-n}
\end{equation}
bounds the fraction of the two probabilities to be very close to one, hence telling us that Eq.~\eqref{easy_threshold} is satisfied. Therefore, the results obtained by analyzing strings of length smaller than $\mathbf{n}$ by means of the Born rule retain their validity in the RM scenario; in particular, one can test the thermal distribution characterizing the Unruh effect.

Moreover, thanks to Eq.~\eqref{assumption_on_Q} it is possible to evaluate \eqref{StringABorn} numerically; the results of such analysis can be found in Fig.~\ref{fig:Born_and_Almost}, and are in accordance with the above discussion. Indeed, the figures show that the difference between the Born probabilities and those provided by RM is several order of magnitude smaller than $P_{q_i}(B)$. The gray lines in Fig.s~\ref{fig:Inertial_Born_and_Almost} and \ref{fig:g=10_Born_and_Almost} respectively correspond to the Born and RM probabilities of finding $M=(0,0,0,1)$ (Fig.~\ref{fig:Inertial_Born_and_Almost}) and $M=(1,0,0,0)$ (Fig.~\ref{fig:g=10_Born_and_Almost}); these are overlapping and hence appear as a single line in both plots. In Fig.s~\ref{fig:Inertial_Born_and_Almost_Detail} and \ref{fig:g=10_Born_and_Almost_Detail}, these lines are magnified to show how the RM probabilities depend on the previous history: orange squares related to strings having same amounts of ones in different locations have different probabilities of occurring (RM case). On the contrary blue dots related to bit strings with the same number of ones are at the same height (Born case). Whether the RM probabilities are lower or higher than the Born ones is non-trivially regulated by the signs of the corrections applied to the Born probabilities at each step (see Eq.~\eqref{p as sigmas} and App.s~\ref{app:F}-\ref{app:F4}). In particular, in Fig.~\ref{fig:Inertial_Born_and_Almost_Detail}, $\tilde{P}_I(B)\leq P_I(B)$ for all $B$ considered. In Fig.~\ref{fig:g=10_Born_and_Almost_Detail}, $\tilde{P}_A(B)> P_A(B)$ only for $B=3,6,12$ which are those strings with ones located next to each other, and $\tilde{P}_A(B)< P_A(B)$ only for all other $B$. 

Notice that, while Fig.~\ref{fig:g=10_Born_and_Almost} provides numerical evidence that one can uncover the Unruh effect by analyzing strings obtained via RM, testing this effect in a real lab is currently impossible. This is because of two reasons. First, directly testing the Unruh effect itself is considered beyond current experimental capacities due to the required linear size of the laboratory and the fact that a linear acceleration of order $10^{20}m/s^2$ is required to reach a temperature $1K$~\cite{CrispinoEtAl08}; yet, some experimental proposals have been recently put forward for measuring the Unruh effect on circular trajectories~\cite{BunneyL23, BunneyEtAl23, BunneyElAl23b} and in analogue tabletop systems~\cite{WeinfurtnerEtAl11, GoodingEtAl20,  LynchEtAl21}. Second, even assuming the Unruh effect can be observed, Fig.~\ref{fig:g=10_Born_and_Almost} shows that measuring deviations that are sufficiently small to justify the use of the Born rule at first yet large enough to be tested on longer strings requires performing many measurements on the same non-replicable system, hence requiring inaccessibly long coherence times for the global quantum state~\cite{Schlosshauer19, JoosZ85, KieferJ98}.

\section{Conclusions}
\label{sec:5}
In the first part of this article, we introduced the Repeated Measurement (RM) scenario describing the outcomes obtained by measuring many times a system which interacts with an environment in between each measurement. Then, we gave a general formula for calculating the probability of getting any string of outcomes from RM performed on a system where the interaction with the environment is weak and argued that in some cases, an observer applying the Born rule to RM could not realize she is making a mistake, hence making her approach, wrong in principle, correct FAPP. In the second half of the paper, we applied this framework for finding the bit string probabilities from RM performed on a two-level Unruh-DeWitt detector interacting with a scalar field, obtaining both an analytical expression and upper and lower bounds to it. Finally, we studied the explicit cases of a detector moving along an inertial and accelerated trajectory. We found that the results gathered via RM are often indistinguishable from those obtained by the standard approach. This suggests the RM approach can be used as an operationally-based alternative to the idealized one, in which replicas of unreplicable systems are often posited. In particular, we deem the RM approach as suitable for studying all the systems for which making replicas is either unfeasible or impossible. The following paragraphs discuss fundamental aspects of the RM approach in more detail, its significance for quantum cosmology, its relation with the Consistent Histories Approach to QM, and prospects for this work.

Despite being built for non-replicable systems, the RM scenario is testable in standard quantum mechanics. This can be done by taking many copies of the system prepared at each measurement time and applying the frequentist approach to each collection obtained in this way. Hence, testing the RM up to the $k$-th measurement step with the same degree of accuracy one tests a quantum system replicated $N$ times, one needs $Nk$ replicas. This poses a practical limitation but does not invalidate the results. However, the relevance of RM is when it is applied to non-replicable systems: yet, using QM for single instances of systems is a delicate matter. Strictly speaking, the Born rule tells nothing about a single realization of the RM procedure: the string of outcomes can only be read as a time series, which generally does not give information about future probabilities from past outcomes~\cite{PollockEtAl99}. In this sense, RM is never enough to foresee future outcomes. However, if the postulates of QM are to work even when considering a single system, RM can be used as a substitute tool when no replicas are available. Still, it is essential to remark that, when doing so, one is extending QM's domain of validity. In this paper, we argued that when without replicas, one can take RM as a practical tool to foresee future outcomes and often act as if the Born rule was effective (as it holds FAPP).

Let us further comment on the significance of our discussion about RM for Quantum Gravity (QC) and Cosmology (QC). In these disciplines, one is usually interested in computing transition amplitudes for the quantum state of the universe and related quantities. Since the universe is a non-replicable and non-resettable system, it is clear from our discussion that, in general, the outputs of QG and QC are fundamentally inaccessible to experiments (see Sec.~2.2). Strictly speaking, the predictive power of QM does not hold for these systems \textit{even assuming them to behave quantum mechanically}. However, our approach to RM shows that using the Born rule on such quantum systems is sometimes possible (e.g., in the weak coupling regime). Hence, RM may be valuable in many cosmological and quantum gravity-related problems. Specifically, RM gives a prescription for interpreting the data obtained when one does not have complete control over all degrees of freedom of the system under consideration (e.g. the field, in the UDW case) and cannot reset the full system, which is a challenging problem even in non-relativistic quantum mechanics~\cite{Navascues18}. Applying this idea to QG and QC, the probe must be read as a subsystem of a quantum universe in a global pure state which can not be replicated~\cite{DeWitt67,Bojowald15,Vilenkin94}, and even its most classical features (e.g. the metric) are treated as quantum~\cite{Hawking84,HartleH83,Coleman91,deBoerEtAl22,Kiefer08}. While strictly speaking QM is not a theory about systems of this kind, assuming it to be valid can still give sensible results, for example if the probe-universe coupling is weak. In other words, when using the postulates of quantum mechanics is impossible, one should give a new significance to the Born rule or replace it with some effective version of it~\cite{Page22}: the RM scenario provides one such possibility.

Although standard QM needs the assumption that experimenters have many replicas of a given system at their disposal, the question of how to apply it beyond this regime is not new. Amongst the others, our proposal shares this question with the so-called Consistent Histories Approach to QM~\cite{Griffiths84, Omnes95, Halliwell95}. The fact that, in our approach, observers collapse the systems they measure is necessary for RM to work, as it postpones the problem of selecting one particular string of outcomes (or history) to that of describing how measurement outcomes are chosen (i.e. solving the measurement problem). This allows us to take an operational stance without the need to give a solution to the latter conundrum (which also plagues the Consistent Histories Approach~\cite{Omnes04, BassiG99}).  

Finally, let us discuss future prospects. First, instead of describing the problem from an inertial frame, one could consider the point of view of the accelerated observer. In this case, instead of the Minkowski vacuum, one should start from the state obtained by applying the Bogoliubov transformation on it, i.e. the Unruh thermal state~\cite{BirrellD82,ParkerT09}. Despite the technical difficulty of collapsing a much more convoluted state, this approach would allow us to compare the Unruh effect to the RM performed on thermal states. As the latter are testable with the usual Born rule, analysing this problem might allow verifying RM with systems we can test in the lab. 

As discussed at the end of Sec.~\ref{sec:4}, testing our framework in the lab is out of the reach of current experimental capabilities. However, work is proceeding toward realising numerical simulations for RM on standard quantum systems and QFT. Some of the authors are currently investigating RM performed on a spin chain, where one spin acts as a detector testing the rest of the chain. Via this model, classical numerical simulations can be achieved by standard methods, and quantum simulation can be obtained by mapping the spin system to a set of qubits. Furthermore, the spin chain's coupling and topology can be chosen to represent a discretised version of a fermionic field coupled with a UDW detector. Specifically, this is done by realising the circuit in Fig.~\ref{fig:RM_scheme}, where the detector is identified with one qubit of a quantum circuit, the (fermionic) field is discretised and described in terms of a finite collection of fermionic modes, and both the field and the detector-field interaction are read as non-local interactions between qubits via a standard fermion-to-qubit mapping. Once this setup is established, several issues related to measurements and RM in QFT might be easily addressed. For example, one could numerically quantify the variations in entanglement established between the detector and the field after each repetition, study the effects of measurement-induced updates on entanglement harvesting, and quantify the perturbation introduced on the field's state by measurements via the Loschmidt echo or similar tools.

The next question is about the potential alternative ways to describe the update of the state of the quantum field in the future lightcone, if the outcome of the measurement of the detector is known. The updated field state can perhaps be thought of as created by an operator acting on the vacuum at the measurement space-time point, i.e. a local operator quench. In the case when the quantum field theory has a gravitational dual, the local quench at the boundary QFT affects the geometry of the bulk spacetime (see e.g. \cite{NozakiEtAl13,ShimajiEtAl19}). Then one may ask if a bulk observer crossing into the region of the modified geometry could infer that a measurement has been performed at the boundary. 

Another issue to be addressed is that of energy storage, which means asking where the energy extracted from the field by the measurements is conserved, and if and how it must be considered for a complete description of the problem at hand. In fact, for each measurement having outcome $1$, we must dump an energy $\omega$ into a reservoir we did not describe. If considering only a few measurements, one can overlook this storage. However, persisting in not accounting for the dumped energy may pose problems, as it can start to be so big that the spacetime cannot be approximated as flat anymore, causing modifications to both the dynamics of the field and the detector's trajectory. Discussing this could also lead to considering the much more complicated problem in which the detectors trajectory is a quantum degree of freedom~\cite{BarbadoEtAl20,FooEtAl20,FooEtAl21,FooEtAl22,GiacominiK22}, the back reaction of the measurements on the trajectory form a completely quantum perspective~\cite{WoodZ22} and, through RM, describe the trajectory collapse as a semi-classical phenomenon. Finally, applying the tools provided by the RM approach in QFT in curved spacetime could give a new perspective to quantum gravity problems such as the black hole information paradox, leading to consider the quantum measurement postulate where it is often missing.

\section*{Acknowledgments}
We thank Marco Cattaneo and Otto Veltheim for valuable discussions and helpful comments. We also thank two anonymous referees for their remarks and useful observations. N.P. acknowledges financial support from the Magnus Ehrnrooth foundation. N.P. and S.M. acknowledge financial support from the Academy of Finland via the Centre of Excellence program (Project No. 336810 and Project No. 336814). 


\bibliographystyle{quantum}
\bibliography{bib_new}

\begin{thebibliography}{10}

\bibitem{Born26}
M.~Born.
\newblock ``Zur quantenmechanik der sto{\ss}vorg{\"a}nge''.
\newblock \href{https://dx.doi.org/10.1007/BF01397477}{Z. Phys. {\bf 37}, 863--867}~(1926).

\bibitem{NielsenC10}
M.~A. Nielsen and I.~L. Chuang.
\newblock ``Quantum computation and quantum information: 10th anniversary edition''.
\newblock \href{https://dx.doi.org/10.1017/CBO9780511976667}{Cambridge University Press}. ~(2010).

\bibitem{Araki99}
H.~Araki.
\newblock ``{Mathematical Theory of Quantum Fields}''.
\newblock \href{https://dx.doi.org/https://doi.org/10.1093/oso/9780198517733.001.0001}{International series of monographs on physics}. Oxford University Press. ~(1999).

\bibitem{DeWitt67}
B.~S. DeWitt.
\newblock ``{Quantum Theory of Gravity. 1. The Canonical Theory}''.
\newblock \href{https://dx.doi.org/10.1103/PhysRev.160.1113}{Phys. Rev. {\bf 160}, 1113--1148}~(1967).

\bibitem{Hawking84}
S.W. Hawking.
\newblock ``The quantum state of the universe''.
\newblock \href{https://dx.doi.org/https://doi.org/10.1016/0550-3213(84)90093-2}{Nucl. Phys. B. {\bf 239}, 257--276}~(1984).

\bibitem{BirrellD82}
N.~D. Birrell and P.~C.~W. Davies.
\newblock ``{Quantum Fields in Curved Space}''.
\newblock \href{https://dx.doi.org/10.1017/CBO9780511622632}{Cambridge Monographs on Mathematical Physics}. Cambridge Univ. Press. Cambridge, UK~(1984).

\bibitem{ParkerT09}
L.~Parker and D.~Toms.
\newblock ``Quantum field theory in curved spacetime: Quantized fields and gravity''.
\newblock \href{https://dx.doi.org/10.1017/CBO9780511813924}{Cambridge Monographs on Mathematical Physics}. Cambridge University Press. ~(2009).

\bibitem{BreuerP02}
H.~P. Breuer and F.~Petruccione.
\newblock ``{The Theory of Open Quantum Systems}''.
\newblock \href{https://dx.doi.org/https://doi.org/10.1093/acprof:oso/9780199213900.001.0001}{Oxford University Press}. Great Clarendon Street~(2002).

\bibitem{PlenioV01}
M.~B. Plenio and V.~Vitelli.
\newblock ``The physics of forgetting: Landauer's erasure principle and information theory''.
\newblock \href{https://dx.doi.org/10.1080/00107510010018916}{Contemp. Phys. {\bf 42}, 25--60}~(2001).

\bibitem{vonToussaint11}
U.~von Toussaint.
\newblock ``Bayesian inference in physics''.
\newblock \href{https://dx.doi.org/10.1103/RevModPhys.83.943}{Rev. Mod. Phys. {\bf 83}, 943--999}~(2011).

\bibitem{StanislavW08}
S.~Volkov and T.~Wong.
\newblock ``A note on random walks in a hypercube''.
\newblock Pi Mu Epsilon Journal {\bf 12}, 551--557~(2008).

\bibitem{Unruh76}
W.~G. Unruh.
\newblock ``Notes on black-hole evaporation''.
\newblock \href{https://dx.doi.org/10.1103/PhysRevD.14.870}{Phys. Rev. D {\bf 14}, 870--892}~(1976).

\bibitem{DeWitt80}
B.~S. DeWitt.
\newblock ``{Quantum Gravity: the new synthesis}''.
\newblock In S.~W. Hawking and W.~Israel, editors, {General Relativity}: {An Einstein Centenary Survey}.
\newblock Pages 680--745.
\newblock Cambridge University Press, Cambridge~(1980).

\bibitem{CrispinoEtAl08}
L.~C.~B. Crispino, A.~Higuchi, and G.~E.~A. Matsas.
\newblock ``{The Unruh effect and its applications}''.
\newblock \href{https://dx.doi.org/10.1103/RevModPhys.80.787}{Rev. Mod. Phys. {\bf 80}, 787--838}~(2008).

\bibitem{Wald95}
R.~M. Wald.
\newblock ``Quantum field theory in curved space-time and black hole thermodynamics''.
\newblock Chicago Lectures in Physics. University of Chicago Press. Chicago, IL~(1995).

\bibitem{Schlicht04}
S.~Schlicht.
\newblock ``{Considerations on the Unruh effect: causality and regularization}''.
\newblock \href{https://dx.doi.org/10.1088/0264-9381/21/19/011}{Class. Quantum Gravity {\bf 21}, 4647–4660}~(2004).

\bibitem{LoukoS06}
J.~Louko and A.~Satz.
\newblock ``{How often does the Unruh--DeWitt detector click? Regularization by a spatial profile}''.
\newblock \href{https://dx.doi.org/10.1088/0264-9381/23/22/015}{Class. Quantum Gravity {\bf 23}, 6321--6343}~(2006).

\bibitem{Satz07}
A.~Satz.
\newblock ``{Then again, how often does the Unruh–DeWitt detector click if we switch it carefully?}''.
\newblock \href{https://dx.doi.org/10.1088/0264-9381/24/7/003}{Class. Quantum Gravity {\bf 24}, 1719}~(2007).

\bibitem{Takagi86}
S.~Takagi.
\newblock ``{Vacuum Noise and Stress Induced by Uniform Acceleration: Hawking-Unruh Effect in Rindler Manifold of Arbitrary Dimension}''.
\newblock \href{https://dx.doi.org/10.1143/PTP.88.1}{Prog. Theor. Phys. Supp. {\bf 88}, 1--142}~(1986).

\bibitem{BarbadoEtAl20}
L.~C. Barbado, E.~Castro-Ruiz, L.~Apadula, and C.~Brukner.
\newblock ``Unruh effect for detectors in superposition of accelerations''.
\newblock \href{https://dx.doi.org/10.1103/PhysRevD.102.045002}{Phys. Rev. D {\bf 102}, 045002}~(2020).

\bibitem{FooEtAl20}
J.~Foo, S.~Onoe, and M.~Zych.
\newblock ``{Unruh-deWitt detectors in quantum superpositions of trajectories}''.
\newblock \href{https://dx.doi.org/10.1103/PhysRevD.102.085013}{Phys. Rev. D {\bf 102}, 085013}~(2020).

\bibitem{Foo21}
J.~Foo, S.~Onoe, R.~B. Mann, and M.~Zych.
\newblock ``{Thermality, causality, and the quantum-controlled Unruh--deWitt detector}''.
\newblock \href{https://dx.doi.org/10.1103/PhysRevResearch.3.043056}{Phys. Rev. Research {\bf 3}, 043056}~(2021).

\bibitem{BenattiF04}
F.~Benatti and R.~Floreanini.
\newblock ``{Entanglement generation in uniformly accelerating atoms: Reexamination of the Unruh effect}''.
\newblock \href{https://dx.doi.org/10.1103/PhysRevA.70.012112}{Phys. Rev. A {\bf 70}, 012112}~(2004).

\bibitem{SvaiterS92}
B.~F. Svaiter and N.~F. Svaiter.
\newblock ``Inertial and noninertial particle detectors and vacuum fluctuations''.
\newblock \href{https://dx.doi.org/10.1103/PhysRevD.46.5267}{Phys. Rev. D {\bf 46}, 5267--5277}~(1992).

\bibitem{JunkerS02}
W.~Junker and E.~Schrohe.
\newblock ``Adiabatic vacuum states on general spacetime manifolds: Definition, construction, and physical properties''.
\newblock \href{https://dx.doi.org/10.1007/s000230200001}{Ann. Henri Poincar{\'e} {\bf 3}, 1113--1181}~(2002).

\bibitem{HellwigK70}
K.~E. Hellwig and K.~Kraus.
\newblock ``Formal description of measurements in local quantum field theory''.
\newblock \href{https://dx.doi.org/10.1103/PhysRevD.1.566}{Phys. Rev. D {\bf 1}, 566--571}~(1970).

\bibitem{Sorkin93}
R.~D. Sorkin.
\newblock ``{Impossible measurements on quantum fields}''.
\newblock In {Directions in General Relativity: An International Symposium in Honor of the 60th Birthdays of Dieter Brill and Charles Misner}.
\newblock ~(1993).

\bibitem{Lin14}
S.-Y. Lin.
\newblock ``Notes on nonlocal projective measurements in relativistic systems''.
\newblock \href{https://dx.doi.org/10.1016/j.aop.2014.08.018}{Ann. Phys. {\bf 351}, 773--786}~(2014).

\bibitem{BorstenEtAl21}
L.~Borsten, I.~Jubb, and G.~Kells.
\newblock ``Impossible measurements revisited''.
\newblock \href{https://dx.doi.org/10.1103/PhysRevD.104.025012}{Phys. Rev. D {\bf 104}, 025012}~(2021).

\bibitem{Garcia-ChungEtAl21}
Angel Garcia-Chung, Benito~A. Ju\'arez-Aubry, and Daniel Sudarsky.
\newblock ``What happens once an accelerating observer has detected a rindler particle?''.
\newblock \href{https://dx.doi.org/10.1103/PhysRevD.108.025002}{Phys. Rev. D {\bf 108}, 025002}~(2023).

\bibitem{Fewster19}
C.~J. Fewster.
\newblock ``A generally covariant measurement scheme for quantum field theory in curved spacetimes''.
\newblock In Felix Finster, Domenico Giulini, Johannes Kleiner, and J{\"u}rgen Tolksdorf, editors, Progress and Visions in Quantum Theory in View of Gravity.
\newblock \href{https://dx.doi.org/https://doi.org/10.1007/978-3-030-38941-3_11}{Pages 253--268}.
\newblock Cham~(2020). Springer International Publishing.

\bibitem{FewsterV20}
C.~J. Fewster and R.~Verch.
\newblock ``Quantum fields and local measurements''.
\newblock \href{https://dx.doi.org/10.1007/s00220-020-03800-6}{Commun. Math. Phys. {\bf 378}, 851--889}~(2020).

\bibitem{BostelmannEtAl21}
H.~Bostelmann, C.~J. Fewster, and M.~H. Ruep.
\newblock ``Impossible measurements require impossible apparatus''.
\newblock \href{https://dx.doi.org/10.1103/PhysRevD.103.025017}{Phys. Rev. D {\bf 103}, 025017}~(2021).

\bibitem{Polo-GomezEtAl21}
J.~Polo-G\'omez, L.~J. Garay, and E.~Mart\'{\i}n-Mart\'{\i}nez.
\newblock ``A detector-based measurement theory for quantum field theory''.
\newblock \href{https://dx.doi.org/10.1103/PhysRevD.105.065003}{Phys. Rev. D {\bf 105}, 065003}~(2022).

\bibitem{HawkingE73}
S.~W. Hawking and G.~F.~R. Ellis.
\newblock ``The large scale structure of space-time''.
\newblock \href{https://dx.doi.org/10.1017/CBO9780511524646}{Cambridge Monographs on Mathematical Physics}. Cambridge University Press. ~(1973).

\bibitem{SriramkumarP96}
L.~Sriramkumar and T.~Padmanabhan.
\newblock ``{Finite-time response of inertial and uniformly accelerated Unruh--DeWitt detectors}''.
\newblock \href{https://dx.doi.org/10.1088/0264-9381/13/8/005}{Class. Quantum Gravity {\bf 13}, 2061–2079}~(1996).

\bibitem{HiguchiEtAl93}
A.~Higuchi, G.~E.~A. Matsas, and C.~B. Peres.
\newblock ``Uniformly accelerated finite-time detectors''.
\newblock \href{https://dx.doi.org/10.1103/PhysRevD.48.3731}{Phys. Rev. D {\bf 48}, 3731--3734}~(1993).

\bibitem{Tu11}
L.~W. Tu.
\newblock ``An introduction to manifolds''.
\newblock \href{https://dx.doi.org/https://doi.org/10.1007/978-1-4419-7400-6}{Universitext}. Springer. New York~(2011).
\newblock 2nd edition.

\bibitem{Nestruev06}
J.~Nestruev.
\newblock ``Smooth manifolds and observables''.
\newblock \href{https://dx.doi.org/https://doi.org/10.1007/978-3-030-45650-4}{Graduate Texts in Mathematics}. Springer New York. ~(2006).

\bibitem{JonssonEtAl15}
R.~H. Jonsson, E.~Mart\'{\i}n-Mart\'{\i}nez, and A.~Kempf.
\newblock ``Information transmission without energy exchange''.
\newblock \href{https://dx.doi.org/10.1103/PhysRevLett.114.110505}{Phys. Rev. Lett. {\bf 114}, 110505}~(2015).

\bibitem{CzaporML08}
S.~R. Czapor and R.~G. McLenaghan.
\newblock ``Hadamard's problem of diffusion of waves''.
\newblock Acta Physica Polonica Series B, Proceedings Supplement {\bf 1}, 55--75~(2008).

\bibitem{McLenaghan74}
R.~G. McLenaghan.
\newblock ``On the validity of {Huygens'} principle for second order partial differential equations with four independent variables. {Part} {I} : derivation of necessary conditions''.
\newblock Annales de l'I.H.P. Physique th\'eorique {\bf 20}, 153--188~(1974).

\bibitem{Margolius01}
B.~H. Margolius.
\newblock ``Avoiding your spouse at a bridge party''.
\newblock \href{https://dx.doi.org/https://doi.org/10.2307/2691151}{Math. Mag. {\bf 74}, 33--41}~(2001).

\bibitem{Brawner00}
J.~N. Brawner.
\newblock ``Dinner, dancing, and tennis, anyone?''.
\newblock \href{https://dx.doi.org/https://doi.org/10.2307/2691486}{Math. Mag. {\bf 73}, 29--36}~(2000).

\bibitem{OEIS22}
N.~J.~A. Sloane and The OEIS~Foundation Inc.
\newblock ``The on-line encyclopedia of integer sequences''~(2022).

\bibitem{JuarezAubry19}
B.~A. Ju\'arez-Aubry and D.~Moustos.
\newblock ``{Asymptotic states for stationary Unruh-DeWitt detectors}''.
\newblock \href{https://dx.doi.org/10.1103/PhysRevD.100.025018}{Phys. Rev. D {\bf 100}, 025018}~(2019).

\bibitem{BunneyL23}
C.~R.~D. Bunney and J.~Louko.
\newblock ``{Circular motion analogue Unruh effect in a 2+1 thermal bath: robbing from the rich and giving to the poor}''.
\newblock \href{https://dx.doi.org/10.1088/1361-6382/acde3b}{Class. Quantum Gravity {\bf 40}, 155001}~(2023).

\bibitem{BunneyEtAl23}
C.~R.~D. Bunney, L.~Parry, T.~R. Perche, and J.~Louko.
\newblock ``Ambient temperature versus ambient acceleration in the circular motion unruh effect''.
\newblock \href{https://dx.doi.org/10.1103/PhysRevD.109.065001}{Phys. Rev. D {\bf 109}, 065001}~(2024).

\bibitem{BunneyElAl23b}
C.~R.~D. Bunney, S.~Biermann, V.~S. Barroso, A.~Geelmuyden, C.~Gooding, G.~Ithier, X.~Rojas, J.~Louko, and S.~Weinfurtner.
\newblock ``Third sound detectors in accelerated motion''.
\newblock \href{https://dx.doi.org/10.1088/1367-2630/ad5758}{New Journal of Physics {\bf 26}, 065001}~(2024).

\bibitem{WeinfurtnerEtAl11}
S.~Weinfurtner, E.~W. Tedford, M.~C.~J. Penrice, W.~G. Unruh, and G.~A. Lawrence.
\newblock ``{Measurement of Stimulated Hawking Emission in an Analogue System}''.
\newblock \href{https://dx.doi.org/10.1103/PhysRevLett.106.021302}{Phys. Rev. Lett. {\bf 106}, 021302}~(2011).

\bibitem{GoodingEtAl20}
C.~Gooding, S.~Biermann, S.~Erne, J.~Louko, W.~G. Unruh, J.~Schmiedmayer, and S.~Weinfurtner.
\newblock ``{Interferometric Unruh Detectors for Bose-Einstein Condensates}''.
\newblock \href{https://dx.doi.org/10.1103/PhysRevLett.125.213603}{Phys. Rev. Lett. {\bf 125}, 213603}~(2020).

\bibitem{LynchEtAl21}
M.~H. Lynch, E.~Cohen, Y.~Hadad, and I.~Kaminer.
\newblock ``Experimental observation of acceleration-induced thermality''.
\newblock \href{https://dx.doi.org/10.1103/PhysRevD.104.025015}{Phys. Rev. D {\bf 104}, 025015}~(2021).

\bibitem{Schlosshauer19}
M.~Schlosshauer.
\newblock ``Quantum decoherence''.
\newblock \href{https://dx.doi.org/https://doi.org/10.1016/j.physrep.2019.10.001}{Phys. Rep. {\bf 831}, 1--57}~(2019).

\bibitem{JoosZ85}
E.~Joos and H.~D. Zeh.
\newblock ``The emergence of classical properties through interaction with the environment''.
\newblock \href{https://dx.doi.org/10.1007/BF01725541}{Z. Phys. B {\bf 59}, 223--243}~(1985).

\bibitem{KieferJ98}
C.~Kiefer and E.~Joos.
\newblock ``Decoherence: Concepts and examples''.
\newblock In Quantum Future From Volta and Como to the Present and Beyond.
\newblock \href{https://dx.doi.org/https://doi.org/10.1007/BFb0105334}{Pages 105--128}.
\newblock Springer Berlin Heidelberg~(1999).

\bibitem{PollockEtAl99}
D.~S.~G. Pollock, R.~C. Green, and Nguyen T.
\newblock ``Handbook of time series analysis, signal processing, and dynamics.''.
\newblock \href{https://dx.doi.org/https://doi.org/10.1016/B978-0-12-560990-6.X5000-3}{Signal Processing and Its Applications}. Academic Press. ~(1999).

\bibitem{Navascues18}
M.~Navascu\'es.
\newblock ``Resetting uncontrolled quantum systems''.
\newblock \href{https://dx.doi.org/10.1103/PhysRevX.8.031008}{Phys. Rev. X {\bf 8}, 031008}~(2018).

\bibitem{Bojowald15}
Martin B.
\newblock ``Quantum cosmology: a review''.
\newblock \href{https://dx.doi.org/10.1088/0034-4885/78/2/023901}{Rep. Prog. Phys. {\bf 78}, 023901}~(2015).

\bibitem{Vilenkin94}
A.~Vilenkin.
\newblock ``Approaches to quantum cosmology''.
\newblock \href{https://dx.doi.org/10.1103/PhysRevD.50.2581}{Phys. Rev. D {\bf 50}, 2581--2594}~(1994).

\bibitem{HartleH83}
J.~B. Hartle and S.~W. Hawking.
\newblock ``Wave function of the universe''.
\newblock \href{https://dx.doi.org/10.1103/PhysRevD.28.2960}{Phys. Rev. D {\bf 28}, 2960--2975}~(1983).

\bibitem{Coleman91}
S.~R. Coleman, J.~B. Hartle, T.~Piran, and S.~Weinberg, editors.
\newblock ``{Quantum cosmology and baby universes. Proceedings, 7th Winter School for Theoretical Physics}''.
\newblock Jerusalem, Israel~(1991).

\bibitem{deBoerEtAl22}
J.~de~Boer, B.~Dittrich, A.~Eichhorn, S.~B. Giddings, S.~Gielen, S.~Liberati, E.~R. Livine, D.~Oriti, K.~Papadodimas, A.~D. Pereira, M.~Sakellariadou, S.~Surya, and H.~Verlinde.
\newblock ``Frontiers of quantum gravity: shared challenges, converging directions''~(2022) \href{http://arxiv.org/abs/2207.10618}{arXiv:2207.10618}.

\bibitem{Kiefer08}
C.~Kiefer and B.~Sandhöfer.
\newblock ``Quantum cosmology''.
\newblock \href{https://dx.doi.org/doi:10.1515/zna-2021-0384}{Z. Naturforsch. A {\bf 77}, 543--559}~(2022).

\bibitem{Page22}
D.~N. Page.
\newblock ``Possibilities for probabilities''.
\newblock \href{https://dx.doi.org/10.1088/1475-7516/2022/10/023}{J. Cosmol. Astropart. Phys. {\bf 2022}, 023}~(2022).

\bibitem{Griffiths84}
R.~B. Griffiths.
\newblock ``Consistent histories and the interpretation of quantum mechanics''.
\newblock \href{https://dx.doi.org/10.1007/BF01015734}{J. Stat. Phys. {\bf 36}, 219--272}~(1984).

\bibitem{Omnes95}
R.~{Omn{\`e}s}.
\newblock ``{A new interpretation of quantum mechanics and its consequences in epistemology}''.
\newblock \href{https://dx.doi.org/10.1007/BF02059008}{Found. Phys. {\bf 25}, 605--629}~(1995).

\bibitem{Halliwell95}
J.~J. Halliwell.
\newblock ``A review of the decoherent histories approach to quantum mechanicsa''.
\newblock \href{https://dx.doi.org/10.1111/j.1749-6632.1995.tb39014.x}{Ann. N. Y. Acad. Sci. {\bf 755}, 726--740}~(1995).

\bibitem{Omnes04}
R.~Omn\`es.
\newblock ``Model of quantum reduction with decoherence''.
\newblock \href{https://dx.doi.org/10.1103/PhysRevD.71.065011}{Phys. Rev. D {\bf 71}, 065011}~(2005).

\bibitem{BassiG99}
A.~Bassi and G.~Ghirardi.
\newblock ``Can the decoherent histories description of reality be considered satisfactory?''.
\newblock \href{https://dx.doi.org/https://doi.org/10.1016/S0375-9601(99)00303-5}{Phys. Lett. A {\bf 257}, 247--263}~(1999).

\bibitem{NozakiEtAl13}
M.~Nozaki, T~Numasawa, and T.~Takayanagi.
\newblock ``Holographic local quenches and entanglement density''.
\newblock \href{https://dx.doi.org/10.1007/JHEP05(2013)080}{J. High Energy Phys. {\bf 2013}, 80}~(2013).

\bibitem{ShimajiEtAl19}
T.~Shimaji, T.~Takayanagi, and Z.~Wei.
\newblock ``Holographic quantum circuits from splitting/joining local quenches''.
\newblock \href{https://dx.doi.org/10.1007/JHEP03(2019)165}{J. High Energy Phys. {\bf 2019}, 165}~(2019).

\bibitem{FooEtAl21}
J.~Foo, R.~B. Mann, and M.~Zych.
\newblock ``Entanglement amplification between superposed detectors in flat and curved spacetimes''.
\newblock \href{https://dx.doi.org/10.1103/PhysRevD.103.065013}{Phys. Rev. D {\bf 103}, 065013}~(2021).

\bibitem{FooEtAl22}
J.~Foo, C.~S. Arabaci, M.~Zych, and R.~B. Mann.
\newblock ``Quantum superpositions of minkowski spacetime''.
\newblock \href{https://dx.doi.org/10.1103/PhysRevD.107.045014}{Phys. Rev. D {\bf 107}, 045014}~(2023).

\bibitem{GiacominiK22}
F.~Giacomini and A.~Kempf.
\newblock ``{Second-quantized Unruh-DeWitt detectors and their quantum reference frame transformations}''.
\newblock \href{https://dx.doi.org/10.1103/PhysRevD.105.125001}{Phys. Rev. D {\bf 105}, 125001}~(2022).

\bibitem{WoodZ22}
C.~E. Wood and M.~Zych.
\newblock ``{Quantized mass-energy effects in an Unruh-DeWitt detector}''.
\newblock \href{https://dx.doi.org/10.1103/PhysRevD.106.025012}{Phys. Rev. D {\bf 106}, 025012}~(2022).

\bibitem{Fulton96}
W.~Fulton.
\newblock ``Young tableaux: With applications to representation theory and geometry''.
\newblock \href{https://dx.doi.org/10.1017/CBO9780511626241}{London Mathematical Society Student Texts}. Cambridge University Press. ~(1996).

\bibitem{Weisstein22}
E.~W. Weisstein.
\newblock ``Circular permutation. {From MathWorld---A Wolfram Web Resource}''.
\newblock Last visited: 23/01/2023.

\end{thebibliography}

\newpage
\onecolumngrid
\appendix
\section{Tight and loose bounds for $\Sigma(J_1,\dots,J_n)$}
\label{app:F}
In this section, we study the structure of the integrals appearing in Eq.~\eqref{p_N} and use the results we get for finding tight and loose bounds for $\Sigma(J_1,\dots,J_n)$.
\subsection{Tight  bound}
\label{app:F_tight}
Using the notation introduced in Sec.~\ref{sec:3.probabilities}, each integral
\begin{equation}
    \Sigma(J_1,\dots,J_n)=\int_{J_1,\dots,J_n}\mathcal{W}_{2n}(X_n,X'_n,\dots,X_1,X'_1)
    \label{a:e.sigma}
\end{equation}
can be decomposed via Wick's theorem into a sum of $\mathcal{N}(n)$ terms. To discuss these terms, let us introduce a pictorial representation for products of $2$-point Wightman functions. First, we represent the detector's trajectory by a straight horizontal line. Then, a Wightman evaluated across two points of the trajectory is represented by a curved line connecting points of the above line. Points belonging to the same interval are represented as closer on the line than those belonging to separate intervals. Finally, a product of more Wightman functions is represented by one horizontal line with multiple curved lines connecting points on it. It follows from this construction that if one diagram can be cut in two pieces by erasing one point on the horizontal line, then the integrals over the Wightman functions it represents can be factorized. Using the above definitions, out of all the terms appearing from the decomposition of Eq.~\ref{a:e.sigma}:
\begin{itemize}
    \item one is the product of $\mathfrak{C}(0){n \choose 0}=n$ integrals of $2$-point Wightman functions evaluated within the same interval, i.e. $\mathcal{W}_s$. This term is represented as
    \begin{equation}
            \int_{J_1,\dots,J_n}\mathcal{W}(s_1)\dots\mathcal{W}(s_n)=\int_{J_1,\dots,J_n} \left(
            \begin{tikzpicture}[baseline={([yshift=1ex]current bounding box.center)},vertex/.style={anchor=base,
            circle,fill=black!25,minimum size=18pt,inner sep=2pt},scale=0.6]
            \draw (0,0) -- (3.5,0);
            \draw (0.5,0) arc (180:0:0.25cm);
            \draw (2.5,0) arc (180:0:0.25cm);
            \fill (0.5,0) circle[radius=2pt]  node[anchor=north] {$~~J_1$};
            \fill (1,0) circle[radius=2pt]  node[anchor=north] {};
            \fill (2.5,0) circle[radius=2pt]  node[anchor=north] {$~~J_2$};
            \fill (3,0) circle[radius=2pt]  node[anchor=north] {};
            \end{tikzpicture}\dots\right)=\prod_i\left[\int_{J_i} \left(
        \begin{tikzpicture}[baseline={([yshift=1ex]current bounding box.center)},vertex/.style={anchor=base,
        circle,fill=black!25,minimum size=18pt,inner sep=2pt},scale=0.6]
        \draw (0,0) -- (1.5,0);
        \draw (0.5,0) arc (180:0:0.25cm);
        \fill (0.5,0) circle[radius=2pt]  node[anchor=north] {$~~J_i$};
        \fill (1,0) circle[radius=2pt]  node[anchor=north] {};
        \end{tikzpicture}\right)\right]=\mathcal{Q}^n
        \end{equation}
    where the last equality holds whenever the hypothesis \textbf{A0}  (i.e. Eq.~\eqref{A0}) holds.
    \item $\mathfrak{C}(2) {n \choose 2 }=n(n-1)$ are obtained by selecting two intervals, and writing the product of one integral over all possible combinations of $2$-point Wightman functions connecting these intervals and the product of $n-2$ integrals of $\mathcal{W}_s$ functions across the other intervals. These terms are represented as 
    \begin{equation}
        \sum_{j=1}^n\sum_{j'\neq j}\left[\prod_{i\neq j,j'}\int_{J_i} \left(
        \begin{tikzpicture}[baseline={([yshift=1ex]current bounding box.center)},vertex/.style={anchor=base,
        circle,fill=black!25,minimum size=18pt,inner sep=2pt},scale=0.6]
        \draw (0,0) -- (1.5,0);
        \draw (0.5,0) arc (180:0:0.25cm);
        \fill (0.5,0) circle[radius=2pt]  node[anchor=north] {$~~J_i$};
        \fill (1,0) circle[radius=2pt]  node[anchor=north] {};
        \end{tikzpicture}\right)\right]\int_{J_j,J_{j'}}\left(             \begin{tikzpicture}[baseline={([yshift=-.5ex]current         bounding box.center)},vertex/.style={anchor=base,
        circle,fill=black!25,minimum size=18pt,inner sep=2pt},scale=0.6]
        \draw (0,0) -- (3.5,0);
        \draw (1,0) arc (180:0:0.75cm);
        \draw (0.5,0) arc (180:0:1.25cm);
        \fill (0.5,0) circle[radius=2pt]  node[anchor=north] {$~~J_j$};
        \fill (1,0) circle[radius=2pt]  node[anchor=north] {};
        \fill (2.5,0) circle[radius=2pt]  node[anchor=north] {$~~J_{j'}$};
        \fill (3,0) circle[radius=2pt]  node[anchor=north] {};
        \end{tikzpicture}+\begin{tikzpicture}[baseline={([yshift=.05ex]current bounding box.center)},scale=0.6]
        \draw (0,0) -- (3.5,0);
        \draw (1,0) arc (180:0:1cm);
        \draw (0.5,0) arc (180:0:1cm);
        \fill (0.5,0) circle[radius=2pt]  node[anchor=north] {$~~J_j$};
        \fill (1,0) circle[radius=2pt]  node[anchor=north] {};
        \fill (2.5,0) circle[radius=2pt]  node[anchor=north] {$~~J_{j'}$};
        \fill (3,0) circle[radius=2pt]  node[anchor=north] {};
        \end{tikzpicture}\right)=\mathcal{Q}^{n-2}\sum_{j=1}^n\sum_{j'\neq j}\mathcal{C}_2(J_j,J_{j'})~,
    \end{equation}
    where $j$ and $j'$ run over all possible selections of the two intervals, and, without loss of generality, we took $J_{j'}>J_{j}$ (if this is not the case, one just has to switch the labels in the diagram).
    \item $\mathfrak{C}(3){n \choose 3}=\frac{4}{3}n(n-1)(n-2)$ are products of $n-3$ integrals of $\mathcal{W}_s$ and one is an integral of terms having the form
    \begin{equation}
        \begin{tikzpicture}[baseline={([yshift=-0.5ex]current bounding box.center)},vertex/.style={anchor=base,
        circle,fill=black!25,minimum size=18pt,inner sep=2pt},scale=0.6]
        \draw (0,0) -- (5.5,0);
        \draw (1,0) arc (180:0:1cm);
        \draw (2.5,0) arc (180:0:1.25cm);
        \draw (0.5,0) arc (180:90:0.5cm);
        \draw (4.5,0) arc (0:90:0.5cm);
        \draw (1,0.5) -- (4,0.5);
        \fill (0.5,0) circle[radius=2pt]  node[anchor=north] {$~~J_j$};
        \fill (1,0) circle[radius=2pt]  node[anchor=north] {};
        \fill (2.5,0) circle[radius=2pt]  node[anchor=north] {$~~J_{j'}$};
        \fill (3,0) circle[radius=2pt]  node[anchor=north] {};
        \fill (4.5,0) circle[radius=2pt]  node[anchor=north] {$~~J_{j''}$};
        \fill (5,0) circle[radius=2pt]  node[anchor=north] {};
        \end{tikzpicture}~,
    \end{equation}
    hence giving
    \begin{equation}
    \begin{split}
        \sum_{j=1}^n\sum_{\underset{j''\neq j,j'}{ j'\neq j}}\left[\prod_{i\neq j,j',j''}\int_{J_i} \left(
        \begin{tikzpicture}[baseline={([yshift=1ex]current bounding box.center)},vertex/.style={anchor=base,
        circle,fill=black!25,minimum size=18pt,inner sep=2pt},scale=0.6]
        \draw (0,0) -- (1.5,0);
        \draw (0.5,0) arc (180:0:0.25cm);
        \fill (0.5,0) circle[radius=2pt]  node[anchor=north] {$~~J_i$};
        \fill (1,0) circle[radius=2pt]  node[anchor=north] {};
        \end{tikzpicture}\right)\right]\int_{J_j,J_{j'},J_{j''}}\left(
        \begin{tikzpicture}[baseline={([yshift=-0.5ex]current bounding box.center)},vertex/.style={anchor=base,
        circle,fill=black!25,minimum size=18pt,inner sep=2pt},scale=0.6]
        \draw (0,0) -- (5.5,0);
        \draw (1,0) arc (180:0:1cm);
        \draw (2.5,0) arc (180:0:1.25cm);
        \draw (0.5,0) arc (180:90:0.5cm);
        \draw (4.5,0) arc (0:90:0.5cm);
        \draw (1,0.5) -- (4,0.5);
        \fill (0.5,0) circle[radius=2pt]  node[anchor=north] {$~~J_j$};
        \fill (1,0) circle[radius=2pt]  node[anchor=north] {};
        \fill (2.5,0) circle[radius=2pt]  node[anchor=north] {$~~J_{j'}$};
        \fill (3,0) circle[radius=2pt]  node[anchor=north] {};
        \fill (4.5,0) circle[radius=2pt]  node[anchor=north] {$~~J_{j''}$};
        \fill (5,0) circle[radius=2pt]  node[anchor=north] {};
        \end{tikzpicture}+~\textrm{similar~terms}\right)=\\
        =\mathcal{Q}^{n-3}\sum_{j=1}^n\sum_{j'\neq j}\sum_{j''\neq j,j'}\mathcal{C}_3(J_j,J_{j'},J_{j''})~,
    \end{split}
    \end{equation}
    where we took $J_{j''}>J_{j'}>J_{j}$ without loss of generality (again, if this is not the case one has to switch the labels in the diagram).
    \item proceeding in this way, we eventually reach the last set of $\mathfrak{C}(n){n \choose n}=\mathfrak{C}(n)$ terms coming from integrals of a product of Wightman functions none of which is evaluated within the same interval.
\end{itemize}
As a side result, note that this procedure give further meaning to the expression of Eq.~\eqref{decomposition}.

Having classified all terms appearing in $\Sigma(J_1,\dots,J_n)$, we can now search for a tight  bound for $F_{J_1,\dots,J_k}$. The complexity of this task comes from the fact that $F_{J_1,\dots,J_k}$ contains integrals over $2$-point Wightman functions evaluated in points belonging to various combinations of distinct intervals, so that the value of the bound of each term appearing in $F_{J_1,\dots,J_k}$ depends of the specific intervals considered and on how the Wightman functions connect them. In the main body, we obtained a simple bound by replacing all $2$-point Wightman functions with a constant factor bounding the integral of the $2$-point function evaluated across closest-neighbours intervals. In this way, one can bound the absolute value of $\mathcal{C}_{k,i}(J_1,\dots,J_k)$ by some number proportional to $\gamma^k$ regardless of $(J_1,\dots,J_k)$, and obtain a looser but easily accessible result. On the contrary, we here give a more complete (and complicated) strategy to obtain the aforementioned tight er bounds. 

\begin{table}
\centering
\begin{tabular}{| c | c | c | c | c | c | c | c | c | c |} 
\hline
k & ~~0~~ & ~~1~~ & ~~2~~ & ~~3~~ & ~~4~~ & ~~5~~ & ~~6~~ & ~~7~~ & ~~8~~~\\
\hline
$\pi(k)$ & 1 & 1 & 2 & 3 & 5 & 7 & 11 & 15 & 22 \\
\hline
$\pi_r(k)$ & - & - & 1 & 1 & 2 & 2 & 4 & 4 & 7  \\
\hline
\end{tabular}
\caption{Values of $\pi(k)$ and $\pi_r(k)$ up to $k=8$.}
\label{Table_parts}
\end{table}

\begin{table}
\centering
\begin{tabular}{| c | c | c | c | c | c | c | c | c | c |} 
\hline
k & ~~0~~ & ~~1~~ & ~~2~~ & ~~3~~ & ~~4~~ & ~~5~~ & ~~6~~ & ~~7~~ & ~~8~~~\\
\hline
$\mathfrak{C}(k)$ & 0 & 1 & 2 & 8 & ~60~ & ~544~ & ~6040~ & ~79008~ & ~1190672~\\
\hline
$~(2k-1)!!/\sqrt{e}~$ & 0.7 & 0.7 & 1.82 & 9.10 & 63.69 & 573.17 & 6304.89 & 81963.52 & 1229452.81~\\
\hline
\end{tabular}
\caption{Values of $\mathfrak{C}(k)$ and $(2k-1)!!/\sqrt{e}$ up to $k=8$. Note that $\mathfrak{C}(k)$ and $(2k-1)!!/\sqrt{e}$ grow similarly for large $k$, as is was claimed in Eq.~\eqref{c(k) growth}.}
\label{Table_dancing}
\end{table}

First, we classify the $\mathfrak{C}(k)$ terms appearing in $\mathcal{C}_k(J_1,\dots,J_k)$, for any $k\geq2$. As these only contain $\mathcal{W}_d$ functions by construction, to accomplish the task we need the partitions of $k$ that do not contain $1$ as a part, i.e. all the possible ways of writing the natural number $k$ as sum of smaller numbers, $1$ excluded. By denoting $\pi(k)$ the partitions of $k$, and $\pi_r(k)$ those restricted to not contain $1$, it is 
\begin{equation}
    \pi_r(k)=\pi(k)-\pi(k-1)~.
\end{equation}
The first ten values of $\pi(k)$ and $\pi_r(k)$ are listed in Table \ref{Table_parts}. For completeness, the first eight values of $\mathfrak{C}(k)$ and $~(2k-1)!!/\sqrt{e}~$ are listed in Table \ref{Table_dancing}. A strategy for obtaining the restricted partitions of a number is that of finding all the Young diagrams having $n$ blocks and more than one block on each row~\cite{Fulton96}. Then, each diagram represents a partition whose parts are the number of blocks on each row. When read in terms of integrals over $\mathcal{W}_d$ functions, each Young diagram become a product of a number of terms equal to the number of rows, each representing the sum of all integrals that cannot be factored into products of integrals over smaller sets of functions, called Irreducible Integrals (II). The rules relating Young diagrams to sums of II are the following. Let us consider a Young diagram having $k$ blocks. A diagram composed of one row with $k$ blocks represents the sum of $(2k-2))!!$ terms, each being an II over the $k$ intervals; that is because each term is obtained by first choosing one point in one interval and a point in a distinct interval (for which we have $2(k-1)$ options) and writing in the $2k$-integral the Wightman function evaluated between these two points, then taking the other point of the second interval (for which we have no choice) and connecting it to a point in a third interval (for which we have $2(k-2)$ options), and so on until we are forced to connect the $k$-th chosen interval with the first (for which there is no choice left); this procedure gives
\begin{equation}
    \prod_{i=1}^{k-1}(2(k-i))=(2k-2)!!
\end{equation}
IIs over $k$ intervals, as claimed. Next, a row having length $m_r<k$ and coming after other $(r-1)$ rows, each having $\{m_1,\dots,m_{r-1}\}$ blocks each, represents all possible ways of choosing the $m_r$ intervals to be integrated amongst the all possible $k$ minus those already considered in the rows above it, meaning that it accounts for
\begin{equation}
    {k-(m_1+\dots+m_{r-1}) \choose m_r}(2m_r-2)!!
\end{equation}
terms. Finally, being $\sigma_c$ the number of rows of length $c$, the total number of terms calculated by the previous rules must be divided by $\sigma_c!$ $\forall c\leq k$ to account for repeated counting; finally, a given Young diagram associated with the partition $P_k$ of $k$ accounts for
\begin{equation}
    N_{P_k}=\prod_{c=2}^k\frac{1}{\sigma_c!}\prod_{r}{k-(m_1+\dots+m_{r-1}) \choose m_r}(2m_r-2)!!
    \label{app:N_P_k}
\end{equation}
terms, where $r$ runs top-down on all the rows of the diagram, and $m_r$ is the length of the $r$-th row. By this technique (and some book keeping) one can easily reconstruct the analytic expression of all integrals appearing in the sum denoted by $\mathcal{C}_k(J_1,\dots,J_k)$. Examples of this procedure are provided in App.s~\ref{app:P2&3} and \ref{app:F4}.

While being useful for writing the explicit integrals appearing in $\mathcal{C}_k(J_1,\dots,J_k)$, the real power of the above procedure is that it easily provides tight  bounds for it. This can be achieved by replacing all $2$-point Wightman functions appearing in $\mathcal{C}_k(J_1,\dots,J_k)$ with those evaluated in the extreme points of the intervals they connect, i.e.
\begin{equation}
    \mathcal{W}(X_i,X_j)\rightarrow\mathcal{W}(T\abs{N_i-N_j}-T_{\textrm{on}})=w_{ij}~:
    \label{app:substitution}
\end{equation}
thanks to hypothesis \textbf{A1} it is
\begin{equation}
    \abs{w_{ij}}>\abs{\mathcal{W}(X_i,X_j)}~,~~\forall i\neq j~.
\end{equation}
Moreover, in order to make the discussion easier, in what follows we will use the shorthand notation 
\begin{equation}
    \mathfrak{I}_{i_1,i_2}\mathfrak{I}_{j_1,j_2}\dots\mathfrak{I}_{k_1,k_2}
    \label{app:products}
\end{equation}
to denote the integral of II evaluated across the intervals $(J_{i_1},J_{i_2})\cup (J_{j_1},J_{j_2})\cup\dots\cup(J_{k_1},J_{k_2})$, in which the $2$-point functions connect $J_{i_1}$ with $J_{i_2}$, $J_{j_1}$ with $J_{j_2}$ and so on. For example
\begin{equation}
    \mathcal{C}_{3,1}(J_j,J_{j'},J_{j''})=\int_{J_j,J_{j'},J_{j''}}\left(
        \begin{tikzpicture}[baseline={([yshift=-0.5ex]current bounding box.center)},vertex/.style={anchor=base,
        circle,fill=black!25,minimum size=18pt,inner sep=2pt},scale=0.6]
        \draw (0,0) -- (5.5,0);
        \draw (1,0) arc (180:0:1cm);
        \draw (2.5,0) arc (180:0:1.25cm);
        \draw (0.5,0) arc (180:90:0.5cm);
        \draw (4.5,0) arc (0:90:0.5cm);
        \draw (1,0.5) -- (4,0.5);
        \fill (0.5,0) circle[radius=2pt]  node[anchor=north] {$~~J_j$};
        \fill (1,0) circle[radius=2pt]  node[anchor=north] {};
        \fill (2.5,0) circle[radius=2pt]  node[anchor=north] {$~~J_{j'}$};
        \fill (3,0) circle[radius=2pt]  node[anchor=north] {};
        \fill (4.5,0) circle[radius=2pt]  node[anchor=north] {$~~J_{j''}$};
        \fill (5,0) circle[radius=2pt]  node[anchor=north] {};
        \end{tikzpicture}\right)\longrightarrow \mathfrak{I}_{jj'}\mathfrak{I}_{j'j''}\mathfrak{I}_{jj''}~.
\end{equation}
Clearly enough, this notation hides the information about which specific points are connected by the Wightman functions, and it is hence inadequate for finding the analytic expression of $F_{j_1,\dots,j_k}$. However, the notation shows its fruitfulness once the Wightman functions have been replaced via \eqref{app:substitution}, as we can formally write
\begin{equation}
    \mathfrak{I}_{jj'}\geq\mathcal{Q}\gamma_{jj'}~,
    \label{app:formal_ineq}
\end{equation}
where
\begin{equation}
    \gamma_{jj'}=\frac{w_{ij}}{\mathcal{W}(T_{\textrm{on}})}~.
\end{equation}
Note that the above inequality only makes sense when considering the products in \eqref{app:products}, as $\mathfrak{I}_{jj'}$ is part of a II and hence represents a part of an integral that is not separable. Also note that both sides in Eq.~\eqref{app:formal_ineq} are negative, i.e.
\begin{equation}
    \abs{\mathfrak{I}_{jj'}}\leq\abs{\mathcal{Q}}\gamma_{jj'}~.
\end{equation}
As a consequence, replacing the integrals with the corresponding product \eqref{app:products} is sufficient for obtaining the bounds we are interested in. Moreover, the fact that this notation assigns the same name to quantities that are not equal, e.g. 
\begin{equation}
    \begin{tikzcd}
    \mathcal{C}_{2,1}(J_j,J_{j'})  \arrow[swap]{d}{}\arrow[equal, "/" marking]{r} & \mathcal{C}_{2,2}(J_j,J_{j'}) \arrow{d}{} \\%
    (\mathfrak{I}_{jj'})^2 & (\mathfrak{I}_{jj'})^2
    \end{tikzcd}~,
\end{equation}
makes the search for bounds easier (while being drastically wrong if one is searching for analytical expressions). In fact, all terms related to one Young diagram with $k$ blocks are easily obtained by:
\begin{itemize}
    \item if the diagram has one row only, consider the
        \begin{equation}
            P^{c}_m=\frac{1}{2}(m-1)!
        \end{equation}
        free cyclic permutations $\sigma$~\cite{Weisstein22} of the indices of the expression
        \begin{equation}
            \mathfrak{I}_{i_1i_2}\mathfrak{I}_{i_2i_3}\dots \mathfrak{I}_{i_ki_1}~,
        \end{equation}
        i.e. by filling the Young diagram's boxes with the label of the integrals composing the II and applying the rule
        \begin{equation}
            \begin{ytableau}
                i_1&i_2&\none[...]&i_k
            \end{ytableau}
            ~~\rightarrow~~\sigma(\mathfrak{I}_{i_1i_2}\mathfrak{I}_{i_2i_3}\dots \mathfrak{I}_{i_ki_1})~.
        \end{equation}
    \item if the diagram has more than one row, one must first fill the boxes with all possible
    \begin{equation}
        \prod_{c=2}^k\frac{1}{\sigma_c!}\prod_{r}{k-(m_1+\dots+m_{r-1}) \choose m_r}
    \end{equation}
    inequivalent fillings (where the notation is as in Eq.~\eqref{app:N_P_k}), and then write all possible permutation for each row as in the above case.
\end{itemize}
The number of terms extracted with this procedure is
\begin{equation}
        B_{P_k}=\prod_{c=2}^k\frac{1}{\sigma_c!}\prod_{r}{k-(m_1+\dots+m_{r-1}) \choose m_r}P^c_{m_r}~,
\end{equation}
and hence much smaller than $N_{P_k}$. Finally, by bounding all $\mathfrak{I}_{ij}$ with the corresponding $\mathcal{Q}\gamma_{ij}$ as prescribed by Eq.~\eqref{app:formal_ineq}, one trades $N_{P_k}$ II represented by the diagram with $B_{P_k}$ tight  bounds for them (whether these are upper or lower bounds depends on the parity of $k$). Examples of this procedure are provided in App.s~\ref{app:P2&3}, \ref{app:F4}.

\subsection{Loose bound}
\label{app:F_loose}
To obtain loose bounds, let us start by considering one $\mathcal{C}_{k,i}(J_1,\dots,J_k)$, with $i\in[0,\mathfrak{C}(k)]$, which is the integral of $k$ Wightman functions of the kind $\mathcal{W}_d$, and has sign $(-1)^{k}$. Therefore, we can write
\begin{equation}
      \mathcal{C}_{k,i}(J_1,\dots,J_k)=(-1)^{k}\int_{J_1,\dots,J_k}\prod_{i=1}^k\abs{\mathcal{W}(\mathfrak{p}_i(\bold{u},\bold{s}))}~,
\end{equation}
where $\mathfrak{p}_i(\bold{u},\bold{s})$ are some linear combination of the components of $\bold{u}=(u_1,\dots,u_k)$ and $\bold{s}=(s_1,\dots,s_k)$. Thanks to \textbf{A2}, all functions appearing in the integral are now definite positive in the integration region, and hence multiplying and dividing in the integral by $\prod_{i=1}^k\abs{\mathcal{W}(s_i)}$ and leaving only the one of the two at numerator inside the integral when maximising gives
\begin{equation}
    0\leq \frac{\mathcal{C}_{k,i}(J_1,\dots,J_k)}{(-1)^k} \leq \mathcal{Q}^k\max_{\cup_i J_i}\left\lbrace\frac{\prod_{i=1}^k\abs{\mathcal{W}(\mathfrak{p}_i(\bold{u},\bold{s}))}}{\prod_{i=1}^k\abs{\mathcal{W}(s_i)}}\right\rbrace ~.
    \label{general bound}
\end{equation}
A general strategy to solve the above bounds tight ly and sum them to obtain $\Sigma(J_1,\dots,J_k)$ is given in App.~\ref{app:F}. Here, we proceed by making the bound expressed in \eqref{general bound} easier to handle, at the cost of it being looser. Since all $\mathfrak{p}_i(\bold{u},\bold{s})$ are some combination of at most $u_j$ and $u_{j'}$ and the relative $s$, we have
\begin{equation}
    \max_{\cup_i J_i}\left\lbrace\prod_{i=1}^k\abs{\mathcal{W}(\mathfrak{p}_i(\bold{u},\bold{s}))}\right\rbrace\leq \abs{\mathcal{W}(T_{\textrm{off}})}^k~,
    \label{max_to_gamma}
\end{equation}
meaning that 
\begin{equation}
     \gamma^k \geq\frac{(-1)^k\mathcal{C}_{k,i}(J_1,\dots,J_k)}{\mathcal{Q}^k}\geq0
    \label{C_bound}
\end{equation}
where we used the $\gamma$ defined in Eq.~\eqref{gamma_loose}, which assuming \textbf{B1} and \textbf{B2} satisfies $\gamma\ll 1$. Moreover, since all $\mathcal{C}_{k,i}(J_1,\dots,J_k)$ within the same $F_{J_1,\dots,J_k}$ have the same sign, it follows from Eq.~\eqref{C_bound} that
\begin{equation}
    \mathfrak{C}(k)\gamma^k\geq \abs{F_{J_1,\dots,J_k}}\geq 0 ~;
\end{equation}
therefore, following Eq.~\eqref{p as sigmas}, and defining 
\begin{equation}
    \mathfrak{L}(n;p)=\begin{dcases}
        \sum_{k~\textrm{even}}^n{n \choose k}\mathfrak{C}(k)\gamma^k~&\textrm{if}~p=+\\
        \sum_{k~\textrm{odd}}^n{n \choose k}\mathfrak{C}(k)\gamma^k~&\textrm{if}~p=-
    \end{dcases}
\end{equation}
we obtain the loose upper and lower bounds
\begin{equation}
    \frac{1+\mathfrak{L}(n;+)}{1-\mathfrak{L}(n-1;-)} \geq\frac{\mathcal{P}_n}{q}\geq \frac{1-\mathfrak{L}(n;-)}{1+\mathfrak{L}(n-1;+)} ~.
\end{equation}
Note that the above expression is independent of the specific intervals $(N_1,\dots,N_{n-1},L)$ for which the observations giving the outcomes $1$ happened. As mentioned in the main text, as for large $k$ we have~\cite{OEIS22}
\begin{equation}
    \mathfrak{C}(k)\sim \frac{(2k-1)!!}{\sqrt{e}}~,
    \label{c(k) growth}
\end{equation}
the above bounds become meaningless for large $n$. In fact, the highest value of $n$ for which the bounds are valid is
\begin{equation}
    \mathcal{N}(q,\gamma)=\min\left\lbrace \mathcal{N}_1,\mathcal{N}_2,\mathcal{N}_3 \right\rbrace~.
\end{equation}
where $\mathcal{N}_1$, $\mathcal{N}_2$ and $\mathcal{N}_3$ are the largest numbers such that
\begin{equation}
\begin{dcases}
    1-\mathfrak{L}(\mathcal{N}_1;-)>0\\
    1-\mathfrak{L}(\mathcal{N}_2-1;-)>0
\end{dcases}
\end{equation}
and
\begin{equation}
    q\left[\frac{1+\mathfrak{L}(\mathcal{N}_3;+)}{1-\mathfrak{L}(\mathcal{N}_3-1;-)}\right]<1~.
\end{equation}
Then, for $\mathbf{n}$ sufficiently smaller than $\mathcal{N}(q,\gamma)$ we have
\begin{equation}
    \mathfrak{L}(n;\pm)\ll 1~,~\forall n\leq \mathbf{n}~,
    \label{small_M}
\end{equation}
and hence there exist $\epsilon_n,\delta_n\ll 1$ such that
\begin{equation}
     q(1+\epsilon_n)\geq\mathcal{P}_n(L;N_1,\dots,N_{n-1})\geq q(1-\delta_n)~,
    \label{app:useful_bound}
\end{equation}
independently of the specific $(N_1,\dots,N_{n-1},L),~\forall n\leq \mathbf{n}$.

\section{Explicit calculation of tight  and loose bounds for $\mathcal{P}_2$ and $\mathcal{P}_3$}
\label{app:P2&3}
In this appendix, we study the probability of observing the detector in the excited state during the $L$-th interval, given that it was previously found in the excited state in the $N_1$-th one (part 1) and in both the $N_1$-th and $N_2$-th ones (part 2). In other words, we study the quantities
\begin{equation}
    P(L|N_1)=\frac{\lambda^6}{q}\int_{N_1L} \mathcal{W}_4(X_2,X'_2,X_1,X'_1)~,
    \label{app:p_2}
\end{equation}
and
\begin{equation}
    P(L|N_1,N_2)=\frac{\lambda^6}{q P(L|N_1)}\int_{N_1N_2L} \mathcal{W}_6(X_3,X'_3,X_2,X'_2,X_1,X'_1)~,
    \label{app:p_3}
\end{equation}
where we used the shorthand notation defined in Eq.~\eqref{int} and Eq.~\eqref{p_N}. Throughout this appendix, hypotheses \textbf{A0}-\textbf{2} and \textbf{B1}-\textbf{2} are always assumed to hold. Moreover, we recall from Eq.~\eqref{F_repeat} that the first transition probability equals $q$ regardless of the interval in which it occurs. 

\subsection{Second transition}
By expanding the $4$-points Wightman function via the Wick's theorem we get
\begin{equation}
    \mathcal{N}(2)=\frac{(2\times 2)!}{2^2\times 2!}=3
\end{equation}
terms. In fact, these were already found in Eq.~\eqref{p_2_Wick}, which resulted in Eq.~\eqref{p_2_with_defs}, i.e. 
\begin{equation}
    \mathcal{P}_2=q\left(1+\frac{\mathcal{C}_2(N_1,L)}{\mathcal{Q}^2}\right)~,
\end{equation}
where 
\begin{equation}
    \mathcal{C}_2(N_1,L)=\mathcal{C}_{2,1}(N_1,L)+\mathcal{C}_{2,2}(N_1,L)~.
\end{equation}
While we already know that in this case tight  and loose bounds are equal,  because $\gamma_{ij}=\gamma$ for all $i,j$, let us proceed with the procedure defined in App.~\ref{app:F}: doing so will clarify the strategy to be applied in more complicated cases. The only restricted partition of $2$ is represented by 
\begin{equation}
    \ydiagram{2}
\end{equation}
describing
\begin{equation}
    N_2=(4-2)!!=2
\end{equation}
integrals, namely $\mathcal{C}_{2,1}(N_1,L)$ and $\mathcal{C}_{2,2}(N_1,L)$. Performing the substitution \eqref{app:products}, both integrals are mapped into $(\mathfrak{I}_{N_1,L})^2$. This can also be seen by means of the Young tableau~\footnote{The difference between the names Young diagram and tableau is only a matter of whether the boxes are empty or filled with labels.}
\begin{equation}
    \begin{ytableau}
            N_1&L
    \end{ytableau}
\end{equation}
for which the only cyclic permutation is $\mathfrak{I}_{N_1,L}\mathfrak{I}_{L,N_1}=(\mathfrak{I}_{N_1,L})^2$. Finally, using Eq.~\eqref{app:formal_ineq} we get
\begin{equation}
    0\leq\mathcal{C}_2(N_1,L)\leq2\mathcal{Q}^2\gamma_{N_1,L}^2=2\mathcal{Q}^2\gamma^2~;
\end{equation}
when read on $P(L|N_1)$, the above inequality gives
\begin{equation}
    q\leq P(L|N_1) \leq q(1+2\gamma^2)~.
\end{equation}

\subsection{Third transition}
For the sake of simplicity, let us call $L=N_3$. By expanding the $6$-points Wightman function via the Wick's theorem we get
\begin{equation}
    \mathcal{N}(3)=\frac{(2\times 3)!}{2^3\times 3!}=15
\end{equation}
term; these can be written in a compact fashion by expressing Eq.~\eqref{app:p_3} as
\begin{equation}
    P(N_3|N_1,N_2)=\mathcal{Q}^3+\mathcal{Q}\sum_{j=1}^{2}\sum_{j'>j}\mathcal{C}_2(N_j,N_{j'})+\mathcal{C}_3(N_1,N_2,N_3)~,
    \label{app:expanded_p3}
\end{equation}
where 
\begin{equation}
    \mathcal{C}_{3}(J_1,J_2,J_3)=\sum_{i=1}^{\mathfrak{C}(3)}\mathcal{C}_{3,i}(J_1,J_2,J_3)~,
\end{equation}
and where $\mathfrak{C}(3)=8$. Therefore, we can rewrite $P(N_3|N_1,N_2)$ as
\begin{equation}
    P(N_3|N_1,N_2)=q\left(\frac{1+\sum_{j=1}^{2}\left(\sum_{j'>j}F_{N_jN_j'}\right)+F_{N_1N_2N_3}}{1+F_{N_1N_2}}\right)~,
\end{equation}
where the $F_{J_1,\dots,J_n}$ are those defined in Eq.~\eqref{Fn}. From the above section, we know that
\begin{equation}
    0\leq F_{ij}\leq 2\gamma_{ij}^2~.
    \label{app:Fij}
\end{equation}
To bound $F_{N_1N_2N_3}$, we write its Young diagram
\begin{equation}
    \ydiagram{3}~,
\end{equation}
and fill it with the labels of the integrated intervals to get the corresponding Young tableau. From the latter, we extract the only possible cyclic permutation
\begin{equation}
    \begin{ytableau}
            N_1&N_2&N_3
    \end{ytableau}~\longrightarrow~\mathfrak{I}_{N_1N_2}\mathfrak{I}_{N_2N_3}\mathfrak{I}_{N_3N_1}~.
\end{equation}
It is worth noting that under the substitution \eqref{app:products} all eight terms of $\mathcal{C}_3(N_1,N_2,N_3)$ are denoted by the above expression. Using Eq.~\eqref{app:formal_ineq} gives
\begin{equation}
    8\gamma_{N_1N_2}\gamma_{N_2N_3}\gamma_{N_3N_1} \leq F_{N_1N_2N_3}\leq 0~,
\end{equation}
which, when put in Eq.~\eqref{app:expanded_p3} together with Eq.~\eqref{app:Fij} (and substituting back $L$), gives the tight  bound
\begin{equation}
    q\frac{1-8\abs{\gamma_{N_1N_2}\gamma_{N_2L}\gamma_{N_1L}}}{1+2\gamma_{N_1N_2}^2}\leq  P(L|N_1,N_2)\leq q(1+2\gamma_{N_1N_2}^2+2\gamma_{N_2L}^2+2\gamma_{N_1L}^2)~.
\end{equation}
Finally, we can obtain its looser version by simply noting that
\begin{equation}
    \gamma=\gamma_{01}\geq \gamma_{ij}~,~~\forall i,j~.
\end{equation}
Therefore, the loose bound is
\begin{equation}
    q\frac{1-8\abs{\gamma}^3}{1+2\gamma^2}\leq P_3 \leq q(1+6\gamma^2)~.
\end{equation}
Once again, we stress that the loose bound is independent from the specific intervals during which the excited outcomes are obtained (as expected).

\section{Bounds for $F_{J_iJ_jJ_kJ_l}$ from the restricted Young diagrams of $n=4$}
\label{app:F4}
In this appendix, we follow the steps prescribed in App.~\ref{app:F} to evaluate and bound $F_{J_iJ_jJ_kJ_l}$, for any choice of intervals $J_i,J_j,J_k,J_l$. The Young diagrams describing the partitions of $4$ restricted to not contain $1$ are \begin{equation}
\begin{split}
    4=\scriptsize{\ydiagram{0+4}}~
\end{split}
    \label{app:young:4}
\end{equation}
and
\begin{equation}
\begin{split}
    2+2=\scriptsize{\ydiagram{0+2,0+2}}~,
\end{split}
    \label{app:young:2+2}
\end{equation}
respectively accounting for
\begin{equation}
    N_4=6!!=48
\end{equation}
and
\begin{equation}
    N_{2+2}=(2!!)^2\times{4 \choose 2}\times\frac{1}{2!}=12
\end{equation}
II. The procedure for obtaining analytical expressions for all II is the following:
\begin{itemize}
\item for the  $48$ II over the four intervals labelled by $\{J_1,J_2,J_3,J_4\}$ corresponding to the diagram \eqref{app:young:4}, let us start by taking any integration variable in any interval to begin the procedure, say $u_1$ in $J_1$, and select one out of the other possible six points belonging to the other three intervals, say $u_2$ in $J_2$. This corresponds to pick the first and third points of
    \begin{equation}
        \begin{tikzpicture}[baseline={([yshift=-0.5ex]current bounding box.center)},vertex/.style={anchor=base,
        circle,fill=black!25,minimum size=18pt,inner sep=2pt},scale=0.6]
        \draw (0,0) -- (7.5,0);
        \fill (0.5,0) circle[radius=2pt]  node[anchor=north] {$~~J_1$};
        \fill (1,0) circle[radius=2pt]  node[anchor=north] {};
        \fill (2.5,0) circle[radius=2pt]  node[anchor=north] {$~~J_2$};
        \fill (3,0) circle[radius=2pt]  node[anchor=north] {};
        \fill (4.5,0) circle[radius=2pt]  node[anchor=north] {$~~J_3$};
        \fill (5,0) circle[radius=2pt]  node[anchor=north] {};
        \fill (6.5,0) circle[radius=2pt]  node[anchor=north] {$~~J_4$};
        \fill (7,0) circle[radius=2pt]  node[anchor=north] {};
        \end{tikzpicture}
    \end{equation}
    and connect them with an arc, i.e.
    \begin{equation}
        \begin{tikzpicture}[baseline={([yshift=-0.5ex]current bounding box.center)},vertex/.style={anchor=base,
        circle,fill=black!25,minimum size=18pt,inner sep=2pt},scale=0.6]
        \draw (0,0) -- (7.5,0);
        \draw (0.5,0) arc (180:0:1cm);
        \fill (0.5,0) circle[radius=2pt]  node[anchor=north] {$~~J_1$};
        \fill (1,0) circle[radius=2pt]  node[anchor=north] {};
        \fill (2.5,0) circle[radius=2pt]  node[anchor=north] {$~~J_2$};
        \fill (3,0) circle[radius=2pt]  node[anchor=north] {};
        \fill (4.5,0) circle[radius=2pt]  node[anchor=north] {$~~J_3$};
        \fill (5,0) circle[radius=2pt]  node[anchor=north] {};
        \fill (6.5,0) circle[radius=2pt]  node[anchor=north] {$~~J_4$};
        \fill (7,0) circle[radius=2pt]  node[anchor=north] {};
        \end{tikzpicture}=\mathcal{W}(u_1-u_2)\times (3~\textrm{more}~\mathcal{W}~\mathrm{functions})~,\label{example:app:first:step}~.
    \end{equation}
    Once the first point is chosen (in our example, $u_1$), there are six possible points to pick, hence giving six possible expressions of the form \eqref{example:app:first:step}. Then, we  select one point in one of the intervals we did not already pick (in our example, $J_3$ or $J_4$), and connect it with the free point in the interval we choose at the step before (for us, $J_2$); for example, starting from \eqref{example:app:first:step} and picking pick the point $u_3$ we get
    \begin{equation}
    \begin{tikzpicture}[baseline={([yshift=-0.5ex]current bounding box.center)},vertex/.style={anchor=base,
        circle,fill=black!25,minimum size=18pt,inner sep=2pt},scale=0.6]
        \draw (0,0) -- (7.5,0);
        \draw (0.5,0) arc (180:0:1cm);
        \draw (3,0) arc (180:0:0.75cm);
        \fill (0.5,0) circle[radius=2pt]  node[anchor=north] {$~~J_1$};
        \fill (1,0) circle[radius=2pt]  node[anchor=north] {};
        \fill (2.5,0) circle[radius=2pt]  node[anchor=north] {$~~J_2$};
        \fill (3,0) circle[radius=2pt]  node[anchor=north] {};
        \fill (4.5,0) circle[radius=2pt]  node[anchor=north] {$~~J_3$};
        \fill (5,0) circle[radius=2pt]  node[anchor=north] {};
        \fill (6.5,0) circle[radius=2pt]  node[anchor=north] {$~~J_4$};
        \fill (7,0) circle[radius=2pt]  node[anchor=north] {};
        \end{tikzpicture}=\mathcal{W}(u_1-u_2)\mathcal{W}(u_2-s_2-u_3)\times (2~\textrm{more}~\mathcal{W}~\mathrm{functions})~.
    \end{equation}
    Clearly, as we have four options to do so this steps gives four different expressions for each of the six we already had, for a total of $4\times6=24$. Next, we only have to pick one of the two points in the remaining interval; continuing the example, if we select the point $u_4$ we get
    \begin{equation}
    \begin{tikzpicture}[baseline={([yshift=-0.5ex]current bounding box.center)},vertex/.style={anchor=base,
        circle,fill=black!25,minimum size=18pt,inner sep=2pt},scale=0.6]
        \draw (0,0) -- (7.5,0);
        \draw (0.5,0) arc (180:0:1cm);
        \draw (3,0) arc (180:0:0.75cm);
        \draw (5,0) arc (180:0:0.75cm);
        \fill (0.5,0) circle[radius=2pt]  node[anchor=north] {$~~J_1$};
        \fill (1,0) circle[radius=2pt]  node[anchor=north] {};
        \fill (2.5,0) circle[radius=2pt]  node[anchor=north] {$~~J_2$};
        \fill (3,0) circle[radius=2pt]  node[anchor=north] {};
        \fill (4.5,0) circle[radius=2pt]  node[anchor=north] {$~~J_3$};
        \fill (5,0) circle[radius=2pt]  node[anchor=north] {};
        \fill (6.5,0) circle[radius=2pt]  node[anchor=north] {$~~J_4$};
        \fill (7,0) circle[radius=2pt]  node[anchor=north] {};
        \end{tikzpicture}=\mathcal{W}(u_1-u_2)\mathcal{W}(u_2-s_2-u_3)\mathcal{W}(u_3-s_3-u_4)\times (\mathrm{one~more}~\mathcal{W})~.
    \end{equation}
    This steps gives two expressions for each of the $24$ of above, for a total of $48$. Finally, there is no other choice left other than inserting the Wightman function evaluated between the two remaining points; in our example these are $u_4-s_4$ and $u_1-s_1$, hence giving
    \begin{equation}
        \begin{tikzpicture}[baseline={([yshift=-0.5ex]current bounding box.center)},vertex/.style={anchor=base,
        circle,fill=black!25,minimum size=18pt,inner sep=2pt},scale=0.6]
        \draw (0,0) -- (7.5,0);
        \draw (0.5,0) arc (180:0:1cm);
        \draw (3,0) arc (180:0:0.75cm);
        \draw (5,0) arc (180:0:0.75cm);
        \draw (1,0) arc (180:90:0.5cm);
        \draw (7,0) arc (0:90:0.5cm);
        \draw (1.5,0.5) -- (6.5,0.5);
        \fill (0.5,0) circle[radius=2pt]  node[anchor=north] {$~~J_1$};
        \fill (1,0) circle[radius=2pt]  node[anchor=north] {};
        \fill (2.5,0) circle[radius=2pt]  node[anchor=north] {$~~J_2$};
        \fill (3,0) circle[radius=2pt]  node[anchor=north] {};
        \fill (4.5,0) circle[radius=2pt]  node[anchor=north] {$~~J_3$};
        \fill (5,0) circle[radius=2pt]  node[anchor=north] {};
        \fill (6.5,0) circle[radius=2pt]  node[anchor=north] {$~~J_4$};
        \fill (7,0) circle[radius=2pt]  node[anchor=north] {};
        \end{tikzpicture}=\mathcal{W}(u_1-u_2)\mathcal{W}(u_2-s_2-u_3)\mathcal{W}(u_3-s_3-u_4)\mathcal{W}(u_4-s_4-u_1+s_1)~.
    \end{equation}
    as there is no freedom in the last step, this do not increase the number of expressions. Finally, all expressions obtained in this way must be integrated as prescribed by Eq.~\eqref{p_N}. In our example this gives
    \begin{equation}
    \begin{split}
        \int_{J_1,J_2,J_3,J_4}&\begin{tikzpicture}[baseline={([yshift=-0.5ex]current bounding box.center)},vertex/.style={anchor=base,
        circle,fill=black!25,minimum size=18pt,inner sep=2pt},scale=0.6]
        \draw (0,0) -- (7.5,0);
        \draw (0.5,0) arc (180:0:1cm);
        \draw (3,0) arc (180:0:0.75cm);
        \draw (5,0) arc (180:0:0.75cm);
        \draw (1,0) arc (180:90:0.5cm);
        \draw (7,0) arc (0:90:0.5cm);
        \draw (1.5,0.5) -- (6.5,0.5);
        \fill (0.5,0) circle[radius=2pt]  node[anchor=north] {$~~J_1$};
        \fill (1,0) circle[radius=2pt]  node[anchor=north] {};
        \fill (2.5,0) circle[radius=2pt]  node[anchor=north] {$~~J_2$};
        \fill (3,0) circle[radius=2pt]  node[anchor=north] {};
        \fill (4.5,0) circle[radius=2pt]  node[anchor=north] {$~~J_3$};
        \fill (5,0) circle[radius=2pt]  node[anchor=north] {};
        \fill (6.5,0) circle[radius=2pt]  node[anchor=north] {$~~J_4$};
        \fill (7,0) circle[radius=2pt]  node[anchor=north] {};
        \end{tikzpicture}=\\
        &=\int_{J_1,J_2,J_3,J_4}\mathcal{W}(u_1-u_2)\mathcal{W}(u_2-s_2-u_3)\mathcal{W}(u_3-s_3-u_4)\mathcal{W}(u_4-s_4-u_1+s_1)~.
    \end{split}
    \end{equation}
\item for the  $12$ II over two pairs of intervals labelled by corresponding to the diagram \eqref{app:young:2+2}, given that any row of the Young diagram having two blocks gives an integral that can be factorized into two II, the procedure to build the integrals is as follows. First, we fill the first row of the diagram with all inequivalent choices of two intervals' labels, i.e.
\begin{equation}
    \begin{ytableau}
            J_1&J_2\\
            &
    \end{ytableau}~,~
    \begin{ytableau}
            J_1&J_3\\
            &
    \end{ytableau}~,~
    \begin{ytableau}
            J_1&J_4\\
            &
    \end{ytableau}~,~
    \begin{ytableau}
            J_2&J_3\\
            &
    \end{ytableau}~,~
    \begin{ytableau}
            J_2&J_4\\
            &
    \end{ytableau}~,~
    \begin{ytableau}
            J_3&J_4\\
            &
    \end{ytableau}
\end{equation}
out of the possible four, and write their $2!!$ IIs, e.g.
\begin{equation}
    \begin{ytableau}
            J_1&J_2\\
            &
    \end{ytableau}\longrightarrow\mathcal{C}_2(J_1,J_2)~.
\end{equation}
Next, for all Young diagram we fill the remaining boxes with the left intervals' labels, and multiply the corresponding II to the $2!!$ new ones, i.e.
\begin{equation}
    \begin{ytableau}
            J_1&J_2\\
            J_3&J_4
    \end{ytableau}~,~
    \begin{ytableau}
            J_1&J_3\\
            J_2&J_4
    \end{ytableau}~,~
    \begin{ytableau}
            J_1&J_4\\
            J_2&J_3
    \end{ytableau}~,~
    \begin{ytableau}
            J_2&J_3\\
            J_1&J_4
    \end{ytableau}~,~
    \begin{ytableau}
            J_2&J_4\\
            J_1&J_3
    \end{ytableau}~,~
    \begin{ytableau}
            J_3&J_4\\
            J_1&J_2
    \end{ytableau}
\end{equation}
and
\begin{equation}
    \begin{ytableau}
            J_1&J_2\\
            J_3&J_4
    \end{ytableau}~=~\mathcal{C}_2(J_1,J_2)\mathcal{C}_2(J_3,J_4)~.
\end{equation}
Finally, we keep only one copy of each identical Young tableau (where two tableau are said to be identical if they become equal by permuting their rows), hence obtaining
\begin{equation}
    \begin{ytableau}
            J_1&J_2\\
            J_3&J_4
    \end{ytableau}~=~\mathcal{C}_2(J_1,J_2)\mathcal{C}_2(J_3,J_4)~,~
    \begin{ytableau}
            J_1&J_3\\
            J_2&J_4
    \end{ytableau}~=~\mathcal{C}_2(J_1,J_3)\mathcal{C}_2(J_2,J_4)~,~
    \begin{ytableau}
            J_1&J_4\\
            J_2&J_3
    \end{ytableau}~=~\mathcal{C}_2(J_1,J_4)\mathcal{C}_2(J_2,J_3)~;
    \label{app:YT2}
\end{equation}
this removal corresponds to the multiplication by $1/2$ of Eq.~\eqref{app:N_P_k}, accounting for double counting. As expected, this procedure gives
    \begin{equation}
    \frac{1}{2}{4 \choose 2}\times 2!!\times 2!!=12
    \end{equation}
    inequivalent products of IIs. Finally, the procedure to explicitly write these II follows the one presented in the above case giving, for example,
    \begin{equation}
    \mathcal{C}_2(J_1,J_2)=\int_{J_1,J_2}\mathcal{W}(u_1-u_2)\mathcal{W}(u_1-s_1-u_2+s_2)+\int_{J_1,J_2}\mathcal{W}(u_1-s_1-u_2)\mathcal{W}(u_2-s_2-u_1)~.
    \end{equation}
\end{itemize}
Finally, it is
\begin{equation}
\begin{split}
     F_{J_i,J_j,J_k,J_l}&=\frac{\mathcal{C}_4(J_i,J_j,J_k,J_l)}{\mathcal{Q}^4}=\\
    &=\frac{1}{\mathcal{Q}^4}\biggl(48~\textrm{IIs~coming~from~}\scalebox{0.7}{\ydiagram{0+4}}~\biggr)~+\frac{1}{\mathcal{Q}^4}\biggl(12~\textrm{IIs~coming~from~}\scalebox{0.7}{\ydiagram{0+2,0+2}}~\biggr)~.
\end{split}
\label{F_4}
\end{equation}
As discussed in App.~\ref{app:F}, the advantage of expressing $F_{J_i,J_j,J_k,J_l}$ explicitly in terms of Young diagrams is that it directly allows one to find a tight bound for it. Indeed, once the substitution \eqref{app:substitution} is performed all integrals in Eq.~\eqref{F_4} give place to products of $\mathfrak{I}_{ij}$ that are much easier to evaluate. In particular, from the free cyclic permutations of four intervals (i.e.  $P^{c}_4=3$) we get
\begin{equation}
    \mathfrak{I}_{ij}\mathfrak{I}_{jl}\mathfrak{I}_{lk}\mathfrak{I}_{ki},~\mathfrak{I}_{il}\mathfrak{I}_{lk}\mathfrak{I}_{kj}\mathfrak{I}_{ji},~\mathfrak{I}_{ik}\mathfrak{I}_{kj}\mathfrak{I}_{jl}\mathfrak{I}_{li}~,
\end{equation}
and from the ${4 \choose 2 }\frac{1}{2!}=3$ inequivalent combinations of the products in Eq.~\eqref{app:YT2} we get
\begin{equation}
    \mathfrak{I}_{ij}^2 \mathfrak{I}_{kl}^2,~ \mathfrak{I}_{ik}^2\mathfrak{I}_{jl}^2,~\mathfrak{I}_{il}^2\mathfrak{I}_{jk}^2~.
\end{equation}
Hence, using Eq.~\eqref{app:formal_ineq} it is easy to show that $F_{J_i,J_j,J_k,J_l}$ is upper bounded by
\begin{equation}
\begin{split}
    F_{J_i,J_j,J_k,J_l}&\leq \frac{ N_{4}}{P^{c}_4}\left(\gamma_{ij}\gamma_{jl}\gamma_{lk}\gamma_{ki}+\gamma_{il}\gamma_{lk}\gamma_{kj}\gamma_{ji}+\gamma_{ik}\gamma_{kj}\gamma_{jl}\gamma_{li}\right) + \frac{ N_{2+2}}{{4 \choose 2}\times\frac{1}{2!}}\left(\gamma_{ij}^2 \gamma_{kl}^2 + \gamma_{ik}^2\gamma_{jl}^2+\gamma_{il}^2\gamma_{jk}^2\right)=\\
    &=\underbrace{16\left(\gamma_{ij}\gamma_{jl}\gamma_{lk}\gamma_{ki}+\gamma_{il}\gamma_{lk}\gamma_{kj}\gamma_{ji}+\gamma_{ik}\gamma_{kj}\gamma_{jl}\gamma_{li}\right)}_{N_4=48~\textrm{terms}} + \underbrace{4\left(\gamma_{ij}^2 \gamma_{kl}^2 + \gamma_{ik}^2\gamma_{jl}^2+\gamma_{il}^2\gamma_{jk}^2\right)}_{N_{2+2}=12~\textrm{terms}}~;
\end{split}
\label{F4_bounds}
\end{equation}
and lower bounded by 0. Since $\gamma_{ij}\leq\gamma$ $\forall i,j$, this is a tighter bound than the one obtained by the strategy provided in the main body of the article, i.e. 
\begin{equation}
    F_{J_i,J_j,J_k,J_l}\leq60\gamma^4~.
\end{equation}

Finally, assuming that we calculated $F_{J_iJ_j}$ and $F_{J_iJ_jJ_k}$ in a way similar to the above (see App.~\ref{app:P2&3}), and obtained tight bounds for them, we can obtain a most stringent bound for $P(L|J_1,J_2,J_3)$. By calling $L=J_4$, from Eq.s~(\ref{sigma definition}) and (\ref{p as sigmas}) it is
\begin{equation}
    P(J_4|J_1,J_2,J_3)=q\left[\frac{1+\sum_{j=1}^4\sum_{j'>j}F_{J_jJ_{j'}}+\sum_{j=1}^4\sum_{j'>j}\sum_{j''>j'}F_{J_jJ_{j'}J_{j''}}+F_{J_1J_2J_3J_4}}{1+F_{J_1J_2}+F_{J_1J_3}+F_{J_2J_3}+F_{J_1J_2J_3}}\right]~.
\end{equation}
Hence, recalling that any $F_{i_1,\dots,i_{2n}}$ is positive and any $F_{i_1,\dots,i_{2n+1}}$ is negative, we can lower bound $P(J_4|J_1,J_2,J_3)$ by
\begin{equation}
   q\left[\frac{1+F_{J_1J_2J_3}+F_{J_1J_2J_4}+F_{J_1J_3J_4}+F_{J_2J_3J_4}}{1+F_{J_1J_2}+F_{J_1J_3}+F_{J_2J_3}}\right]~,
\end{equation}
and upper bound it with
\begin{equation}
    q\left[\frac{1+F_{J_1J_2}+F_{J_1J_3}+F_{J_1J_4}+F_{J_2J_3}+F_{J_2J_4}+F_{J_3J_4}+F_{J_1J_2J_3J_4}}{1+F_{J_1J_2J_3}}\right]~.
\end{equation}
Moreover, by using the ratio defined in Eq.~\eqref{gamma_loose}, we get the realization-independent bounds
\begin{equation}
\begin{dcases}
    0\leq F_{J_iJ_j}\leq 2\gamma^2\\
    8\gamma^3\leq F_{J_iJ_jJ_k}\leq0\\
    0\leq F_{J_iJ_jJ_kJ_l}\leq 60\gamma^4
\end{dcases}
\end{equation}
which, when read on $P(J_4|J_1,J_2,J_3)$, give
\begin{equation}
    \frac{q}{1+6\gamma^2}\leq P_4\leq q\frac{1+12\gamma^2+32\gamma^3+60\gamma^4}{1-8\abs{\gamma}^3}~,
\end{equation}
regardless of the specific $L=J_4$ and $(J_1,J_2,J_3)$ considered.
\end{document}